\documentclass[aps,reprint,superscriptaddress,eqsecnum,nofootinbib,amsmath,amssymb,prd,floatfix]{revtex4-2}

\usepackage[utf8]{inputenc}
\usepackage[T1]{fontenc}
\usepackage[dvipsnames]{xcolor}
\usepackage{graphicx}
\usepackage[colorlinks=true,allcolors=blue]{hyperref}
\usepackage[normalem]{ulem}

\newcommand{\ud}{\mathrm{d}}
\newcommand{\acc}{\mathrm{acc}}

\newcommand{\cut}{\mathrm{cut}}
\newcommand{\DF}{\mathrm{DF}}
\newcommand{\DM}{\mathrm{DM}}

\newcommand{\enc}{\mathrm{enc}}

\newcommand{\I}{\mathrm{i}}
\newcommand{\IN}{\mathrm{in}}
\newcommand{\MIN}{\mathrm{min}}
\newcommand{\MAX}{\mathrm{max}}
\newcommand{\ISCO}{\mathrm{ISCO}}

\newcommand{\SP}{\mathrm{sp}}

\newcommand{\tot}{\mathrm{tot}}
\newcommand{\fy}{\mathrm{2,4y}}
\newcommand{\radsch}{R_\mathrm{s}}
\newcommand{\Msolar}{\mathrm M_\odot}
\newcommand{\Jmin}{\mathcal J_\MIN}

\newcommand{\UVA}{Department of Physics, University of Virginia, P.O.~Box 400714, Charlottesville, Virginia 22904-7414, USA}

\begin{document}

\title{Intermediate mass-ratio inspirals in a dense dark-matter environment:\\ Effects of the initial dark-matter distribution}

\author{Benjamin A.\ Wade}
\email{baw8td@virginia.edu}
\affiliation{\UVA}%

\author{David A.\ Nichols}%
\email{david.nichols@virginia.edu}
\affiliation{\UVA}

\date{\today}

\begin{abstract}
Recent work has shown the possibility of detecting dense dark-matter distributions surrounding intermediate or extreme mass-ratio inspirals through gravitational waves using LISA. 
Modeling these systems requires evolving the coupled dynamics of the binary and the dark matter. 
This also requires setting reasonable initial conditions for the dark-matter distribution, which itself relies upon understanding the formation history of these systems.
In this paper, we investigate how two aspects of these systems' formation histories shape the dark-matter distribution: accretion onto the primary and prior merger events. 
We model accretion by introducing a minimum allowed angular momentum of dark-matter particles, which removes such particles that would have been accreted by the primary.
When simulating an inspiral within such a distribution, we find a smaller dephasing of the gravitational-wave signal from a vacuum binary as compared to an inspiral without such a cutoff, particularly for more extreme mass-ratios. 
We also simulate an inspiral which takes place within a dark-matter distribution that remains after a prior merger.
We find that the decrease in dephasing from vacuum binaries when compared to the prior inspiral is most significant for less extreme mass-ratios. 
Nevertheless, the environmental effects from the dark matter for these different cases of initial data are still expected to be measurable by future space-based detectors.
\end{abstract}

\maketitle

\tableofcontents

\section{Introduction}

Observations over a wide range of astrophysical length and time scales, from the cosmic microwave background in the early universe to the rotation curves of individual nearby galaxies in the present era all suggest that there must be an abundance of cold, weakly-interacting matter known as \textit{dark matter} (see, e.g.,~\cite{Bertone:2004pz,Garrett:2010hd,Bertone:2016nfn}).
The nature of the fields, objects or particles that compose dark matter (DM) remains unknown, and there are viable dark-matter candidates that range in masses from ultralight particles with galactic-scale Compton wavelengths to primordial black holes~\cite{Bertone:2018krk}. 
This is the state of the field despite a long period of dedicated research efforts to detect the underlying dark-matter particle through its production at colliders, via direct-detection experiments, and in a variety of indirect astrophysical and cosmological searches over many scales~\cite{Feng:2010gw}.
The lack of a detection has motivated exploring ``far and wide'' for new detection techniques and dark-matter models~\cite{Bertone:2018krk}.

Following the detection of gravitational waves in 2015~\cite{LIGOScientific:2016aoc} by the LIGO-Virgo Collaboration and the rapid growth in the number of gravitational-wave (GW) detections~\cite{LIGOScientific:2018mvr,LIGOScientific:2020ibl,KAGRA:2021vkt}, the possibility of learning about dark matter through GW observations began to be investigated more vigorously~\cite{Bertone:2019irm}. 
With new gravitational-wave bands at lower frequencies opening with the pulsar timing arrays nearing a confident detection of a stochastic GW background~\cite{NANOGrav:2023gor,EPTA:2023fyk,Reardon:2023gzh,Xu:2023wog} and with space-based interferometers preparing to operate in the next decade~\cite{Amaro-Seoane:2017ADS,Baker:2019nia, Hu:2017,TianQin:2015yph}, there is an increasing discovery space not only for GWs, but also potentially for dark matter.

\subsection{Dark-matter effects on the gravitational-wave emission from black-hole binaries}

For astrophysical sources of gravitational waves to be detectable by existing or future GW detectors, the sources typically need to be compact objects---white dwarfs, neutron stars (NSs) or black holes (BHs)---moving at relativistic velocities.
For dark matter to have an impact on the emitted gravitational waves, it needs to be present in sufficiently high densities so that it (perturbatively) modifies the structure of or spacetime around a compact object (see, e.g.,~\cite{Cardoso:2022whc,Figueiredo:2023gas,Speeney:2024mas}) or leads to modifications of the equations of motion for the compact object from phenomena such as dynamical friction~\cite{Chandrasekhar1943a} or accretion~\cite{Unruh:1976fm} (see, e.g.,~\cite{Cardoso:2019rou}, or conversely~\cite{Barausse:2014tra} for results showing that lower dark-matter densities are unlikely to have any measurable effect).
Thus, a mechanism to obtain higher dark-matter densities around compact objects typically is a requirement to have observable GW signals that are influenced by the presence of dark matter.

In ultralight dark-matter models, for example, high densities of dark matter naturally arise through superradiance (see, e.g., the review~\cite{Brito:2015oca}).
For more massive dark-matter particles, such as those that are the target of many direct-detection experiments~\cite{Feng:2010gw}, other mechanisms are needed, such as adiabatic growth of a seed black hole within a dark-matter halo~\cite{Gondolo:1999ef}.
While Ref.~\cite{Gondolo:1999ef} also considered adiabatic growth for self-annihilating dark matter (which could produce electromagnetic or neutrino emission), the growth mechanism works for many types of particle dark matter, and, in fact, it can result in higher densities when there is not a mechanism for DM annihilation~\cite{Gondolo:1999ef}. 
The Newtonian calculations in~\cite{Gondolo:1999ef} were generalized to fully relativistic ones~\cite{Sadeghian:2013laa,Ferrer:2017xwm}, which determined even higher densities were possible.
Astrophysical processes, including galactic mergers, off-center growth in the DM halo, or stellar heating, however, can decrease the large densities formed through adiabatic growth~\cite{Ullio:2001fb,Bertone:2005hw}.
As discussed in~\cite{Eda:2013gg}, if the seed BH grew to an intermediate-mass black hole (IMBH) with a mass in the range $10^2$--$10^5\,\Msolar$, the IMBH is more likely to have a dense dark-matter ``spike'' (namely, a cuspy, DM distribution) than a heavier supermassive BH, which may have undergone a merger with another supermassive BH and disrupted its surrounding DM distribution in the process. 
Because of their lower masses, IMBHs are less likely to have undergone such a major merger that could have disrupted an adiabatically grown dark-matter spike around the IMBH.

An IMBH with a surrounding DM spike does not produce measurable GW emission; however, stellar-mass compact objects (specifically a NS or BH) that are gravitationally bound to the IMBH, can experience gravitational interactions with other astronomical bodies around the IMBH to form a tightly-bound binary called an intermediate mass-ratio inspiral (IMRI)\footnote{Throughout this paper, when we refer to IMRIs, we are discussing \emph{light} IMRIs with a stellar-mass secondary. There also are \emph{heavy} IMRIs formed from a binary composed of a supermassive BH and an IMBH, which we will not consider in this paper.} as described, e.g., in~\cite{LISA:2022yao}.
There are large uncertainties in the rate of formation of IMRIs, and the similar extreme mass-ratio inspirals (EMRIs) consisting of a stellar-mass compact object and a supermassive BH; however, there have been calculations that estimate the rate of IMRI mergers for an individual IMBH primary to be $10^{-8}$--$10^{-7}$\,yr$^{-1}$~\cite{Pestoni:2020alb} (see also~\cite{Fragione:2017blf}).
The gravitational waves emitted during the late inspiral of an IMRI will be in the few to hundreds of mHz range (depending on the IMBH mass), which is well within the frequency band of space-based interferometers, such as LISA~\cite{Amaro-Seoane:2017ADS}.
Estimates of the detection rate of IMRIs with LISA range from less than one to nearly one-hundred per year~\cite{Arca-Sedda:2020lso}.

To measure the presence of dark matter around an IMRI or EMRI, the dark matter must have a sufficiently large influence on the dynamics of the IMRI that the emitted GWs differ by a signal-to-noise-dependent number of cycles from an equivalent system without environmental effects.
Furthermore, this change or ``dephasing'' should not be mimicked by a vacuum system with slightly different mass, spins, or orbital parameters.
In the context of IMRIs with a surrounding ``dress'' of dark matter, there have been a number of works in the past several years showing that it is indeed possible to measure the effects of dark matter in these systems, and in many cases it is necessary to compute the feedback of the inspiral on the DM distribution coupled with the orbital evolution of the IMRI. 
We review this past work next.

The effects of a DM dress on IMRIs has largely considered three types of effects (similarly to what has been studied in the EMRI case~\cite{Barausse:2014tra}): changes in the orbital frequency from the DM mass enclosed within the secondary's orbit, dynamical friction, and accretion onto the secondary.\footnote{More recent work has also shown that a ``slingshot mechanism'' that arises from three-body interactions~\cite{Mukherjee:2023lzn} and a ``stirring'' effect \cite{Kavanagh:2024lgq} related to time-dependence of the potential for the IMRI could be important.
Although we will not investigate these effects in this paper, they both deserve further consideration.}
Reference~\cite{Eda:2013gg} first considered the enclosed-mass effect, which proved to be the smallest of the three for IMRIs.
Dynamical friction (DF) proved to be the largest effect~\cite{Eda:2014kra}, so much so that the calculations in~\cite{Eda:2014kra} that ignored DF feedback onto the DM spike significantly overestimated the amount of energy dissipated through dynamical friction and thus the amount of dephasing~\cite{Kavanagh:2020cfn}.
The importance of dynamical friction motivated additional studies of DF in these systems~\cite{Becker:2021ivq,Dosopoulou:2023umg,Berezhiani:2023vlo}, as well as studies that determined how well the DM distribution could be inferred from GW measurements with LISA~\cite{Coogan:2021uqv} and whether they could be distinguished from other ``environmental'' effects~\cite{Becker:2022wlo,Cole:2022fir}.
Accretion onto the secondary was also considered in~\cite{Yue:2017iwc} and~\cite{Nichols:2023ufs}.
The latter study showed that neglecting feedback from accretion also overestimated the amount of mass captured (and thus its effects on the emitted GWs.
Reference~\cite{Nichols:2023ufs} also computed feedback onto the DM distribution from this secondary accretion (SA) for IMRIs on circular orbits, which was subsequently generalized to eccentric orbits in~\cite{Karydas:2024fcn}.
The combined effects of dynamical friction and secondary accretion with feedback could lead to thousands of cycles of dephasing over an inspiral that encompassed hundreds of thousands to millions of cycles during a four-year LISA mission.

\subsection{Discussion of the effects of the initial dark-matter distribution}

In~\cite{Nichols:2023ufs}, accretion onto the primary was not considered, because during the inspiral, accretion would produce a relative effect on the emitted GWs that would be of order the mass enclosed within the secondary's orbit divided by the primary's mass, which was shown to be small in~\cite{Nichols:2023ufs}.
However, accretion onto the primary is important for determining the correct initial conditions for the DM distribution through which the secondary inspirals.
Both the Newtonian calculations of the adiabatically compressed DM distribution~\cite{Gondolo:1999ef} and the relativistic ones~\cite{Sadeghian:2013laa,Ferrer:2017xwm} take into account that there is a minimum angular momentum that a DM particle can have before it plunges into the primary and is accreted (the treatment is approximate in the Newtonian case and exact in the relativistic ones).
This creates a ``loss cone'' in the space of angular momentum and energy of the DM orbits; in position space, this produces a DM distribution that smoothly goes to zero at a minimum radius.
The simple power-law distribution used in~\cite{Eda:2014kra} or~\cite{Kavanagh:2020cfn} (and other works) with a sharp cutoff in position space at the minimum radius does not have this loss cone, and is not consistent with the capture of dark matter onto the primary.
Moreover, the value of the density close to the innermost stable circular orbit (ISCO) for the position-space cutoff tends to be significantly higher than those with the physical angular-momentum cutoff.
Given that the dynamical friction and accretion forces are both proportional to the density of DM at the location of the secondary, this could have a large impact on the dephasing induced by DM. 

While there have been studies of IMRIs in DM distributions that were fit to those with angular-momentum cutoffs~\cite{Speeney:2022ryg,Mitra:2025tag}, these simulations did not include feedback onto the DM distribution. 
They also did not explicitly compare how different initial dark-matter distributions with a cutoff in position space versus a cutoff in angular-momentum space affect the dynamics of the IMRI.
Thus, one main focus of this paper is to determine the relative amount of GW dephasing that arises in DM distributions with the more physical cutoff in angular-momentum space versus those with a cutoff in position space.
To do so, we first adapt the formalism for computing feedback to the DM distribution function from dynamical-friction and secondary-accretion to be consistent with an angular-momentum cutoff.
The way that we implemented this change also had an ancillary benefit of producing a large speed-up in the \textsc{HaloFeedback} code~\cite{HaloFeedback} that was used to perform the simulations.
We find that for IMRIs with less extreme mass ratios, the GW dephasing between the cases with a position-space cutoff and the angular-momentum cutoff is a small fraction of the total DM-induced dephasing.
This small change occurs because most of the dephasing accumulates at larger radii, where the two dark-matter distributions are very similar.
For more extreme mass ratios, the GW dephasing between the inspirals in the two DM distribution functions is comparable to the total DM-induced dephasing: namely, it can be thousands of GW cycles over a four-year LISA mission.

As different initial conditions can change the total GW dephasing by a multiplicative factor of a few, understanding the dependence of the GW phase for different initial DM distributions will be important for analyzing these IMRIs with a DM dress. 
A simple case study that we perform in this paper is computing the difference in the GW phase from a ``first-generation'' IMRI, in which the primary had not merged with a stellar-mass compact object in the past and a ``second-generation'' IMRI, in which one IMRI merger had preceded it.
Given the rate of mergers in the range $10^{-8}$--$10^{-7}$\,yr$^{-1}$ predicted in~\cite{Pestoni:2020alb}, and the fact that IMRIs will be measurable by LISA mostly at low redshifts (e.g.,~\cite{Coogan:2021uqv}), it is likely that at least one compact object will have merged with the IMBH prior to the inspiral that occurs during the LISA mission lifetime.
Given that the DM distribution was shown in~\cite{Kavanagh:2020cfn, Nichols:2023ufs, Karydas:2024fcn} to change nontrivially from feedback over the course of an inspiral, we also investigate the difference in dephasing between a first- and second-generation IMRI merger.
Here, the result is the opposite of that related to the comparison of the position-space and angular-momentum cutoffs: namely, there is a larger difference between first- and second-generation mergers for IMRIs with less extreme mass ratios than for more extreme mass ratios.
This occurs because the effects of feedback on the DM distribution are larger at the less extreme mass ratios than at more extreme ones.

\subsection{Organization of this paper}

The remainder of the paper is structured as follows. 
We review our notation and discuss the formalism that we use to evolve the coupled binary--dark-matter system in Sec.~\ref{sec:binary_DM_formalism}.
Specifically, Sec.~\ref{sec:binary_formalism} describes the evolution equations for the binary and Secs.~\ref{sec:DM_formalism}, \ref{sec:DF_genform} and \ref{sec:SA_genform} cover the dark-matter distribution (including the cutoff in the space of angular momentum and the impact of a generic angular-momentum cutoff on the the evolution equations for the dark-matter distribution with feedback from the IMRI). 
In Sec.~\ref{sec:cut_cases}, we consider two specific cutoffs in angular momentum: one with no cutoff (which recovers the earlier results in~\cite{Kavanagh:2020cfn,Nichols:2023ufs}), and one that matches features of a fully relativistic calculation when applied in the Newtonian context. 
We discuss in Sec.~\ref{sec:cut_result} simulations with both DF and SA feedback for the two cases of the angular-momentum cutoff.
Our focus is on the GW dephasing during the inspiral of the secondary in two different dark-matter densities with different cutoffs against vacuum binaries. 
Section~\ref{sec:successive_merger} contains a comparison of first- and second-generation mergers. 
Our conclusions are in Sec.~\ref{sec:conclusions}, and there are two Appendices~\ref{sec:DF_feedback_app} and \ref{app:efficiency}, which contain technical results related to the evaluation of the DF-feedback rates and the code efficiency.

While this work was in progress, a formation scenario for a dressed IMBH was put forward in~\cite{Bertone:2024wbn}. 
This work modeled the DM distribution generated when a supermassive star directly collapses to a black hole within an initial Navarro-Frenk-White~\cite{Navarro:1995iw} DM halo.
Reference~\cite{Bertone:2024wbn} also permitted the IMBH to undergo a slow adiabatic growth stage afterwards. 
The resulting distributions were consistent with having the angular-momentum cutoff, and the shape of the distribution differed sufficiently from that of single-power law DM spikes that these new DM structures were referred to as DM ``mounds'' in~\cite{Bertone:2024wbn}.
It would be interesting to consider the DM mounds as initial conditions for the DM distribution in future work.

\section{Evolution equations for the binary and dark matter} \label{sec:binary_DM_formalism}

Modeling an IMRI within a distribution of dark matter requires jointly evolving both the binary and the dark matter; the two are coupled and in many cases it is a poor approximation to evolve them independently. 
We first review the evolution equations for the IMRI in Sec.~\ref{sec:binary_formalism}.
The dark-matter distribution is described by its phase-space distribution function, which is reviewed in Sec.~\ref{sec:DM_formalism}.
The evolution of the distribution function is discussed in Secs.~\ref{sec:DF_genform} and~\ref{sec:SA_genform}.

\subsection{Review of the evolution equations for the binary} \label{sec:binary_formalism}

The evolution equations for the binary and relevant notation will be the same as those presented in~\cite{Nichols:2023ufs}.
As in~\cite{Nichols:2023ufs}, we model the inspiral as an adiabatic evolution between circular orbits driven by effective forces that decrease the orbital separation.
We summarize the results that we will need in this paper below.

We consider a binary consisting of an IMBH of mass $m_1$ (the ``primary'') in the center of a dark-matter halo, and a stellar-mass compact object of mass $m_2$ (the ``secondary'') orbiting on a quasicircular orbit at an orbital separation $r_2$. 
In this paper, we specialize to values of $m_1$ in the range $10^3$--$10^5\,\Msolar$, and we fix the initial mass $m_2$ to be $10\,\Msolar$.
We also introduce a radius $r_\fy$, which is the orbital separation, $r_2$, from which the binary will inspiral to the ISCO radius, $r_\ISCO$, in four years (in the absence of dark matter).
The values of $r_\fy$ are on the order $10^{-8}$--$10^{-7}$\,pc for $m_1$ in the range $10^3$--$10^5\,\Msolar$.

We assume a DM distribution that is centered around the primary, spherically symmetric, time dependent, and has a density denoted by $\rho_\DM(r,t)$.
The form of the evolution equations for the DM density are discussed in Secs.~\ref{sec:DM_formalism}, \ref{sec:DF_genform} and \ref{sec:SA_genform}.
We will use $m_\enc(r_2)$ to denote the amount of DM mass enclosed within the orbital radius $r_2$ for the DM density $\rho_\DM(r,t)$.
For a general $r$, the enclosed mass can be computed from the density as
\begin{equation} \label{eq:enc_mass}
    m_\enc(r) = \int^{r}_{r_\mathrm{min}} \ud^3 \mathbf r' \rho_\DM(r') \, ,
\end{equation}
where $r_\mathrm{min}$ is the smallest radius at which dark matter is present.
Given the typical orbital separations and DM densities that we consider in this paper, $m_\enc(r_2)$ will be less than one solar mass.

We find it to be useful to introduce two mass ratios: $q = m_2/m_1$ (the binary's mass ratio) and $q_\enc(r_2) = m_\enc(r_2)/m_1$.
Both $q$ and $q_\mathrm{enc}$ are small for the systems that we consider, and we will work to leading order in $q$ and $q_\mathrm{enc}$ in this paper.
For example, if we denote the total mass within $r_2$ by $m_\tot(r_2)$, then we will make the approximation that 
\begin{equation}
    m_\tot(r_2) = m_1 (1 + q + q_\mathrm{enc}) \approx m_1 , 
\end{equation}
as was done in~\cite{Nichols:2023ufs}.
Similarly, the velocity $v_2$ of the reduced mass $\mu \approx m_2$ on a circular orbit is given by Kepler's Law
\begin{equation}
    v_2 = \sqrt{\frac{G m_\tot(r_2)}{r_2}} \approx \sqrt{\frac{G m_1}{r_2}} ,
\end{equation}
and the orbital angular frequency is $\Omega = v_2/r_2$.

We focus on the evolution of the orbital separation $r_2$, because given $r_2(t)$, the orbital phase can be inferred from Kepler's third law.
The orbital separation will shrink through gravitational radiation-reaction (RR) as well as its interactions with the dark-matter distribution---specifically, dynamical friction and secondary accretion.\footnote{Radiation reaction and dynamical friction can be explained in the Newtonian context in terms of energy losses. 
Gravitational waves carry away gravitational binding energy from the system, which causes the orbital separation to decrease; dynamical friction transfers binding energy from the IMRI to the dark-matter particles (effectively ``heating up'' the DM distribution).
Accretion, however, increases the mass of the secondary while the angular momentum remains an adiabatic invariant; this requires that the orbital separation decreases~\cite{Hughes:2018qxz}.}
The evolution equation for $r_2$ is given by the sum of the contributions from these processes:
\begin{subequations} \label{eq:dot_r}
\begin{equation}
    \dot r_2 = -\dot r_2^\mathrm{RR} -\dot r_2^\DF - \dot r_2^\mathrm{SA} ,
\end{equation}
where
\begin{align} \label{eq:r_dot_RR}
    \dot r_2^\mathrm{RR} & \approx q \frac{64 }{5c^5} \left( \frac{G m_1}{r_2} \right)^3 ,\\
    \label{eq:r_dot_DF}
    \dot r_2^\DF & \approx q \, 8\pi \sqrt{\frac{G}{m_1}} r_2^{5/2} \log\Lambda \rho_\DM(r_2, t; v < v_2) ,\\
    \label{eq:r_dot_SA}
    \dot r_2^\mathrm{SA} & \approx 2 r_2 \dot m_2/m_2 \, .
\end{align}
\end{subequations}
Equation~\eqref{eq:r_dot_RR} accounts for gravitational radiation-reaction from the quadrupole formula.
In Eq.~\eqref{eq:r_dot_DF}, we introduced the notation $\rho_\DM(r_2, t; v < v_2)$ to indicate that the dynamical-friction force arises from DM particles moving more slowly than the orbital speed of the secondary ($v<v_2$).\footnote{\label{fn:heating}We restrict to $v<v_2$, because as is described in detail in~\cite{BinneyAndTremaine}, the more rapidly moving particles contribute to dynamical heating.
However, dynamical heating is smaller than dynamical friction by a factor of the ratio of the DM particle mass to the mass of the secondary black hole (i.e., $m_\DM/m_2$).
For particle DM, as we consider in this paper, this ratio is minuscule.
Dynamical heating could be relevant if the distribution function were such that the ratio of the density of particles moving faster than the orbital speed to that of the more slowly moving particles were of the order of $m_2/m_\DM$.
However, for the DM spike density that we consider in this paper, the ratio of these densities is of order one (see, e.g.,~\cite{Kavanagh:2020cfn,Nichols:2023ufs}); thus, dynamical heating will be negligible, and we do not consider it.
While there have been results in the literature that state that dynamical heating can be significant (e.g.,~\cite{Dosopoulou:2023umg}), Ref.~\cite{Dosopoulou:2023umg}, for example, neglects the suppression of dynamical heating by $m_\DM/m_2$, thereby implicitly assuming the ratio is order one. 
This could be appropriate for dark matter formed from primordial black holes, but this is not the scenario which we consider in this paper.}
We also introduced the Coulomb logarithm $\log \Lambda$, which arises because it is assumed that there are minimum and maximum impact parameters of DM particles that scatter with the secondary.
As in \cite{Kavanagh:2020cfn}, we use the value $\Lambda \approx \sqrt{m_1/m_{2,0}}$, where $m_{2,0}$ is the initial value of $m_2$ (the secondary's mass is time dependent, because of accretion).

Equation~\eqref{eq:r_dot_SA} captures the effects of accretion onto the secondary, but it requires a supplementary equation that prescribes the evolution of $m_2$. 
We focus on the case in which the secondary is a black hole. 
The accretion cross-section for a black hole, $\sigma(v_2)$, was computed in~\cite{Unruh:1976fm}.
The ``Newtonian'' expression for the cross section was used in~\cite{Nichols:2023ufs}, and is given by
\begin{equation} \label{eq:sigma-of-v}
    \sigma(v_2) = \frac{16\pi (G m_2)^2}{(c v_2)^2} \, .
\end{equation}
Equation~\eqref{eq:sigma-of-v} assumes the typical relative velocity between the secondary and a DM particle is the secondary's velocity $v_2$ (see, e.g.,~\cite{Nichols:2023ufs} and references therein).
Under this assumption, the cross section can be obtained from a calculation of the critical impact parameter for a Schwarzschild black hole to capture particles on geodesics with an asymptotic velocity of $v_2$ (see, e.g., Exercise~25.22 of~\cite{Misner:1973prb}).
Thus, Eq.~\eqref{eq:sigma-of-v} is appropriate for quantifying the accretion of locally isotropic distributions of collisionless particle dark matter; it is not obviously applicable to more complex accretion scenarios.

As we will discuss in Sec.~\ref{sec:SA_genform}, we will modify the cross section from the expression in Eq.~\eqref{eq:sigma-of-v} by an energy-dependent amount in the context of SA feedback to the distribution function; however, we will continue to use the expression in Eq.~\eqref{eq:sigma-of-v} in the evolution equations for the binary.
Next, using that $\dot m_2 = \rho_\DM(r_2,t)\sigma(v_2) v_2$, the evolution equation for $m_2$ is given by
\begin{equation} \label{eq:m2_dot}
    \dot m_2 \approx 16\pi (Gm_2)^2 \rho_\DM(r_2, t) \frac{1}{c^2} \sqrt{\frac{r_2}{Gm_1}} \, ,
\end{equation}
so that the SA-induced evolution of $r_2$ is given by
\begin{equation}
    \dot r_2^\mathrm{SA} \approx q \frac{32\pi}{c^2} (G r_2)^{3/2} \sqrt{m_1} \rho_\mathrm{DM}(r_2, t) \, .
\end{equation}
Secondary accretion depends on the total DM density at the location of the secondary, unlike dynamical friction which only depends on the density of DM particles moving more slowly than the secondary [see the $v < v_2$ in the density in Eq.~\eqref{eq:r_dot_DF}].

In the next three subsections, we review the formalism and evolution equations for the DM spike. 
We introduce a minimum value for the angular momentum of a DM particle's orbit, and determine its effects on the position-space density and the density of states. 
Next, we summarize the changes in the formalism of~\cite{Kavanagh:2020cfn,Nichols:2023ufs} for dynamical-friction and secondary-accretion feedback with an angular-momentum cutoff.

\subsection{Dark-matter distribution function and density}
\label{sec:DM_formalism}

We first review the dark-matter distribution when there is a cutoff in position space, as in~\cite{Kavanagh:2020cfn,Nichols:2023ufs} before we discuss how we treat the distribution function when a cutoff is imposed in the space of angular momentum instead.
The main difference between these cases arises from the region of the configuration space in which dark-matter particles are permitted in these two cases.
In both cases, we will assume that the distribution function $f$, the mass density on configuration or phase space, will be a function of the relative specific energy
\begin{equation} \label{eq:calEdef}
    \mathcal E = \Phi(r) - \frac{v^2}{2} ,
\end{equation}
where $\Phi(r) \approx G m_1/r$ is the Newtonian potential.

\subsubsection{Review of the dark-matter distribution with a position-space cutoff}

In \cite{Kavanagh:2020cfn,Nichols:2023ufs}, the initial dark-matter density was taken to follow a spherically symmetric, power-law profile:
\begin{equation} \label{eq:PL_dens}
    \rho_\mathrm{PL}(r) = \left\{ \begin{array}{ll} 
    \rho_\SP (r_\SP/r)^{\gamma_\SP} & r_\IN \leq r \leq r_\SP \\
    0 & r < r_\IN
    \end{array} \right. .
\end{equation}
The power law $\gamma_\SP$ was shown in~\cite{Gondolo:1999ef} to fall in the interval $[9/4, 5/2]$ if $\rho_\mathrm{PL}(r)$ formed from the adiabatic growth of a seed BH in the center of a similar spherically symmetric initial power-law DM distribution with a shallower power law than $\gamma_\SP$.
A relativistic calculation in~\cite{Sadeghian:2013laa} that generalized the Newtonian calculation of~\cite{Gondolo:1999ef} showed that the spike truncated smoothly at a radius of $r_\IN = 4 G m_1/c^2$, because of the presence of a minimum angular momentum required for a DM particle to avoid plunging into the primary black hole.
To approximate this sharp cutoff in angular-momentum space, Refs.~\cite{Kavanagh:2020cfn,Nichols:2023ufs} instead put a sharp cutoff in position space at $r_\IN$.
The last new parameter $r_\SP$ in Eq.~\eqref{eq:PL_dens}  was fixed in terms of $m_1$, $\gamma_\SP$ and $\rho_\SP$ in~\cite{Kavanagh:2020cfn,Nichols:2023ufs} to be
\begin{equation}
    r_\SP \approx \left[ \frac{(3-\gamma_\SP) 0.2^{3-\gamma_\SP} m_1}{2\pi \rho_\SP} \right]^{1/3} 
\end{equation}
by using the relationship proposed in~\cite{Eda:2014kra} that $r_\SP \approx 0.2 r_\mathrm{h}$.
The new radius $r_\mathrm{h}$, is defined as the upper limit of the following integral:
\begin{equation}
    \int_{r_\IN}^{r_\mathrm{h}}  \ud^3 \mathbf r' \rho_\DM(r') = 2 m_1 \, .
\end{equation}

The feedback prescriptions in~\cite{Kavanagh:2020cfn,Nichols:2023ufs} applied to the distribution function $f(\mathbf{r}, \mathbf{v})$. 
It was assumed that the distribution function was isotropic and spherically symmetric, so that $f(\mathbf{r}, \mathbf{v}) = f(r,v)$, and moreover it was a function of only the specific energy, $f(\mathcal E)$.
An isotropic phase-space distribution can be constructed from the density in Eq.~\eqref{eq:PL_dens} using the Eddington inversion procedure.
It was shown in~\cite{Edwards:2019tzf} that this distribution function, which we denote $f_\mathrm{PL}(\mathcal E)$, has the following form: 
\begin{equation} \label{eq:static_f}
    f_\mathrm{PL}(\mathcal E) = \frac{\gamma_\SP(\gamma_\SP-1)}{(2\pi )^{3/2}} \frac{\Gamma(\gamma_\SP-1)}{\Gamma(\gamma_\SP-1/2)} \left(\frac{r_\SP\mathcal E}{Gm_1}\right)^{\gamma_\SP} \! \rho_\SP \mathcal E^{-3/2} .
\end{equation}
The energies $\mathcal E$ of bound DM particles at a distance $r$ range from $\mathcal E_\MIN = 0$ to $\mathcal E_\MAX = Gm_1/r$.
To ensure that the distribution function~\eqref{eq:static_f} is consistent with the density in Eq.~\eqref{eq:PL_dens}, energies up to $c^2/4$ (up to corrections of order $q$) must be included.
In terms of the speed $v$, its minimum value is $v_\MIN=0$, whereas its maximum at a given radius $r$ and energy $\mathcal E$ is $v_\MAX = \sqrt{2(Gm_1/r-\mathcal E)}$.
Thus, the allowed region for DM particles in the space $(\mathbf{r},\mathbf{v})$ is a spherical shell in position space  with an inner radius $r_\IN$, and spheres in velocity space with radii $v_\MAX$ that depend on $\mathcal E$ and $r$.

\subsubsection{Dark-matter distribution with an angular-momentum cutoff}

To implement a cutoff in the angular momentum of DM particles, we roughly will follow a Newtonian approach first given in~\cite{Gondolo:1999ef}, which was also discussed in~\cite{Sadeghian:2013laa}. 
We also will continue to assume that the distribution function can be written in the form $f(\mathcal E)$, but we will restrict to a different region of the configuration space.\footnote{Alternately, we could write the distribution function as a product of $f(\mathcal E)$ with a unit step function that vanishes below the angular-momentum cutoff. If one assumes that there is no evolution to the angular-momentum dependence of the distribution function (as we do in this paper), then the two approaches are equivalent. To avoid writing additional theta functions in many expressions, we instead restrict the region of the configuration space and work with the spherically symmetric distribution function $f(\mathcal E)$.}
To describe this region, it will be helpful to use the relative specific energy $\mathcal E$ in Eq.~\eqref{eq:calEdef}, the specific angular momentum $\mathcal J$, and the $z$ component of this angular momentum $\mathcal J_z$.
The Newtonian angular momentum $\mathcal J$ can be written in terms of $r$ and the magnitude of the transverse velocity $v_t$, whereas $\mathcal E$ can be expressed in terms of $r$, $\mathcal J$ and the radial component of the velocity $v_r$:
\begin{equation} \label{eq:specE_J}
    \mathcal E = \frac{G m_1}{r} - \frac{(v_r)^2}{2} - \frac{\mathcal J^2}{2 r^2} , \qquad 
    \mathcal J = r v_t .
\end{equation}
Imposing a minimum angular momentum $\mathcal J_\MIN$ gives rise to a minimum velocity of $v_\MIN = \mathcal J_\MIN/r$ at each radius.
Note that the maximum velocity, at a given $r$ is still $v_{\mathrm{max}} = \sqrt{2 (G m_1/r-\mathcal E)}$.
The minimum angular momentum does change the maximum energy that a DM particle can have at a given $r$.
It occurs when the radial velocity vanishes and the angular momentum is at a minimum:
\begin{equation} \label{eq:calEmax}
    \mathcal E_\MAX(r) = \frac{Gm_1}r - \frac{\mathcal J_\MIN^2}{2 r^2} .
\end{equation}
Finally, at a given position $r$ and for a given energy $\mathcal E$, there is a maximum angular momentum that can be realized (namely, when $v_r = 0$).
It is given by 
\begin{equation}
    \mathcal J_\MAX(r,\mathcal E) = \sqrt{2r(Gm_1 - \mathcal{E}r)} .
\end{equation}
The $z$ component of the angular momentum is bounded by $-\mathcal J_\MAX$ for the minimum and $\mathcal J_\MAX$ for the maximum.
The minimum angular momentum also imposes a minimum radius for bound orbits: namely, that $r$ must be greater than $r_{\mathcal J}/2$, where we have defined
\begin{equation}
    r_{\mathcal J} = \frac{\Jmin^2}{Gm_1} .
\end{equation}
Thus, the region of $(\mathbf{r},\mathbf{v})$ space remains a spherical shell in terms of position space, but now the region of velocity space is restricted to be spherical shells with an inner radius that depends on $r$ and $\mathcal J_\MIN$ and an outer radius that depends on $r$ and $\mathcal E$.

The density $\rho(r)$ can be computed from the distribution function by integrating it over all velocity space where DM particles are permitted:
\begin{equation}
    \rho(r) = \int \ud^3 v f(\mathcal E) .
\end{equation}
We follow the approach outlined in~\cite{Sadeghian:2013laa} to transform the integral over $\ud^3 v$ to an integral over $\ud \mathcal E\ud \mathcal J \ud \mathcal J_z$. 
The Jacobian determinant of this transformation depends on both the radial and polar components of the velocity, when it is expressed in a polar coordinate system.
However, for a distribution function that is independent of $\mathcal J_z$, the integral can be performed analytically and the dependence of the Jacobian determinant on the polar component cancels.
Because the magnitude of the radial velocity is given by
\begin{equation}
   v_r = \sqrt{2\left(\frac{Gm_1}r - \mathcal{E}\right) - \frac{\mathcal J^2}{r^2}} ,
\end{equation}
the integral for the density can be written in the form
\begin{equation} \label{eq:density_Jvmin}
        \rho(r) = 4 \pi \int_{0}^{\mathcal E_{\mathrm{max}}} \ud \mathcal E 
        \int_{\mathcal J_{\MIN}}^{\mathcal J_{\mathrm{max}}} \frac{\ud \mathcal J \, \mathcal J f(\mathcal E)}{r \sqrt{ 2r \left(G m_1 - \mathcal{E}r \right) - \mathcal J^2}} .
\end{equation}
In addition, the integral over angular momentum can be evaluated when $f$ is independent of $\mathcal J$.
The result is simplest when written in terms of $v$.
Using the fact that $\ud \mathcal E = -v \ud v$ at fixed $r$, we find
\begin{equation} \label{eq:rho-v-integral}
    \rho(r) = 4\pi \int_{v_\MIN}^{v_\MAX} \ud v \, v \sqrt{v^2 - v_\MIN^2} f\boldsymbol{(}\mathcal E(r,v) \boldsymbol{)} .
\end{equation}
The dependence of the energy on velocity is given in Eq.~\eqref{eq:calEdef}.
If we substitute the distribution function in Eq.~\eqref{eq:static_f} into Eq.~\eqref{eq:rho-v-integral}, we find that the integral can be evaluated, and is given by
\begin{equation}
    \rho(r) = \rho_\SP \left(\frac{r_\SP}{r}\right)^{\gamma_\SP} \left( 1 - \frac{r_{\mathcal J}}{2r} \right)^{\gamma_\SP}  .
\end{equation}

In the calculations in subsequent sections, we will make use of the density of states $g(\mathcal E)$.
As discussed in~\cite{Kavanagh:2020cfn}, it can be computed from the expression
\begin{equation} \label{eq:DoSdef}
    g(\mathcal E) = \int \ud^3 r \int \ud^3 v \, \delta\boldsymbol( \mathcal E - \mathcal E(r,v)\boldsymbol) \, .
\end{equation}
We continue to assume the system is spatially isotropic, such that $\ud^3 r = 4 \pi r^2 \ud r$. 
The integral over velocity can be performed using a similar approach to compute the density in Eq.~\eqref{eq:density_Jvmin}.
The result is
\begin{equation}
    g(\mathcal E) = 16 \pi^2 \int_{r_{\MIN}}^{r_{\mathrm{max}}} \ud r \, r \sqrt{2 r\left(G m_1 - \mathcal{E} r \right) - \mathcal J_{\MIN}^2 }.
\end{equation}
The bounds of integration in $r$ are the maximum and minimum radii for a given energy. 
Solving $\mathcal E = Gm_1/r_\pm -  \mathcal J_\MIN^2/(2r_\pm^2)$ yields two solutions.
They can be written concisely after defining the velocity
\begin{equation}
    v_{\mathcal J} = \frac{G m_1}{\Jmin} = \sqrt{\frac{Gm_1}{r_{\mathcal J}}}.
\end{equation}
The solutions $r_\pm$ are
\begin{equation} \label{eq:r_pm}
    r_\pm(\mathcal E) = \frac{Gm_1}{2\mathcal E} \left(1 \pm \sqrt{1 - \dfrac{2 \mathcal E}{v_{\mathcal J}^2}} \right) .
\end{equation}
Because $r_-$ is less than or equal to $r_+$, we have that $r_{\MIN} = r_-$ and $r_{\mathrm{max}} = r_+$. 
The radial integral in the calculation of the density of states can be rewritten using $r_\pm$ as
\begin{equation} \label{eq:DoS-r-int}
    g(\mathcal E) = 16 \pi^2 \int_{r_-}^{r_+} \ud r \, r \sqrt{2 \mathcal{E} (r_+ - r) (r - r_-)} .
\end{equation}
The integral can be evaluated analytically: 
\begin{equation} \label{eq:DoS_J}
    g(\mathcal E) = 
    \sqrt{2} (\pi G m_1)^3 \mathcal E^{-5/2} \left( 1 - \frac{2 \mathcal E}{v_{\mathcal J}^2} \right) .
\end{equation}
The density of states is nonzero for bound particles with energies $\mathcal E$ less than $(Gm_1)^2 / (2 \mathcal J_\MIN^{2}) = Gm_1/(2r_{\mathcal J})$.
It has a divergence as $\mathcal E$ approaches zero, which was discussed in~\cite{Kavanagh:2020cfn}.
The vanishing of the density of states for energies greater than $v_{\mathcal J}^2/2$ is consistent with the expression for $\mathcal E_\MAX$ as a function of $r$, which has its maximum at $r_{\mathcal J}$, where it equals $v_{\mathcal J}^2/2$.
The $\mathcal J_\MIN = 0$ limit of Eq.~\eqref{eq:DoS_J} is also consistent with the density of states computed in~\cite{Kavanagh:2020cfn}, as will be discussed in more detail in Sec.~\ref{sec:case_nocut}.

The next two subsections (Secs.~\ref{sec:DF_genform} and~\ref{sec:SA_genform}) will discuss the evolution of the DM distribution function in response to the energy injected into the distribution from dynamical friction, and from DM particles removed from the distribution function when the secondary is a black hole.
The form of the evolution equations that we use will be the same as those in~\cite{Kavanagh:2020cfn,Nichols:2023ufs}.
Differences from the work of~\cite{Kavanagh:2020cfn,Nichols:2023ufs} will appear in the rate coefficients for scattering and accretion from the changes in the regions of configuration space consistent with a minimum angular momentum $\mathcal J_\MIN$.
This will change the functional form of these rate coefficients, which we discuss next.

\subsection{Dynamical-friction feedback with an angular-momentum cutoff} \label{sec:DF_genform}

The formalism that we use for incorporating feedback from dynamical friction onto the DM distribution function was presented in~\cite{Kavanagh:2020cfn}, which characterized the scattering processes in terms of ``influxes'' and ``outfluxes'' of particles at a given specific energy. 
We use a notation similar to that used in~\cite{Nichols:2023ufs}.
A key element of the formalism is the differential rate (probability per orbital period) for particles with energy $\mathcal E - \Delta \mathcal E$ to scatter to an energy $\mathcal E$.
Its expression was given in~\cite{Kavanagh:2020cfn}:
\begin{align} \label{eq:DFscatterrate}
    \mathcal R_{\mathcal E}(\Delta \mathcal E) = \frac 1{T_2 g(\mathcal E)} \int \ud^3 r \int \ud^3 v & \, \delta\boldsymbol( \mathcal E - \mathcal E(r,v)\boldsymbol) \nonumber \\
    & \times \delta\boldsymbol( \Delta\mathcal E(b) - \Delta \mathcal E\boldsymbol) \, ,
\end{align}
where $T_2 = 2 \pi r_2 / v_2$ is the orbital period of the secondary. The change in energy $\Delta \mathcal E$ is parameterized in terms of the impact parameter $b$ of a DM particle scattering with the secondary. 
The total scattering rate at a given energy is obtained by integrating over all allowed changes in energy $\Delta \mathcal E$,
\begin{equation}
    R_{\mathcal E} = \int \ud\Delta\mathcal E \, \mathcal R_{\mathcal E}(\Delta \mathcal E) \, .
\end{equation}

The evolution of the distribution function due to DF feedback is governed by the balance of particles scattering out of and into each specific energy:
\begin{align} \label{eq:fIPDE}
    & \frac{\partial f(\mathcal E,t)}{\partial t} = -R_{\mathcal E} f(\mathcal E,t)   \nonumber \\
    &+ \int \ud\Delta\mathcal E \, h(\mathcal E, \Delta \mathcal E) \mathcal R_{\mathcal E-\Delta\mathcal E}(\Delta \mathcal E) f(\mathcal E - \Delta\mathcal E,t) .
\end{align}
We introduced the notation $h(\mathcal E, \Delta \mathcal E)$ for the ratio of the density of states at energy $\mathcal E - \Delta \mathcal E$ to that at $\mathcal E$:
\begin{equation} \label{eq:ratio_DoS}
    h(\mathcal E, \Delta \mathcal E) \equiv \frac{g(\mathcal E - \Delta \mathcal E)}{g(\mathcal E)} = \frac{\mathcal E^{5/2}[v_{\mathcal J}^2 - 2(\mathcal E - \Delta\mathcal E)]}{(\mathcal E- \Delta\mathcal E)^{5/2} (v_{\mathcal J}^2 - 2\mathcal E )}.
\end{equation}
The first term on the right-hand side of Eq.~\eqref{eq:fIPDE} characterizes the evolution of the distribution function from particles at a specific energy $\mathcal E$ that scatter to a different energy due to the secondary.
The second term characterizes the evolution from particles that scatter into an energy $\mathcal E$ from an energy $\mathcal E - \Delta \mathcal E$, weighted by an appropriate scaling coming from the density of states, which is integrated over the allowed $\Delta\mathcal E$.

We briefly comment on the factor of the density of states in the denominator in Eqs.~\eqref{eq:DFscatterrate}--\eqref{eq:ratio_DoS} above.
The scattering rate is nonzero for positive energies $\mathcal E$ which are less than the maximum allowed energy $\mathcal E_\MAX(r_{\mathcal J}) = Gm_1/(2r_{\mathcal J})$.
As the energy approaches the maximum, both the integral in Eq.~\eqref{eq:DFscatterrate} and the normalization of $1/g(\mathcal E)$ approach zero.
Note that in the evolution equation for $f(\mathcal E,t)$ in Eq.~\eqref{eq:fIPDE}, a factor of $h(\mathcal E,\Delta \mathcal E)$ multiplies $\mathcal R_{\mathcal E - \Delta \mathcal E}(\Delta \mathcal E)$, which leads to a factor of $g(\mathcal E)$, not $g(\mathcal E - \Delta \mathcal E)$ in the denominator.
As we discuss in Appendix~\ref{sec:DF_feedback_app}, the value of $\Delta\mathcal E$ that is closest to zero is $-2qv_2^2$.
Thus, when $\mathcal E - \Delta\mathcal E = v_{\mathcal J}^2 /2$, the largest possible energy, the factor of $g(\mathcal E)$ in the denominator will be proportional to $q$, but nonzero, so the influx term remains finite.

The limit of $\mathcal E$ approaching $v_{\mathcal J}^2 /2$ in the outflux term proportional to $R_{\mathcal E}$ is somewhat more subtle.
It can be shown from the results in Appendix~\ref{sec:DF_feedback_app} that in the limits as $r_2$ approaches $r_{\mathcal J}$ and $\mathcal E$ approaches $v_{\mathcal J}^2 /2$, the expression for the rate $R_{\mathcal E}$ goes to a constant.
Whereas $\mathcal R_{\mathcal E}(\Delta \mathcal E)$ is proportional to $b_{90}^2$ (which scales with the mass ratio as $q^2$), in these limits, $R_{\mathcal E}$ scales with a lower power of $q$.
This implies that the scattering probability increases, though the density of states of particles with the maximum energy also goes to zero.
A similar phenomenon will occur with the secondary accretion rate, which we discuss in more detail in Sec.~\ref{sec:SA_genform}.

We can compute the integral in Eq.~\eqref{eq:DFscatterrate} using an approach similar to that in~\cite{Kavanagh:2020cfn} (though now allowing for a nonzero $\mathcal J_\MIN$). 
We relate the change in the relative energy of a DM particle over the course of a single scattering event to the impact parameter by
\begin{equation} \label{eq:Delta-cal-E}
    \Delta \mathcal E(b) = -\dfrac{\Delta E_{\mathrm{CO}}}{m_{\mathrm{DM}}} = -2 v_0^2 \left( 1 + \dfrac{b^2}{b_{90}^2} \right)^{-1} .
\end{equation}
Here $v_0$ is the relative speed of the encounter of the DM particle with the secondary, which we approximate by $v_2$, the orbital speed of the secondary, as in~\cite{Kavanagh:2020cfn}. 
The constant $b_{90}$ is the impact parameter that produces a $90^\circ$ deflection of the DM particle; it is given by $b_{90} = G m_2/(v_0)^2 \approx (m_2/m_1) r_2 = q r_2$. 

We evaluate the integral over velocity space in Eq.~\eqref{eq:DFscatterrate} using techniques similar to those we used in calculating the density and the density of states.
First we transform the integral over velocity to one over $\mathcal E$, $\mathcal J$ and $\mathcal J_z$.
As with the density, we perform the integral over $\mathcal J_z$ first, because it is bounded by $\pm\mathcal J$, and then over $\mathcal J$, because it is bounded above by $\mathcal J_\MAX(r,\mathcal E)$.
As with the integral of the density, we find it convenient to reexpress the integral in terms of velocity, so that it is given by
\begin{equation} \label{eq:calR-delta-v}
    \begin{split}
        \mathcal R_{\mathcal E}(\Delta \mathcal E) =  \frac{4\pi}{T_2 g(\mathcal E)} \int \ud^3 r \int \ud v & \sqrt{v^2 - v_\MIN^2} \delta \boldsymbol(v - v(r,\mathcal E) \boldsymbol)  \\
        & \times \delta(\Delta \mathcal E(b) - \Delta \mathcal E \boldsymbol) 
    \end{split}
\end{equation}
The integral over $v$ runs from $v_\MIN(r) = \mathcal J_\MIN /r$ to $v_\MAX$.
We choose $v_\MAX = v_2 = \sqrt{Gm_1/r_2}$ to be the orbital speed of the secondary.

The integral involving the delta function is nonzero for $v \in [v_\MIN(r),v_2]$.
For any energy $\mathcal E$, the radii $r$ for which no particles can scatter, occur when $v_\MIN(r) > v_2$ or when $r < \sqrt{r_2 r_{\mathcal J}}$.
While we will always need to restrict our integration region in $r$ to be $r \geq \sqrt{r_2 r_{\mathcal J}}$, there are additional constraints on the integration domain in $r$ that arise from the fact that the velocity satisfies $v(r,\mathcal E) \in [v_\MIN(r), v_2]$, which can be more restrictive.
For example, the condition $v_\MIN(r) \leq v(r,\mathcal E)$ gives rise to the condition $r \in [r_-,r_+]$ for $r_\pm$ defined in Eq.~\eqref{eq:r_pm}.
However, the condition $v(r,\mathcal E) \leq v_2$ also gives rise to a condition $r \geq r_\cut(\mathcal E,r_2)$, where we have defined 
\begin{equation} \label{eq:r-cut-def}
    r_{\cut}(\mathcal E,r_2) = \frac{2G m_1 r_2}{2\mathcal E r_2 + G m_1} . 
\end{equation}

Unlike in~\cite{Kavanagh:2020cfn} (where $\Jmin = 0$ implies $r_- = \sqrt{r_2 r_{\mathcal J}} = 0$ and the smallest radius is always determined by $r_\cut$), here we need to determine the regions of $r_2$ and the energies $\mathcal E$ where the three different lower bounds $r_-$, $r_\cut$ and $\sqrt{r_2 r_{\mathcal J}}$ give rise to the most restrictive constraint, which we will call $r_\MIN$.
Namely, this $r_\MIN$ is given by
\begin{equation}
    r_\MIN(\mathcal E, r_2) = \MAX(r_-, r_\cut, \sqrt{r_2 r_{\mathcal J}}) .
\end{equation}
Once $r_\MIN$ is determined, we can find the conditions for which $r_+ \geq r_\MIN$, so that there is a nontrivial region of integration in $r$.

Computing $r_\MIN$ requires considering the relative size of $r_-$, $r_\cut$ and $\sqrt{r_2 r_{\mathcal J}}$.
An energy scale that will come up frequently in these comparisons is
\begin{equation}
    \mathcal E_{2,\mathcal J} = \frac{v_2}2 \left( 2v_{\mathcal J} - v_2 \right) .
\end{equation}
This energy reaches the largest possible energy $G m_1/(2r_{\mathcal J})$ when $v_2 = v_{\mathcal J} = \sqrt{Gm_1/r_{\mathcal J}}$.

Checking the condition $r_\cut \geq \sqrt{r_2 r_{\mathcal J}}$ we find that this is satisfied when $0 < \mathcal E \leq \mathcal E_{2,\mathcal J}$.
A minimum value of permitted $r_2$ comes from determining when $\mathcal E_\MAX(r_2)$ in Eq.~\eqref{eq:calEmax} is positive. 
This gives the constraint that $r_2 > r_{\mathcal J}/2$.
The inequality $r_\cut \geq r_-$ also is satisfied when $0 < \mathcal E \leq \mathcal E_{2,\mathcal J}$ (and again with $r_2 > r_{\mathcal J}/2$).
Finally, we consider when $r_- \geq \sqrt{r_2 r_{\mathcal J}}$ is satisfied.
The occurs when $\mathcal E_{2,\mathcal J} \leq \mathcal E \leq G m_1/(2\sqrt{r_2 r_{\mathcal J}})$.
The energy interval is nontrivial for $r_2 \leq r_{\mathcal J}$, which constrains the upper limit to be $Gm_1/(2r_{\mathcal J})$ instead.
Together, these constraints imply that $r_\MIN$ is given by
\begin{equation} \label{eq:r_min}
    r_\MIN = 
    \begin{cases}
        r_\cut & \mbox{for } \ 0 \leq \mathcal E \leq \mathcal E_{2,\mathcal J}, \ \ r_2 \geq \dfrac{r_{\mathcal J}}2 , \\
        r_- & \mbox{for } \ \mathcal E_{2,\mathcal J} < \mathcal E \leq \dfrac{Gm_1}{2r_{\mathcal J}} , \ \ r_2 < r_{\mathcal J} .
    \end{cases}
\end{equation}
There is not a region of allowed energies $\mathcal E$ and radii $r_2$ for which $\sqrt{r_2 r_{\mathcal J}}$ is larger than both $r_-$ and $r_\cut$.

Checking when $r_+$ is greater than or equal to $r_\MIN$ does not provide any new constraints, because the inequality $r_+ \geq r_\cut$ leads to equivalent conditions to $r_\cut \geq r_-$ and it is always the case that $r_+$ is larger or equal to $r_-$.

Because we now have determined how the delta function in velocity in Eq.~\eqref{eq:calR-delta-v} restricts the domain of integration in $r$, we turn to the integration over position.
To do so, we follow~\cite{Kavanagh:2020cfn} and transform the Dirac delta function in $\Delta \mathcal E$ to one in $b$.
This allows us to write the integral as 
\begin{align} \label{eq:DF_rate_dr3}
     \mathcal R_{\mathcal E}(\Delta \mathcal E) = & {}
        \dfrac{\pi b_{90}^2}{T_2 g(\mathcal E) v_2^2} \displaystyle\int_{r_{\MIN}}^{r_+} d^3 r \dfrac{1}{b} \left( 1 + \dfrac{b^2}{b_{90}^2} \right)^2 \nonumber \\
        & \times \sqrt{2 \left(\dfrac{G m_1}{r} - \mathcal E \right) - \dfrac{\mathcal J_{\MIN}^2}{r^2}}  \delta\boldsymbol( b - b_\star(\Delta \mathcal E)\boldsymbol) ,
\end{align}
where $T_2$ is the orbital period of the secondary. We also introduced the notation
\begin{equation} \label{eq:b-star-def}
    b_\star(\Delta\mathcal E) = b_{90} \sqrt{\frac{2 v_2^2 }{|\Delta \mathcal E|} - 1} 
\end{equation} 
in Eq.~\eqref{eq:DF_rate_dr3}.
We have not yet fully specified the upper and lower range of the energies for which the integral is nonzero, because this will depend on the choice of the possible impact parameters which scatter with the secondary.
The value of impact parameter will also influence whether the DM particle resides within the radial integration domain of the spherical shell with inner radius $r_\MIN$ and outer radius $r_+$ or not.

For a given $\Delta\mathcal E$, the integral can be considered as an integral over the surface of a torus with major radius $r_2$ and minor radius $b_\star (\Delta \mathcal E)$. 
Following~\cite{Kavanagh:2020cfn}, we can evaluate the integral over this torus by switching to coordinates centered on the secondary, $(b, \alpha)$, with $b$ the impact parameter and $\alpha$ the poloidal angle (see Fig.~7 of~\cite{Kavanagh:2020cfn}).
As in~\cite{Kavanagh:2020cfn}, by expanding $r$ about $r_2$ to linear order in $b_\star/r_2$, the integral can be expressed in terms of incomplete elliptic integrals.
We give the details of this calculation in Appendix~\ref{sec:DF_feedback_app} for $\mathcal J_\MIN > 0$.

\subsection{Secondary-accretion feedback with an angular-momentum cutoff} \label{sec:SA_genform}

The form of SA feedback onto the DM distribution shares several key similarities and differences with that of DF feedback.
As discussed further in~\cite{Nichols:2023ufs}, secondary accretion removes dark matter from the distribution function with an outflux term in the dark-matter evolution equations, but there is no corresponding influx term as with dynamical friction. 
Specifically, the outflux term takes the form of an additional term on the right-hand side of Eq.~\eqref{eq:fIPDE}, of the form $-R_{\mathcal E}^\acc f(\mathcal E,t)$.
In the approximations used in~\cite{Nichols:2023ufs}, it could be shown analytically that the decrease in the mass of the dark-matter distribution was balanced by an increase in mass of the secondary black hole, so that mass was conserved. 

The per-period rate coefficient, $R_{\mathcal E}^\acc$, for accreting dark-matter particles with energy $\mathcal E$ was computed in~\cite{Nichols:2023ufs} from an integral of the form 
\begin{equation}
    R_{\mathcal E} = \frac 1{T_2 g(\mathcal E)} \int_{\boldsymbol{r} \in T^2} d^3 r \int d^3 v \, \delta\boldsymbol( \mathcal E - \mathcal E(r,v)\boldsymbol) .
\end{equation}
The velocity-space integral can be performed exactly, as it could with the integral for the density of states, to give
\begin{equation}
    R^\mathrm{acc}_{\mathcal E} = \frac {4 \pi}{T_2 g(\mathcal E)} \int_{\mathbf{r} \in T^2} d^3 r \sqrt{2 \left(\frac{G m_1}{r} - \mathcal E \right) - \frac{\mathcal J_{\MIN}^2}{r^2}} \\
\end{equation}
The region $T^2$ is a torus surrounding the secondary's circular orbit, with major radius $r_2$ and a cross-sectional area determined by the scattering cross section $\sigma(v_2)$.
In~\cite{Nichols:2023ufs}, the cross section was based on the ``Newtonian-order'' limit of an expression due to Unruh~\cite{Unruh:1976fm}, which was a circle of radius $b_\acc = \sqrt{\sigma (v_2)/\pi}$ given by
\begin{equation}
    b_\acc = \frac{4 G m_2}{c v_2} = \frac{4qr_2}{c} \sqrt{\frac{Gm_1}{r_2}} \, .
\end{equation}
The integral was performed over a torus in the small $q$ approximation in~\cite{Nichols:2023ufs}.
Given that $r \in \left[ r_2 - b_\acc, r_2 + b_\acc \right]$ and that $b_\acc$ is of order $q$, we also make the same small-$q$ approximation here, so that we can write
\begin{equation}
    \frac{G m_1}{r} - \frac{\mathcal J_\MIN^2}{2 r^2} \approx \left( \frac{G m_1}{r_2} - \frac{\mathcal J_\MIN^2}{2 r_2^2} \right) \left[ 1 + \mathcal O(q) \right].
\end{equation}
To leading order in $q$, the integrand is constant in $r$ on the domain of integration, which makes the rate coefficient equal to the integrand times the volume of the torus.
This gives for the rate coefficient
\begin{align} \label{eq:LcutSAprob}
        & R^\mathrm{acc}_{\mathcal E} =  \nonumber \\
        & \begin{cases} 
            8\pi^2\dfrac{r_2 \sigma(v_2,\mathcal E)}{T_2 g(\mathcal E)}  \sqrt{2 [\mathcal E_\MAX(r_2)- \mathcal E ]} & \! \! \mbox{for} \ \mathcal E < \mathcal E_\MAX(r_2) \\
            0 & \! \! \mbox{for} \ \mathcal E \geq \mathcal E_\MAX(r_2) \, ,
        \end{cases}
\end{align}
which has the same form as that in~\cite{Nichols:2023ufs}, which assumed $\Jmin=0$.

Note, however, that unlike in~\cite{Nichols:2023ufs}, here we are using a cross section $\sigma(v_2,\mathcal E)$ that depends on energy.
This is necessary because the density of states $g(\mathcal E)$ vanishes at $\mathcal E_\MAX(r_{\mathcal J}) = Gm_1/(2r_{\mathcal J})$.
If $\sigma(v_2)$ is given by $\pi b_\acc^2$, then as $r_2$ approaches $r_{\mathcal J}$ and $\mathcal E$ approaches $\mathcal E_{\mathcal J}$, the rate coefficient the rate coefficient will diverge.
By definition, the rate coefficient is a ratio of phase-space volumes that span different regions of position space (and the corresponding allowed regions of velocities at those positions) normalized by the orbital period of the secondary.
This implies that the rate coefficient should be a finite constant even in the limit as the energy approaches $\mathcal E_\MAX(r_{\mathcal J})$.

To make the rate coefficient have a finite limit, we eliminate regions of the torus that fall outside the allowed radii $r_\pm$ for a particular energy.
Specifically, if we define an angle $\theta$ by
\begin{equation}
    \theta = \cos^{-1}[ \MIN\boldsymbol ( (r_+-r_-)/(2b_\acc),1 \boldsymbol ) ] , 
\end{equation}
then we will write the cross section as
\begin{equation}
    \sigma(v_2,\mathcal E) = b_\acc^2 \left[ 2 \left( \frac \pi 2 - \theta\right) + \sin(2\theta) \right] .
\end{equation}
Because $r_+ - r_-$ is proportional to $\sqrt{1-\mathcal E/\mathcal E_\MAX(r_{\mathcal J})}$ then its product with $\sqrt{ [\mathcal E_\MAX(r_2)- \mathcal E ]}$ in the limit $r_2 \rightarrow r_{\mathcal J}$ will cancel the term in $1/g(\mathcal E)$ that diverges as $\mathcal E$ approaches $\mathcal E_\MAX(r_{\mathcal J})$.

Note also that there is an enhancement of the accretion rate as the energy approaches $\mathcal E_\MAX(r_{\mathcal J})$.
At most energies, the rate $R_{\mathcal E}^\acc$ scales with the mass ratio as $q^2$, because it is proportional to the cross section, which scales as $b_\acc^2$.
However, when $r_+-r_-$ becomes of order $b_\acc$ at energies near $\mathcal E_\MAX(r_{\mathcal J})$, one linear dimension of the domain for the integral for $g(\mathcal E)$ in Eq.~\eqref{eq:DoS-r-int} takes place over a region of size $b_\acc$, which cancels one factor of $b_\acc$ in the cross section in the numerator of $R_{\mathcal E}^\acc$.
Thus, $R_{\mathcal E}^\acc$ scales as $q$ rather than $q^2$ in this limit.
Although the density of states is going to zero, the total rate with which these particles are accreted is significantly higher.
This property of the accretion rate will be relevant for understanding how the density evolves when the secondary is near $r_{\mathcal J}$.

Finally, we note that with both the DF and SA feedback terms, the integral partial differential equation describing the evolution of the distribution function has a similar form as in~\cite{Nichols:2023ufs}:
\begin{align} \label{eq:fIPDEcap}
    & \frac{\partial f(\mathcal E,t)}{\partial t} = -(R_{\mathcal E} + R_{\mathcal E}^\acc ) f(\mathcal E,t) \nonumber \\
    &+ \int d\Delta\mathcal E h(\mathcal E, \Delta \mathcal E) \mathcal R_{\mathcal E-\Delta\mathcal E}(\Delta \mathcal E) f(\mathcal E - \Delta\mathcal E,t) .
\end{align}
The function $h(\mathcal E, \Delta \mathcal E)$ is given in Eq.~\eqref{eq:ratio_DoS}.

\section{Special cases of the minimum angular momentum} \label{sec:cut_cases}

In this section, we discuss some features of the formalism in Secs.~\ref{sec:DM_formalism}--\ref{sec:SA_genform} for specific values of the angular-momentum cutoff.
We will introduce the notation
\begin{equation} \label{eq:jmin}
    \Jmin = j_\MIN \left(\frac{G m_1}{c} \right) 
\end{equation}
for convenience.
We focus on when $j_\MIN = 0$ and $j_\MIN = \sqrt{8}$. 
The former was explored previously in~\cite{Kavanagh:2020cfn,Nichols:2023ufs}.
In most of the expressions in Secs.~\ref{sec:DM_formalism}--\ref{sec:SA_genform}, taking the $j_\MIN = 0 = \Jmin$ limit straightforwardly reproduces the results in~\cite{Kavanagh:2020cfn,Nichols:2023ufs}; thus, we discuss it briefly in Sec.~\ref{sec:case_nocut}.
The $j_\MIN = \sqrt 8$ case is discussed in more detail in Secs.~\ref{sec:case_cutform} and~\ref{sec:case_cutfeedback}.

\subsection{Review of having no angular-momentum cutoff} \label{sec:case_nocut}

In Refs.~\cite{Kavanagh:2020cfn, Coogan:2021uqv, Nichols:2023ufs}, there was no cutoff in angular momentum ($j_\MIN = 0$). 
This implies that
\begin{equation}
    \mathcal J_\MIN = 0 , \quad v_\MIN = 0, \quad \mathcal E_\MAX(r) = \frac{Gm_1}r.
\end{equation}
The expression for the density in Eq.~\eqref{eq:density_Jvmin} is more easily written as an integral over $v$ of the distribution function times $4\pi v^2$, as is typical for isotropic distributions.
The density extends to $r=0$ in this case, which is why a minimum radius at $r_\IN$ was imposed by hand in~\cite{Kavanagh:2020cfn}.
The density of states reduces to the expression in~\cite{Kavanagh:2020cfn}, which is proportional to $\mathcal E^{-5/2}$ without the factor of $1-2\mathcal E/v_{\mathcal J}^2$ [and which is consistent with taking the $v_{\mathcal J} \rightarrow\infty$ limit of expression in Eq.~\eqref{eq:DoS_J}].
Without a cutoff in $r$ at $r_\IN$, any energy would have been permitted.
However, with $r_\IN$ as the smallest radius, the largest $\mathcal E$ used in~\cite{Kavanagh:2020cfn, Nichols:2023ufs} was restricted to $G m_1 / r_\IN = c^2/4$.

The results for the DF feedback scattering probabilities are given in Eq.~\eqref{eq:DF_rate_dr3} and Appendix~\ref{sec:DF_feedback_app}.
With $\mathcal J_\MIN = 0$ and $j_\MIN=0$, they recover the expressions in~\cite{Kavanagh:2020cfn} (see also Appendix~\ref{app:efficiency}).
In addition, the function $h(\mathcal E, \Delta\mathcal E)$ in Eq.~\eqref{eq:ratio_DoS} reduces to $[\mathcal E/(\mathcal E - \Delta\mathcal E)]^{5/2}$, which appears in the evolution equation for the distribution function.
In principle, dynamical friction and its feedback could occur for any orbital separation $r_2$ of the binary, without the position-space cutoff at $r_\IN$ (though given that the inspirals in~\cite{Kavanagh:2020cfn,Nichols:2023ufs} were truncated at $r_\ISCO > r_\IN$, this did not arise in~\cite{Kavanagh:2020cfn,Nichols:2023ufs}). 
The secondary accretion rate in Eq.~\eqref{eq:LcutSAprob} reduces to the result in~\cite{Nichols:2023ufs}, because $\mathcal E_\MAX(r_2) = Gm_1/r_2$ when $j_\MIN = 0$.

\subsection{Properties of the dark matter distribution for a nonzero angular-momentum cutoff} \label{sec:case_cutform}

We now discuss the effects of having a nonzero value of the angular-momentum cutoff $\mathcal J_\MIN$. 
In the Newtonian analysis in~\cite{Gondolo:1999ef}, a minimum angular momentum of $j_\MIN = 4$ was chosen, which causes the density to vanish at a radius $8 G m_1/c^2=4\radsch$, where $\radsch$ is the Schwarzschild radius. 
In the relativistic calculation of~\cite{Sadeghian:2013laa}, the density was found to vanish at a radius of $4 G m_1/c^2 = 2\radsch$ (there also were higher densities in the inner regions of the spike than those computed in~\cite{Gondolo:1999ef}). 
To perform a Newtonian analysis in which the density vanishes at the same radius as in relativistic simulations, we can choose $j_\MIN= \sqrt 8$.\footnote{This can be seen from the integration bounds in Eq.~\eqref{eq:density_Jvmin}.
The square of the maximum speed at a given $r$ is $v_\mathrm{max}^2 = 2Gm_1/r$, whereas the square of the minimum speed is $\Jmin^2/r^2$.
When the integration bounds are equal, the integral will vanish.
Equating the two shows that the radius is $2\radsch = r_{\mathcal J}/2$.}
Thus, in the remainder of this work, we will specialize the results in Secs.~\ref{sec:DM_formalism}--\ref{sec:SA_genform} in which $\mathcal J_\MIN$ and thus $v_\MIN(r)$ have the following values:
\begin{equation}
    \Jmin = \frac{\sqrt{8} G m_1}{c}, \qquad v_{\MIN}^2 = \frac{2 G m_1}{r} \frac{2 \radsch}{r} .
\end{equation}
We kept as $\Jmin$ an arbitrary positive constant in Secs.~\ref{sec:DM_formalism}--\ref{sec:SA_genform}, because the inner radius or $2\radsch$ was appropriate for Schwarzschild black holes.
For Kerr black holes, the inner radius will typically be smaller (see~\cite{Ferrer:2017xwm}); thus, the more general formalism could be applied to this case.
For this paper, we will consider nonspinning primary black holes, however.

\begin{figure*}[t!]
    \centering
    \includegraphics[width=0.49\textwidth]{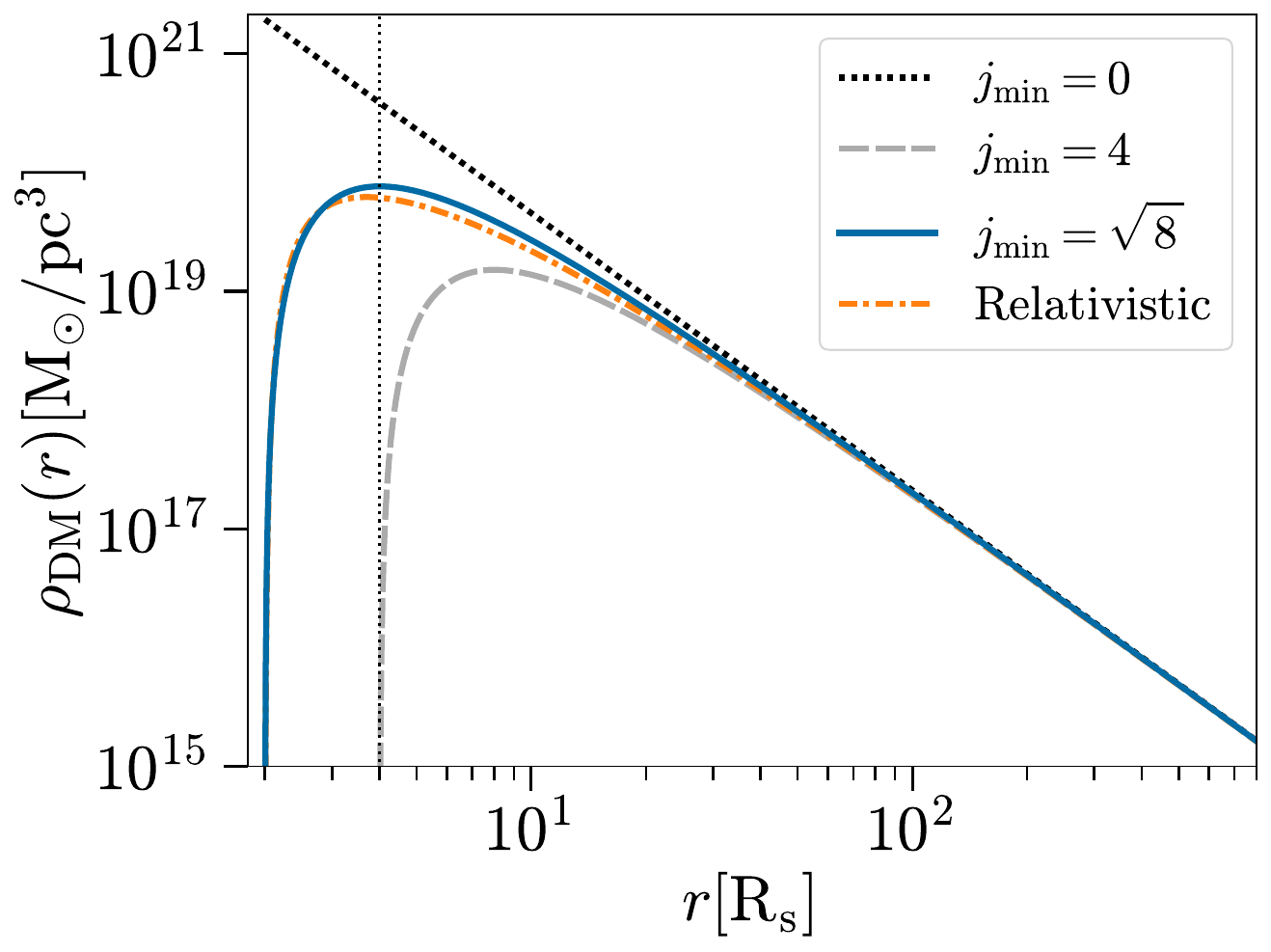} \
    \includegraphics[width=0.49\textwidth]{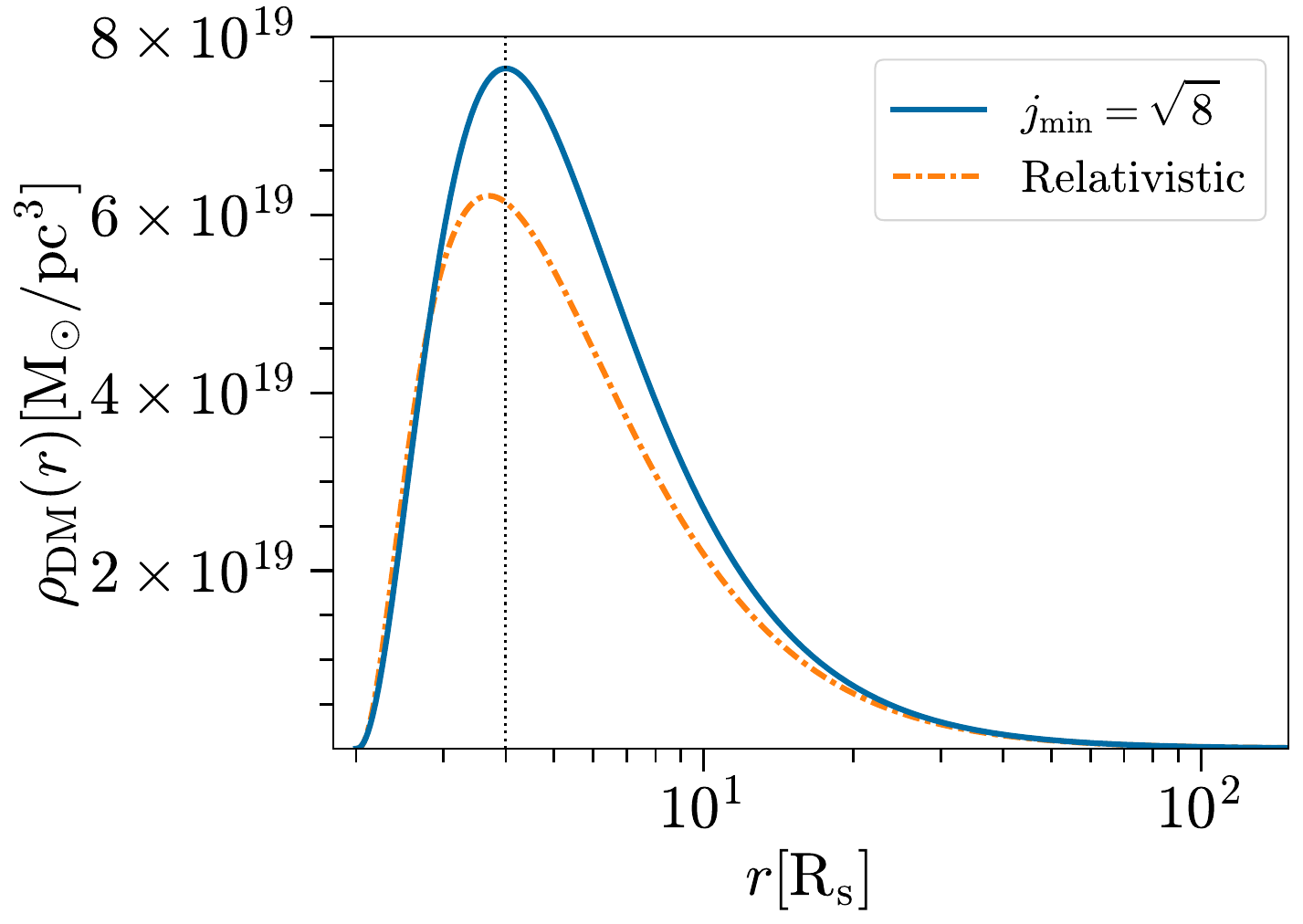}
    \caption{\textbf{Dark-matter density with different angular-momentum cutoffs}. In both panels, we show the density calculated from the distribution function given by Eq.~\eqref{eq:static_f}. 
    We consider here a primary mass $m_1 = 10^5 \, \Msolar$, $\rho_\SP = 200 \, \Msolar/\mathrm{pc}^{3}$ and $\gamma_\SP = 7/3$.
    We plot a dotted, black, vertical line at $r_{\mathcal J} = 4\,\radsch =  8Gm_1/c^2$ for reference.
    \textit{Left}: The analytic power law density ($j_\MIN= 0$) in dotted black, the fully Newtonian result ($j_\MIN= 4$) in dashed light-gray, the fully relativistic result in dash-dotted orange, and the modified Newtonian result ($j_\MIN=\sqrt 8$) in solid blue. \textit{Right}: The latter two densities, on linear scale on the vertical axis. The figure is discussed in more detail in the text of Sec.~\ref{sec:case_cutform}.}
    \label{fig:Jcut_density}
\end{figure*}

We plot several density profiles in Fig.~\ref{fig:Jcut_density} that are computed from the power-law distribution function in Eq.~\eqref{eq:static_f}.
The three curves with different values of $j_\MIN$ are computed from Eq.~\eqref{eq:density_Jvmin}; they are the power law density profile from Eq.~\eqref{eq:PL_dens} ($j_\MIN=0$, dotted black), the case in~\cite{Gondolo:1999ef} ($j_\MIN=4$, dashed light-gray), and the value with the relativistic-inspired cutoff ($j_\MIN=\sqrt 8$, solid blue). 
We additionally plot the density computed using the full relativistic method given in~\cite{Sadeghian:2013laa} as the dash-dotted orange curve.\footnote{In the relativistic case, the density is defined by $-\sqrt{-g^{00}} J_0$, where $g^{00}$ is the timelike component of the metric in Schwarzschild coordinates and $J_0$ is the timelike component of a mass current density in stationary spacetimes, again in Schwarzschild coordinates.
The integral to compute $J_0$ is given by Eq.~(3.15) in \cite{Sadeghian:2013laa}. The integration bounds are as given in Eqs.~(3.16)--(3.18) from \cite{Sadeghian:2013laa} with the upper bound in $\mathcal E$ of $1$ (in units with $G=c=1$). }
The right panel compares the density with $j_\MIN=\sqrt{8}$ and the fully relativistic density when using a linear scale on the vertical axis.
The $j_\MIN=\sqrt 8$ curve peaks at a slightly larger distance and has a larger density at the respective peaks.
The Newtonian $j_\MIN=\sqrt 8$ otherwise matches the relativistic result well.

Note that the higher density for the $j_\MIN=0$ case also produces a significant increase in the amount of mass enclosed within a given radius. 
For example, for the densities given in Fig.~\ref{fig:Jcut_density} ($m_1 = 10^5~\Msolar$, $\rho_\SP = 200~\Msolar \, \mathrm{pc}^{-3}$ and $\gamma_\SP = 7/3$), by using Eq.~\eqref{eq:enc_mass}, we determined that the mass enclosed within $\sim 17\,\radsch$ (the reason for this choice will be discussed in Sec.~\ref{sec:cut_result}) in the $j_\MIN=\sqrt{8}$ case contains about $45\%$ of the mass in the $j_\MIN=0$ case.\footnote{Note also that the $j_\MIN=\sqrt 8$ case has about about $19\%$ more enclosed mass than the fully relativistic result.}

We also briefly comment on the density of states, $g(\mathcal E)$. 
By substituting $j_\MIN= \sqrt{8}$ in Eq.~\eqref{eq:DoS_J} the factor in the parenthesis becomes $1-16\mathcal E/c^2$.
Thus, the largest energy $\mathcal E$ is given by $c^2/16$, 
which is a factor of four smaller than in the $j_\MIN=0$ case with the position-space cutoff at $r_\IN = 2\radsch$.

\subsection{Feedback formalism with a minimum angular momentum} \label{sec:case_cutfeedback}

The effects of the angular-momentum cutoff on the Chandrasekhar expression for the dynamical friction force will be described first.
Because the force depends on the density of particles, then the lower density for $j_\MIN = \sqrt 8$ than for $j_\MIN=0$ implies that the effects of dynamical friction on the binary will be smaller with a nonzero $j_\MIN$ than with $j_\MIN = 0$.
In addition, the density in the Chandrasekhar formula is that of particles moving more slowly than the orbital speed of the secondary.
The integral for the density of such particles has the same form as in Eq.~\eqref{eq:density_Jvmin}, but the maximum velocity will now be $v_2$.
Because $v_2 = \sqrt{Gm_1/r_2}$, solving $v_2 = v_\MIN(r_2)$ gives that the two are equal at a radius $r = 4\radsch$.
There are no particles with speeds slower than the orbital speed at radii smaller than $4\radsch$; thus, the Chandrasekhar formula for dynamical friction implies that the effects of dynamical friction will stop before the secondary reaches the ISCO, for $j_\MIN = \sqrt 8$.

There also will be differences in both DF and SA feedback with a nonzero $j_\MIN$.
With $j_\MIN=0$, the bounds of integration in $r$ for the radial integral in Eq.~\eqref{eq:DF_rate_dr3} are $r_{\mathrm{cut}}$ as the lower and $Gm_1/\mathcal E$ as the upper.
A nonzero $j_\MIN$ restricts the upper bound to be $r_+$, which is smaller.
It does not affect the lower bound $r_{\mathrm{cut}}$ for $r_2 > r_{\mathcal J}$ but the lower bound becomes $r_-$ for $r_2 \leq r_{\mathcal J}$, as discussed in more detail in Sec.~\ref{sec:DF_genform}. 
The smaller range of position space for which particles of a given energy $\mathcal E$ can scatter could lead to a small decrease in the amount of feedback from dynamical friction onto the dark-matter spike.
However, there is also an enhancement of the scattering rate for energies near $\mathcal E_\MAX(r_{\mathcal J})$, which is discussed at the end of Appendix~\ref{app:efficiency}.
The impacts of these differences will be determined most straightforwardly when we discuss the results of numerical simulations of IMRIs in Secs.~\ref{sec:cut_result} and~\ref{sec:successive_merger}.

We briefly comment on dynamical heating, as discussed in Footnote~\ref{fn:heating}, in the context of a nonzero angular-momentum cutoff.
This process could increase a DM particle's relative specific energy in excess of the maximum $c^2 / (2 j_\MIN^2)$, leading it to be captured by the primary.
However, we argued in Footnote~\ref{fn:heating} that dynamical heating is negligible for particle dark matter and for the distribution functions considered in this paper.
Thus, we do not consider such an effect, which is consistent with our (well-justified) neglect of dynamical heating in the dynamics of the binary.

For secondary accretion, the effective force that arises from the increasing mass of the secondary is proportional to the density.
This implies that like the Chandrasekhar dynamical friction force, the amount of secondary accretion will decrease for a nonzero $j_\MIN$ compared to $j_\MIN=0$, simply because the density is smaller in the former case.
The feedback from secondary accretion also decreases somewhat when $j_\MIN$ is nonzero, aside from the subtleties as $\mathcal E$ approaches $v_{\mathcal J}^2/2$ and $r_2$ approaches $r_{\mathcal J}$ (as discussed in more detail in~\ref{sec:SA_genform}).  
As with dynamical friction, the effects of these differences can be established most straightforwardly by performing numerical simulations (which we discuss next).

\section{Simulations and results} \label{sec:cut_result}

In this section, we discuss simulations of IMRIs in a DM distribution with and without an angular-momentum cutoff of $j_\MIN$.
We first review the simulations that we performed, and then present the results for the GW dephasing results and DM distribution during and after the inspiral.

\subsection{Simulations and initial conditions} \label{subsec:simsICs}

We simulate the inspiral of five binaries with different mass ratios using a version of the \textsc{HaloFeedback} code~\cite{HaloFeedback} which was modified to have a nonzero $j_\MIN$.
This code is named \textsc{HaloFeedbackAcc} and is available on GitLab~\cite{HaloFeedbackAcc}.
We initialize the mass of the secondary to be $m_2 = 10\,\Msolar$, and we vary the primary mass.
The five values we choose are $m_1 =$ $10^3 \, \Msolar$, $m_1 =$ $3 \times 10^3 \, \Msolar$, $m_1 = 10^4 \, \Msolar$, $m_1 = 3 \times 10^4 \, \Msolar$ and $m_1 = 10^5 \, \Msolar$.
For each simulation, we initialize the dark-matter distribution function to that given in Eq.~\eqref{eq:static_f}, with $\rho_\SP$ = $200~\Msolar/$pc$^3$ and $\gamma_\SP$ = $7/3$.\footnote{These values of $\rho_\SP$ and $\gamma_\SP$ are comparable to those used in the study~\cite{Eda:2014kra}, which have been used as a benchmark system in several subsequent works on this subject (see, e.g.,~\cite{Yue:2017iwc,Edwards:2019tzf,Kavanagh:2020cfn,Coogan:2021uqv,Nichols:2023ufs,Wilcox:2024sqs}). 
While it would be of interest to study a wider range of this parameter space and to investigate any correlations in this parameter space on the gravitational-wave observables, we leave such studies to future work.} 
We will call an inspiral in such an initial dark-matter distribution a ``first-generation'' inspiral.
The reason for this naming is that the density profile is consistent with a primary black hole that grew adiabatically in mass and did not undergo a previous IMRI inspiral (which would have modified the distribution of dark matter through feedback).

The \textsc{HaloFeedback} code solves the coupled system of the ordinary and integral-partial differential equations that describes the evolution of the orbital separation of the secondary and of the dark-matter distribution function: namely, Eqs.~\eqref{eq:dot_r} and~\eqref{eq:fIPDEcap}.
The code uses an adaptive time step, which is a multiple of the instantaneous orbital period of the binary. 
We used a maximum time step of $50$ orbital periods, meaning our resolution for the number of GW cycles for the quadrupole waves is of order $100$. 
The timestep can be smaller than 50 orbital periods during the inspiral.

There are some subtleties in choosing the initial conditions for the inspiral, discussed in~\cite{Kavanagh:2020cfn,Nichols:2023ufs}, which pertain to how to choose initial binary separations and DM initial conditions that are consistent with an adiabatic inspiral from large separations. 
Because the secondary accretes and redistributes dark matter throughout the inspiral, an inspiral starting from a given radius will not be the same as one that starts from a larger radius and inspirals towards the first, given radius.
However, the effects of DF and SA feedback on the DM distribution are localized around the secondary.
Together these facts imply that while the DM density will depend on the formation history of the binary, if the initial separation of the binary is sufficiently large, then it is possible to have an inspiral at smaller radii that is consistent with an adiabatic inspiral from a much larger radius.
In~\cite{Kavanagh:2020cfn,Nichols:2023ufs}, it was found that an initial binary separation was $r_2 = 3 r_\fy$ (where $r_\fy$ is the radius from which the secondary inspirals to the ISCO in four years for a vacuum binary) was sufficiently large.
We will use this same initial separation in the simulations in this work, though we perform additional simulations to verify that it is still a valid choice for $j_\MIN\neq 0$.

We also note a difference in how we evolve the binary separation $r_2$ that differs from what was performed in~\cite{Nichols:2023ufs}.
In~\cite{Nichols:2023ufs}, the initial value of the mass ratio $q$ was used in the evolution equations for $r_2$ in Eq.~\eqref{eq:dot_r}.
Here, however, we use the time-dependent value of $m_2$ and of the mass ratio in these evolution equations.
This produces an order $m_\acc/m_2$ relative difference in $\dot r_2$, where $m_\acc$ is the mass accreted by the secondary since the beginning of the inspiral (i.e., from $3r_\fy$). 
Upon integrating the equations of motion, the number of GW cycles changes by a relative amount that can be as large as of order $10^{-3}$ (the order of the value of $m_\acc/m_2$).
The term $\dot r_2^\mathrm{RR}$ in the evolution equation for $\dot r_2$ is responsible for most of this change.

The time dependence of $m_2$ in the evolution equations led us to change how we compute the dephasing, which we describe in the next subsection.

We have made the data files for the DM density profiles from each simulation available to download from~\cite{wade_2025_17888476}. 
These data files contain the densities when the secondary is at $3r_\fy$, at $r_\fy$, and at ISCO.
Data for $j_\MIN = 0$ (first-generation inspirals) and $j_\MIN = \sqrt{8}$ (first- and second-generation inspirals) are available at~\cite{wade_2025_17888476}.

\subsection{Gravitational-wave dephasing} \label{sec:cut_result_dephase}

\subsubsection{Definitions and subtleties in computing the dephasing}

\begin{table*}[t!]
    \centering
    \caption{\textbf{Number of gravitational-wave cycles, dephasing, and characteristic masses for inspirals with different $\boldsymbol{j_\MIN}$}. 
    We denote $N_\mathrm{cycles}^{\mathrm{(A)}}$ [see Eq.~\eqref{eq:NcyclesA}] as the number of GW cycles between a given starting radius and ISCO for a simulation corresponding to case $\mathrm{A}$, where $\mathrm{A = 1}$ is the case with no cutoff in angular momentum for a first-generation inspiral, and $\mathrm{A = 2_I}$ is the case with the angular momentum cutoff with $j_\MIN = \sqrt{8}$ for a first-generation inspiral. 
    The dephasing $\Delta N_\mathrm{cycles}^\mathrm{(0-A)}$ [see Eq.~\eqref{eq:DeltaNcyclesAB}] is the difference in cycles between a vacuum inspiral and case $\mathrm{A}$ (both of which are computed from the starting radius of $r_\fy^\mathrm{(A)}$ described in the text of Sec.~\ref{sec:cut_result_dephase}).
    For the vacuum system, we use $m_2 = 10 \, \Msolar + \delta m_\fy^\mathrm{(A)}$ [see Eq.~\eqref{eq:m2A4y}], and list $\delta m^\mathrm{(A)}_\fy$ (again for the cases $\mathrm{A = 1}$ and $\mathrm{A = 2_I}$).
    The dephasing is computed for the last four years of inspiral before the secondary reaches the ISCO, using the initial separations $r_\fy^\mathrm{(A)}$.
    As summarized in Sec.~\ref{subsec:simsICs}, the initial DM distribution is characterized by $\gamma_\SP = 7/3$, and $\rho_\SP = 200 \, \Msolar/\mathrm{pc}^{3}$.}
    \begin{tabular}{ccccccc}
    \hline
    \hline
    $m_1 [\Msolar]$ & $N_\mathrm{cycles}^{(1)}$ & $\Delta N_\mathrm{cycles}^\mathrm{(0-1)}$ & $\delta m_\fy^{(1)} [\Msolar]$ & $N_\mathrm{cycles}^\mathrm{(2_I)}$ & $\Delta N_\mathrm{cycles}^\mathrm{(0-2_I)}$ & $\delta m_\fy^\mathrm{(2_I)} [\Msolar]$ \\
    \hline
    $10^{3}$ & 2,092,200 & 1,300 & 0.0332 & 2,092,300 & 1,200 & 0.0330 \\
    $3\times 10^{3}$ & 1,587,100 & 2,400 & 0.0504 & 1,587,400 & 2,200 & 0.0496 \\
    $10^{4}$ & 1,171,800 & 4,400 & 0.0390 & 1.172,900 & 3,400 & 0.0380 \\
    $3\times 10^{4}$ & 886,100 & 6,200 & 0.0219 & 888,400 & 4,100 & 0.0209 \\
    $10^{5}$ & 649,900 & 4,300 & 0.0128 & 652,000 & 2,200 & 0.0113 \\
    \hline
    \hline
    \end{tabular}
     \label{tab:dephase_VAPLCI}
\end{table*}

Dynamical friction and secondary accretion cause the binary to inspiral more rapidly than without these effects in vacuum. 
This leads to a dephasing of the gravitational-wave signal from an analogous vacuum inspiral, given that the binary undergoes fewer orbital cycles before reaching the ISCO. 
One can compute the number of cycles in the time domain for a case $\mathrm{A}$ by
\begin{equation} \label{eq:NcyclesA}
        N_\mathrm{cycles}^\mathrm{(A)} = \frac 1\pi \int_{t_\mathrm{i}}^{t^\mathrm{(A)}_{\ISCO}} \! \Omega \, dt \, .
\end{equation}
We have used the convention that the time is the same value $t=t_\I$ for all cases when the secondary is at the initial radius, and the time at which the particle reaches the ISCO in a case $\mathrm A$ is $t^\mathrm{(A)}_{\ISCO}$.
As in~\cite{Nichols:2023ufs}, we choose $\mathrm A=0$ to denote the vacuum case; here we choose $1$ to the case with dark matter and with $j_\MIN=0$ and $\mathrm{2_{I}}$ to the be with dark matter, with $j_\MIN=\sqrt 8$ and for a first-generation inspiral.\footnote{In~\cite{Nichols:2023ufs}, the case $\mathrm A = 1$ corresponded to a hypothetical, unphysical scenario in which dynamical friction and feedback occurs, but accretion does not occur. 
We do not consider such a case in this paper, because Ref.~\cite{Nichols:2023ufs} demonstrated that both dynamical friction and accretion (with both types of feedback) should be treated to accurately represent the DM effects on the orbital dynamics, the DM distribution, and the gravitational waves emitted from these systems.
Thus, we will compare simulations with all the relevant DM effects against vacuum systems without any DM effects in this paper.}

The difference in the number of GW cycles between two cases gives the GW dephasing, which we denote by
\begin{equation} \label{eq:DeltaNcyclesAB}
    \Delta N^\mathrm{(A - B)}_\mathrm{cycles} = N^\mathrm{(A)}_\mathrm{cycles} - N^\mathrm{(B)}_\mathrm{cycles} 
\end{equation}
for two cases $\mathrm A$ and $\mathrm B$.
The dephasing gives a simple qualitative assessment of the extent to which two waveforms agree or disagree.
More quantitative assessments of distinguishability require computing Bayes factors or evidence ratios between different models.

Because the secondary accretes mass throughout the inspiral, there are some subtleties that arise when computing the dephasing of a system with dark matter against an ``equivalent'' vacuum system, which we now discuss.

First, note that in Ref.~\cite{Nichols:2023ufs} such subtleties did not arise, because $m_2$ was held fixed in the evolution equations for the secondary, so that the chirp mass remained constant during the inspiral.
However, a fractional change in $m_2$ of order $10^{-3}$ produces a comparable fractional change in the chirp mass (and thus also a comparable fractional change in the number of gravitational-wave cycles).
For systems that undergo of order $10^6$ cycles, this is of order $10^3$ cycles.
Thus, it is important to account for the changing secondary mass during the inspiral in the equations of motion, so we do not repeat the approach of~\cite{Nichols:2023ufs} here.

Next, as discussed in Sec.~\ref{subsec:simsICs}, to obtain appropriate initial conditions at the radius $r_\fy$, we begin evolving the binary at a separation of $3r_\fy$.
As it evolves between these radii, the mass changes from the initial mass $m_{2,0} = 10 \, \Msolar$ by an amount which we denote $\delta m_\fy$.
Because we aim to quantify the effects of dark matter on the inspiral from a most closely equivalent vacuum system during the final four years of the inspiral, we must consider what is a reasonable choice of ``most closely equivalent.''
A natural definition is a vacuum IMRI that has the same mass $m_2 + \delta m_\fy$ at the radius $r_\fy$ as the IMRI with dark matter.
Such a vacuum system will have its gravitational radiation reaction be equivalent at the radius $r_\fy$ to that of the IMRI with dark matter, so any changes to the inspiral from that radius inward will arise from DM effects only.

Finally, there are yet more subtle aspects to the comparison against an ``equivalent vacuum IMRI'' that we now discuss:
(i) The change in the mass $\delta m_\fy$ should, in fact, be labeled $\delta m_\fy^\mathrm{(A)}$, because it will depend on which case $\mathrm A$ we are considering. 
(ii) We do not know the value of $m_\fy^\mathrm{(A)}$ \emph{a priori}; it is determined after running a simulation from the radius $3 r_\fy$ to $r_\fy$ 
(iii) We determine $r_\fy$ from the equivalent vacuum binary with a secondary mass that we denote by $m^\mathrm{(A)}_\fy$, which is defined from
\begin{equation} \label{eq:m2A4y}
    m_\fy^\mathrm{(A)} = m_{2,0} + \delta m_\fy^\mathrm{(A)} 
\end{equation}
(with $m_{2,0} = 10\, \Msolar$, as before). 
This implies that there is also an \emph{a priori} unknown radius $r_\fy^\mathrm{(A)}$, the initial radius to inspiral to the ISCO in four years of the equivalent vacuum IMRI with mass $m_\fy^\mathrm{(A)}$.

We determine $m_\fy^\mathrm{(A)}$ and $r_\fy^\mathrm{(A)}$ approximately by first computing $m_\fy^\mathrm{(A)}$ at the radius $r_\fy$ for a vacuum system with $m_2 = 10 \, \Msolar$, and then we determine $r_\fy^\mathrm{(A)}$ for the vacuum system using the value $m_\fy^\mathrm{(A)}$ from the first step.
Thus, our equivalent vacuum IMRI that we use for computing dephasing numbers and curves will be vacuum IMRI with mass $m_\fy^\mathrm{(A)}$ starting from a radius $r_\fy^\mathrm{(A)}$, with $m_\fy^\mathrm{(A)}$ and $r_\fy^\mathrm{(A)}$ computed via the approximate procedure defined in this paragraph.

\subsubsection{Discussion of dephasing results}

We present in Table~\ref{tab:dephase_VAPLCI} the cycles and dephasing against vacuum for the five simulations described in Sec.~\ref{subsec:simsICs}. 
We also list $\delta m_\fy^{(1)}$ and $\delta m_\fy^\mathrm{(2_I)}$ in Table~\ref{tab:dephase_VAPLCI}.
The results for $\Delta N_\mathrm{cycles}^\mathrm{(0-1)}$ are similar to those in~\cite{Nichols:2023ufs}.
However, because we use a somewhat lower density of $\rho_\SP = 200 \, \Msolar/\mathrm{pc}^{3}$ here and include the evolution of the mass $m_2$ in the equations of motion for $r_2$, the two cases are not directly comparable.
The trend in $\Delta N_\mathrm{cycles}^\mathrm{(0-1)}$ as a function of $m_1$, and the fact that the largest absolute and fractional dephasing against vacuum with $j_\MIN = 0$ happens with a primary mass of $m_1 = 3 \times 10^4~\Msolar$, is the same as in~\cite{Nichols:2023ufs}. 
Therefore, we refer to~\cite{Nichols:2023ufs} for the discussion of this trend.
Here we focus on any differences in the dephasings $\Delta N_\mathrm{cycles}^\mathrm{(0-1)}$ and $\Delta N_\mathrm{cycles}^\mathrm{(0-2_I)}$ in the third and sixth columns, respectively, of Table~\ref{tab:dephase_VAPLCI}.

The values of $\Delta N_\mathrm{cycles}^\mathrm{(0-1)}$ and $\Delta N_\mathrm{cycles}^\mathrm{(0-2_I)}$ were computed from different initial radii (starting from $r_\fy^\mathrm{(A)}$ in each case, respectively), which should be considered when comparing these values for the same primary mass $m_1$.
Nevertheless, we can identify general trends in the effect of introducing the angular-momentum cutoff on the dephasing against vacuum for a given $m_1$.
The largest absolute difference in dephasing between simulations with $j_\MIN=\sqrt 8$ and $j_\MIN = 0$ occurs for two values: $m_1 = 3\times 10^4 \, \Msolar$ and $m_1 = 10^5 \, \Msolar$.
Given the smaller number of cycles for $m_1 = 10^5 \, \Msolar$, the largest fractional difference occurs for $m_1 = 10^5 \, \Msolar$. 
As the mass ratio becomes more extreme, the secondary orbits at a smaller number of gravitational radii during the last four years of the inspiral than it does for less extreme mass ratios.
Because the densities with $j_\MIN = \sqrt 8$ differ most from those with $j_\MIN=0$ near the ISCO, it is reasonable that a nonzero $j_\MIN$ has the largest effect for these more extreme mass ratios.
That this difference is the same at both $m_1 = 3\times 10^4 \, \Msolar$ and $m_1 = 10^5 \, \Msolar$ is likely coincidental.
However, the trend that the total number of orbital cycles between $r_\fy^\mathrm{(A)}$ and ISCO decreases as the mass ratio becomes more extreme means the number of cycles of dephasing also decreases.

In~\cite{Nichols:2023ufs}, it was found that accretion between $r_\fy$ and $r_\ISCO$ increased as the mass ratio became less extreme, whereas here we find that both $\delta m_\fy^{(1)}$ and $\delta m_\fy^\mathrm{(2_I)}$ (the accretion that takes place between $3r_\fy$ and $r_\fy^\mathrm{(A)}$) is largest for $m_1 = 3\times 10^3 \Msolar$ and decreases for both less and more extreme mass ratios.
Although there is no reason that in different spatial regions the accretion should behave similarly as a function of mass ratio, one might have anticipated a common trend.
Unlike the the accretion between $r_\fy$ and $r_\ISCO$ in~\cite{Nichols:2023ufs}, the values of $\delta m_\fy^\mathrm{(A)}$ are affected by the initial conditions at $3r_\fy$ that are inconsistent with an adiabatic inspiral from larger radii.
Thus, we do not dwell on the details of the $m_1$ dependence of the results for $\delta m_\fy^\mathrm{(A)}$ given the unrealistic initial conditions that necessarily arise in their computation.
Nevertheless, we speculate that the $\delta m_\fy^\mathrm{(A)}$ are smaller for $m_1 = 10^3\,\Msolar$ than for $m_1 = 3\times 10^3\,\Msolar$, because DF feedback and SA feedback are more efficient for the $10^3\,\Msolar$ case, and because the $10^3\,\Msolar$ case starts at a larger number of gravitational radii from $m_1$, where the corresponding densities are lower.

\begin{figure}[t!]
    \centering
    \includegraphics[width=\columnwidth]{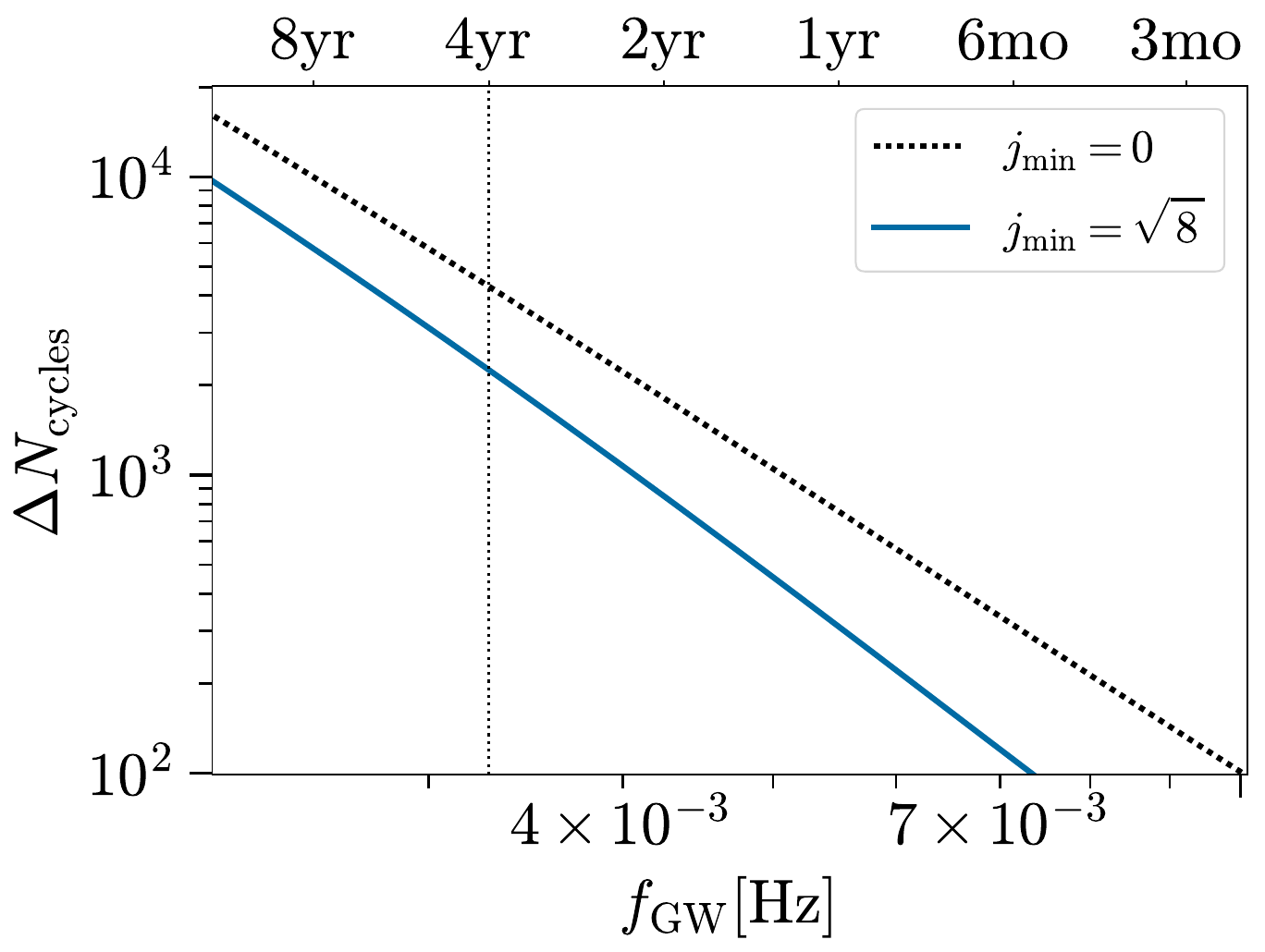} \
    \caption{\textbf{Cycles of dephasing versus gravitational-wave frequency with a zero and nonzero angular-momentum cutoff}. 
    We show the GW dephasing relative to a vacuum system for an inspiral in an initial dark-matter distribution characterized by $\rho_\SP = 200~\Msolar \, \mathrm{pc}^{-3}$ and $\gamma_\SP = 7/3$ (and a primary mass $m_1 = 10^5~\Msolar$). 
    The dotted black line is the dephasing using the power-law density profile without enforcing the cutoff in angular momentum, while the solid blue is the dephasing against vacuum using the density with a nonzero angular-momentum cutoff. 
    A vertical, black, dotted line is plotted at the frequency four years from ISCO, $r_\fy$, which was computed using the prescription in Sec.~\ref{sec:cut_result_dephase} (there is further discussion of this figure there, as well).}
    \label{fig:dephase_PL1_100K}
\end{figure}

A plot of the dephasing as a function of frequency is shown in Fig.~\ref{fig:dephase_PL1_100K} for a primary mass of $m_1 = 10^5\,\Msolar$ and for two different values of $j_\MIN$. 
The dotted black line corresponds to the dephasing against vacuum for the density with $j_\MIN=0$, whereas the solid blue is that for density with $j_\MIN=\sqrt{8}$. 
There is a consistently smaller dephasing at all frequencies for the later case.
We discuss the reason for this behavior in more detail below, as there are, in fact, two competing effects that arise from introducing the cutoff in angular momentum.

First, we compute the dynamical-friction force and the effective force from accretion from expressions that are proportional to the density of particles at the location of the secondary.
The lower density with the angular-momentum cutoff (see Fig.~\ref{fig:Jcut_density}) leads to smaller DM effects on the binary's orbit, and thus smaller dephasing effects.
This is consistent with the results in Fig.~\ref{fig:dephase_PL1_100K}, where the dephasing is smaller at all the frequencies illustrated therein. 
There is, however, a second effect from the angular-momentum cutoff that, in principle, could increase the amount of dephasing: namely, for most energies and radii the angular-momentum cutoff decreases the scattering and accretion probabilities, so that the feedback rate is smaller.
This can be more easily verified for the secondary-accretion rate given its simpler expression, though it is also the case for the dynamical-friction rate.

The difference in the accretion or scattering rates is several percent with and without the cutoff, for most values of $r_2$ and $\mathcal E$.
The net effect of this difference on the dephasing turns out to be similarly small.  
In addition to the dephasing numbers given in Table~\ref{tab:dephase_VAPLCI}, we also simulated binaries in which the density was computed assuming $j_\MIN=\sqrt 8$, but the scattering and accretion probabilities were computed assuming $j_\MIN = 0$.
If we label this case as ``$\mathrm{1B}$'', then the difference in the dephasings $\Delta N_\mathrm{cycles}^\mathrm{(0-2_I)}$ and $\Delta N_\mathrm{cycles}^\mathrm{(0-1B)}$ was less than 100 for these runs and therefore below the accuracy at which we give our results. 
Therefore, the majority of the dephasing between the cases with and without the angular-momentum cutoff arises from the differences in densities, rather than the changes to the accretion or scattering rates.

We conclude this subsection with a remark about the choice of the starting separation $r_2 = 3 r_\fy$, which we selected to obtain reasonable initial conditions for the density at $r_2 = r_\fy^\mathrm{(A)}$, the largest radius where we compute the dephasing against vacuum binaries.
One might wonder whether introducing the cutoff in angular momentum has any effect on this starting separation, which was determined in~\cite{Kavanagh:2020cfn,Nichols:2023ufs} to be an appropriate choice for $j_\MIN=0$.\footnote{Specifically, in~\cite{Kavanagh:2020cfn,Nichols:2023ufs} it was noted that, when $r_2=3r_\fy$, the secondary's influence on the DM distribution was small at radii smaller than $2 r_\fy$, so that the density at $r = r_\fy$ was largely unaffected. For this reason, initializing the secondary at $r_2 = 3r_\fy$ was taken to be consistent with an adiabatic inspiral from further out.}
We performed a somewhat different test here. 
Specifically, we simulated inspirals starting from $r_2 = 2 r_\fy$ and $r_2 = 4 r_\fy$, and we compared the dephasing from these simulations against those given in Table~\ref{tab:dephase_VAPLCI}. 
The magnitude of the change in dephasing depends on the mass ratio. 
Nevertheless, the difference in dephasing, when comparing the simulations starting at $r_2 = 4 r_\fy$ against those at $r_2 = 3 r_\fy$, was $\mathcal O(10)$ cycles or fewer, for the cases that we considered (again below the accuracy at which we present our results). 
For a starting separation of $r_2 = 2 r_\fy$, there were differences in the dephasing of as much as a few hundred cycles, which is larger than the accuracy of our results. 
Thus, the starting separation of $3r_\fy$ is also appropriate for the cutoff of $j_\MIN = \sqrt 8$.

\subsection{Dark-matter density} \label{sec:cut_result_density}

\begin{figure*}[t!]
    \centering
    \includegraphics[width=0.49\textwidth]{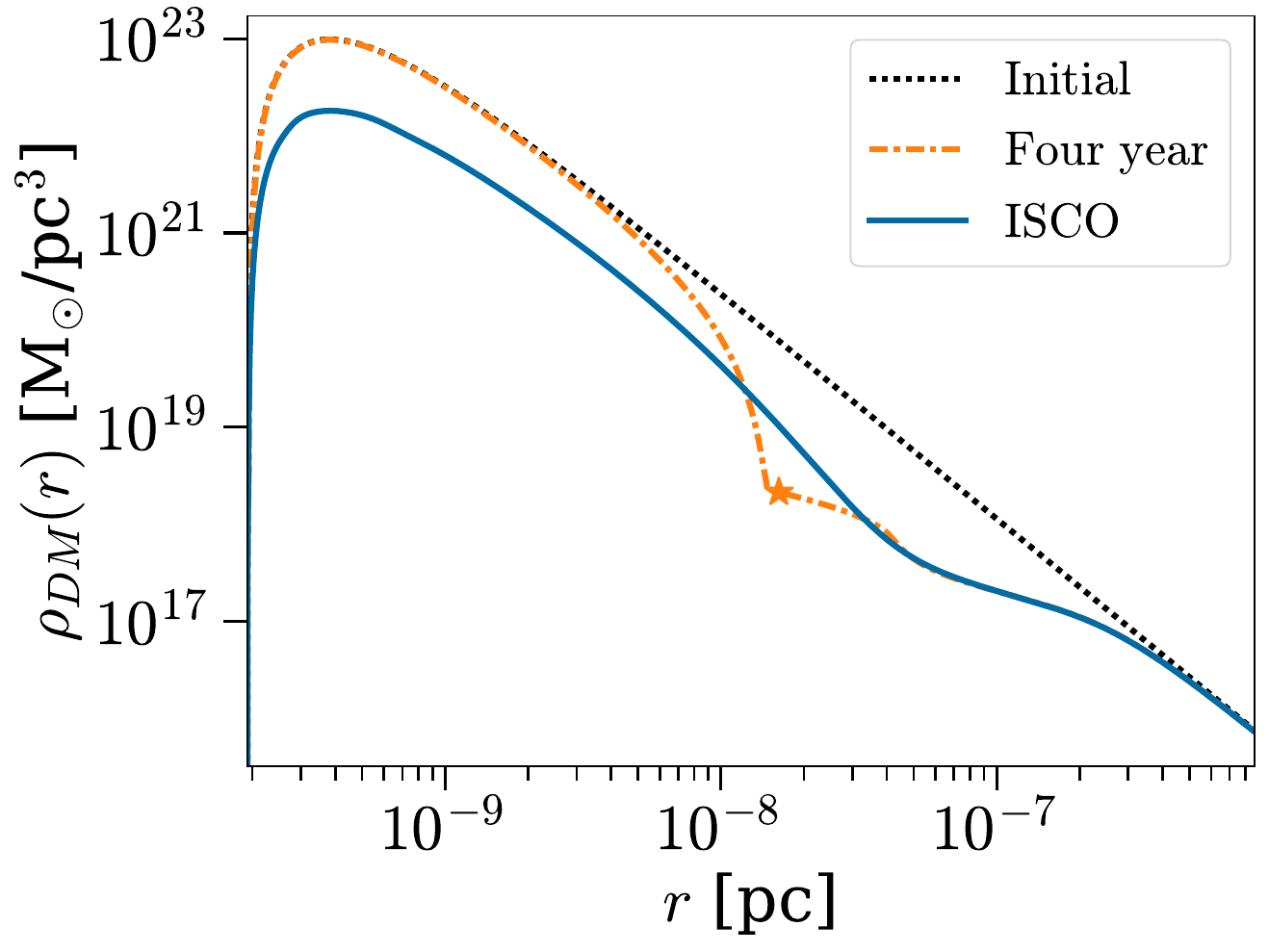} \
    \includegraphics[width=0.49\textwidth]{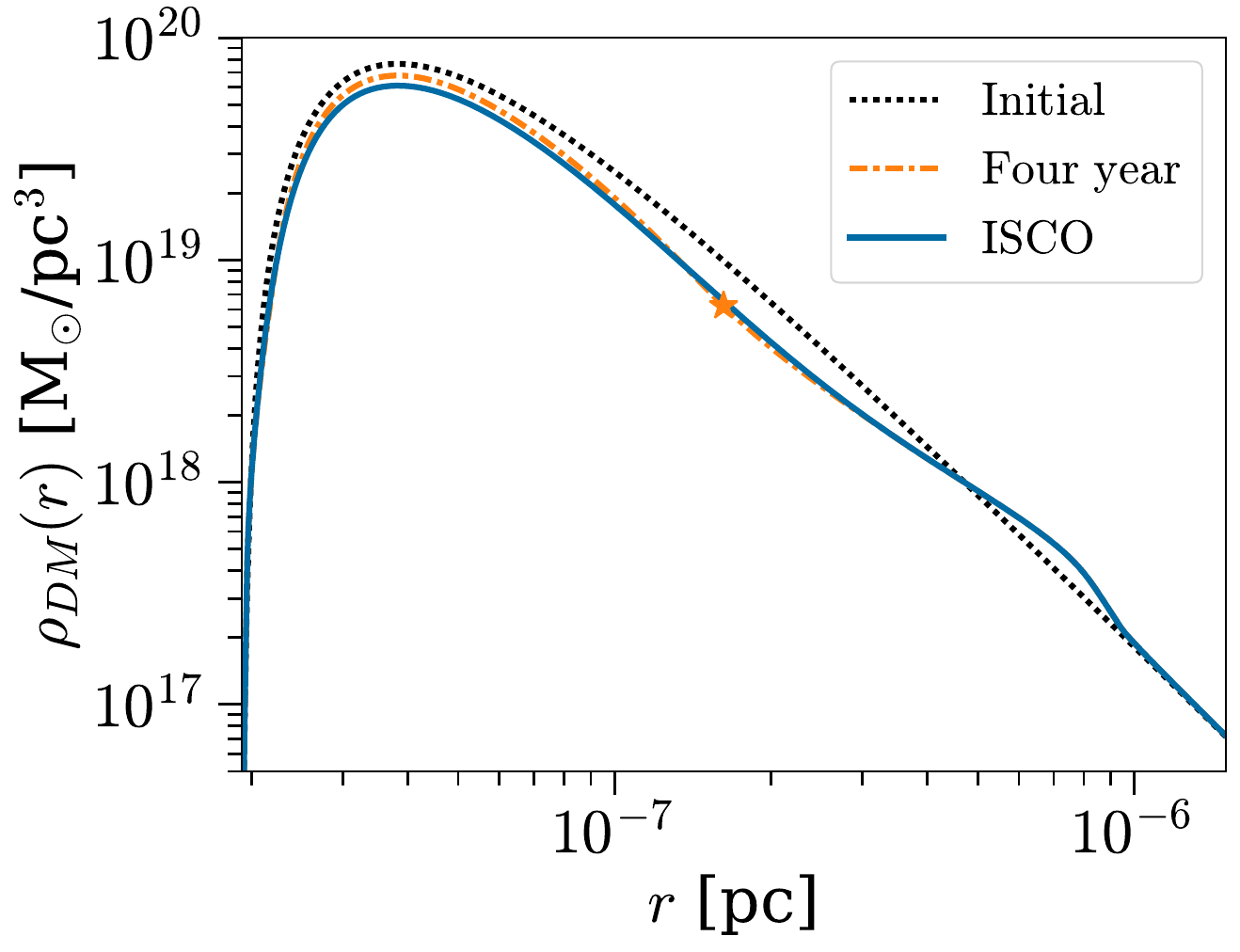}
    \caption{\textbf{Dark-matter density at three times during an inspiral}. 
    In all the times depicted in both panels, the density has a cutoff in the angular momentum ($j_\MIN = \sqrt{8}$) so that it vanishes at $r=2\radsch$. 
    We chose $\rho_\SP = 200\,\Msolar \, \mathrm{pc}^{-3}$ and $\gamma_\SP = 7/3$ for the dark-matter density in both panels and $m_1 = 10^3\,\Msolar$ (\emph{left}), or $m_1 = 10^5\,\Msolar$ (\emph{right}), for the primary mass. 
    The dotted black line is the initial density profile (when $r_2 = 3 r_\fy$, but before the evolution begins), the dash-dotted orange curve is the density when $r_2 = r_\fy$, and the solid blue is the density when the secondary reaches the ISCO. 
    Further discussion of the figure is given in Sec.~\ref{sec:cut_result_density}.}
    \label{fig:density_PL1_1K}
\end{figure*}

Figure~\ref{fig:density_PL1_1K} shows the dark-matter density at three stages during the inspiral of a binary with $m_{2,0} = 10\,\Msolar$ and $m_1 = 10^3\,\Msolar$ (left) or $m_1 = 10^5\,\Msolar$ (right).
The initial density in each panel is the dotted black curve. 
In the left panel, the initial binary separation is $3 r_\fy \approx 5 \times 10^{-8}\,\mathrm{pc}$, whereas in the right panel it is $3 r_\fy \approx 5\times 10^{-7}$\,pc. 
The density for a binary separation of $r_2$ = $r_\fy$ is the dash-dotted orange line, and the solid blue curve is the density when the secondary reaches the ISCO (the end of our simulation) in both panels. 
The location of the secondary at $r_2 = r_\fy$ is marked with an orange star. 

The combination of DF and SA feedback has a significant effect on the DM density, and the effect is stronger for the $m_1=10^3 \Msolar$ case. 
Throughout the inspiral, feedback from dynamical friction depletes the density of particles near the secondary by scattering them to larger radii.
Secondary accretion leads to a decrease in the density by removing DM particles from the distribution function. 
The decrease in the density between the initial time and final time in the simulation (the ISCO case) is around an order of magnitude at radii smaller than $3r_\fy$, and somewhat larger at the larger end of this interval.
At radii larger than $r_\fy$, the densities look qualitatively similar to those with $j_\MIN=0$ (see the right panel of Fig.~6 in~\cite{Nichols:2023ufs}). 
However, there are more significant differences between the $j_\MIN=0$ in~\cite{Nichols:2023ufs} and the $j_\MIN=\sqrt 8$ for radii less than $r_\fy$.

In the left panel of Fig.~\ref{fig:density_PL1_1K} ($m_1 = 10^3\,\Msolar$), the density after the inspiral (the ISCO case) decreases by almost an order of magnitude for $r \lesssim 10^{-9}$\,pc, whereas it was closer to a factor of a few for the $j_\MIN=0$ given in~\cite{Nichols:2023ufs}.
As in the $j_\MIN=0$ case, DF feedback has a larger impact on the density than SA feedback for this region as a whole.
Specifically, the mass enclosed within $r_\fy$ is initially roughly $5 \times 10^{-3}\,\Msolar$, and it changes by roughly $4 \times 10^{-3}\,\Msolar$ between the start of the simulation and the end. 
We can compute how much of the mass is accreted by the secondary (about $30\%$) and how much is ejected from the spike (about $10\%$).
The remaining amount ($60\%$ of the mass decrease) arises from the scattering particles from smaller radii to larger radii, which is characteristic of dynamical friction.

This global measure of the change in mass in this region does not give complete information about how the density changes at a given radius.
It was noted in Sec.~\ref{sec:SA_genform} that particles at energies close to $\mathcal E_\MAX(r_{\mathcal J})$ are accreted much more efficiently.
Particles with energies this large must be located at radii close to $r_{\mathcal J}$, near the peak of the density.
Thus, we would anticipate that the density of particles near the peak would be depleted more significantly than it is at other radii.
This is consistent with the density shown in Fig.~\ref{fig:density_PL1_1K}.

In the right panel of Fig.~\ref{fig:density_PL1_1K} ($m_1 = 10^5\,\Msolar$), we observe that the density is not changed as significantly as it is in the $m_1 = 10^3\,\Msolar$ case.
There still is a decrease in the density because of dynamical friction and accretion onto the secondary. 
Both dynamical friction and accretion affects are proportional to the mass ratio, so it is not surprising that they become less efficient for more-extreme mass ratios. 
The presence of the increased density for $r \gtrsim 4 \times 10^{-7}$\,pc, when the secondary reaches $r_\fy$ and $r_\ISCO$, occurs because dynamical friction scatters particles to larger radii, but secondary accretion is not sufficiently strong to accrete these particles (as it was in the $m_1 = 10^3 \Msolar$ case). 

As in~\cite{Nichols:2023ufs}, we remind the reader that care should be taken when interpreting the density for the larger radii depicted in Fig.~\ref{fig:density_PL1_1K}. 
While the choice of starting the inspiral at $3 r_\fy$ was made to create a density at $r < r_\fy$ consistent with an adiabatic inspiral from larger radii, the form of the density for larger $r$ is significantly affected by the choice of this initial condition.
Thus, the region with $r > r_\fy$ would not be the density obtained from an inspiral from a larger radius (or another formation scenario for the binary).

\section{Simulations of successive mergers} \label{sec:successive_merger}

The results in Sec.~\ref{sec:cut_result} show that the radial dependence of the dark-matter density, even close to the ISCO of the primary black hole, can make a significant difference on the rate of inspiral (and thus the dephasing from a vacuum binary). 
The detailed profile of the initial DM density before the inspiral, however, would depend on the formation scenario of the spike, including if the primary black hole had previously merged with a secondary compact object prior in its history. 
An assumption underlying the initial data in Sec.~\ref{sec:cut_result} was that the secondary inspiraled and circularized from large radii in a dark-matter spike in which there were no prior IMRI or EMRI mergers.
In this section, we relax the assumption that there were no prior mergers and investigate second-generation mergers in which the IMRI is the second IMRI to merge in the dark-matter spike and the DM density through which the second IMRI inspirals is the same as that produced after the first IMRI.

We briefly discuss why this is a probable merger scenario.
First, IMBHs may form in a variety of environments, including within globular clusters or at the center of dwarf galaxies (see, e.g.,~\cite{LISA:2022yao}). 
These environments can host a population of stellar-mass black holes in proximity to the IMBH. 
For clumpy star-forming clusters with a central IMBH surrounded by stellar-mass objects (stars and compact objects), \cite{Pestoni:2020alb} predicts an IMRI rate of up to $10^{-7}$ yr$^{-1}$, while~\cite{Fragione:2017blf} predicts up to $10^{-8}$ yr$^{-1}$ for IMBHs in local-universe globular clusters.
These rates imply that in a $10^8$ yr time period, there is an order one probability of the IMBH capturing a second compact object and undergoing a second IMRI merger.
Each IMRI merger will modify the DM distribution around the IMBH through DF and SA feedback.\footnote{Each merger will also slightly spin up the primary black hole by an amount of order $qGm_1^2/c$; thus, it would take of order $q^{-1}$ mergers to appreciably spin up the primary if it were initially nonspinning. We continue to ignore the spin of the primary, as a result.
The merger will also give rise to a gravitational-wave kick, which scales as $q^2$~\cite{Fitchett:1983qzq}; thus, the linear momentum imparted to the merger remnant is also negligible in our leading-order treatment in $q$.} 
The predicted time between mergers ($10^7$--$10^8$ years) is shorter than the characteristic relaxation (or ``refilling'') timescale of the DM distribution, which had been estimated in~\cite{Kavanagh:2020cfn} to be of order $10^{70}\,$yr for a $100\,$Gev DM particle (the timescale goes inversely with the DM particle mass).
This makes it a reasonable approximation to assume that the DM distribution does not evolve from the post-merger configuration following the prior IMRI merger.

We will work under the assumption that the dark-matter distribution close to the primary (in regions relevant for IMRI inspirals) evolves from just feedback from IMRI mergers and that the density near the primary will remain largely unchanged in the span between such mergers. 
Our focus will be on second-generation mergers (namely, only one prior IMRI merger took place in the spike).
We run simulations similar to those described in Sec.~\ref{sec:cut_result}, and we also focus on the cycles of dephasing and the density of dark matter at different stages in the inspiral.
The simulations are of the same systems listed in Table~\ref{tab:dephase_VAPLCI}, but the initial distribution of the dark matter was taken to be the distribution ``at ISCO'' (i.e., when the secondary reached ISCO) from the first inspiral (as in Fig.~\ref{fig:density_PL1_1K}). 
While this neglects any effects on the DM distribution that take place after the secondary reaches ISCO during the first merger, given the shorter timescale of the transition to plunge and the plunge (see, e.g.,~\cite{Ori:2000zn,Compere:2021zfj,Kuchler:2024esj}) feedback during this stage would be subleading to that during the adiabatic inspiral.
Aside from this initial data, the simulations were performed as those described in Sec.~\ref{sec:cut_result} (including evolving the binary at a separation of $r_2 = 3 r_\fy$, so as to mimic a quasicircular inspiral as described in Sec.~\ref{sec:cut_result}). 
The dephasing results are presented in Sec.~\ref{sec:successive_dephasing} and those for the density are in Sec.~\ref{subsec:2gdensity}.

The data files for the DM density profiles in this section (along with those described in Sec.~\ref{sec:cut_result}) can be downloaded from~\cite{wade_2025_17888476}.

\subsection{Dephasing for second-generation inspirals} \label{sec:successive_dephasing}

\begin{table}[t!]
    \centering
    \caption{\textbf{Gravitational-wave dephasing for a second-generation inspiral relative to vacuum}. 
    We denote $N_\mathrm{cycles}^{\mathrm{2_{II}}}$ [see Eq.~\eqref{eq:NcyclesA}] as the number of GW cycles between a given starting radius and ISCO for a simulation with $j_\MIN=\sqrt 8$ for a second-generation inspiral. The GW dephasing between a vacuum inspiral and one with an angular-momentum cutoff ($j_\MIN=\sqrt 8$) for a second-generation merger is given by $\Delta N_\mathrm{cycles}^\mathrm{(0-2_{II})}$ [see Eq.~\eqref{eq:DeltaNcyclesAB}].
    As discussed further in Sec.~\ref{sec:cut_result_dephase}, we use $m_2 = 10 \, \Msolar + \delta m_\fy^\mathrm{2_{II}}$ [see Eq.~\eqref{eq:m2A4y}] for the vacuum binary when we compute the dephasing.
    The dephasing is computed for the last four years of inspiral before the secondary reaches the ISCO for the same initial radius in vacuum. 
    We choose $\gamma_\SP = 7/3$ and $\rho_\SP = 200\,\Msolar \, \mathrm{pc}^{-3}$ for the dark-matter parameters.}
    \begin{tabular}{cccc}
    \hline
    \hline
    $m_1 [\Msolar]$ & $N_\mathrm{cycles}^\mathrm{2_{II}}$ & $\Delta N_\mathrm{cycles}^\mathrm{(0-2_{II})}$ & $\delta m_\fy^{\mathrm{2_{II}}} [\Msolar]$ \\
    \hline
    $10^{3}$ & 2,095,100 & 300 & 0.0090 \\
    $3\times 10^{3}$ & 1,590,300 & 800 & 0.0242 \\
    $10^{4}$ & 1,175,000 & 1,300 & 0.0367 \\
    $3\times 10^{4}$ & 890,500 & 1,800 & 0.0227 \\
    $10^{5}$ & 652,600 & 1,600 & 0.0110 \\
    \hline
    \hline
    \end{tabular}
     \label{tab:dephase_VACICII}
\end{table}

We label the simulations of second-generation inspirals with a cutoff by $\mathrm{2_{II}}$, so as to distinguish them from the first-generation inspirals (labeled by $\mathrm{2_{I}}$). 
The GW dephasing of the second-generation inspirals from corresponding vacuum systems  are given in Table~\ref{tab:dephase_VACICII}. 
Similar to Table~\ref{tab:dephase_VAPLCI}, there is a peak in the dephasing values against vacuum at a primary mass of $3 \times 10^{4}\,\Msolar$, but the fractional dephasing is largest for $m_1=10^5\,\Msolar$.
As in Sec.~\ref{sec:cut_result_dephase}, we caution that when comparing the numbers between the third column from Table~\ref{tab:dephase_VACICII} and the sixth column from Table~\ref{tab:dephase_VAPLCI} that the systems have different masses $m^\mathrm{(A)}_\fy$ and inspiral from different radii $r^\mathrm{(A)}_\fy$.
Nevertheless, the second-generation mergers have a smaller dephasing from vacuum than the first-generation mergers do, and the dephasing is smaller in second-generation binaries with less extreme mass ratios.
This trend in the dephasing is related to the fact that there is more feedback  for smaller primary masses.
The increased feedback leads to a greater difference in the initial data between the first- and second-generation mergers as the mass ratio becomes less extreme.
However, the feedback from dynamical friction and accretion can be so large during the inspiral that the density is significantly depleted and the dark-matter effects on the inspiral are reduced (which also produces the peak in the dephasing numbers).

Part of the GW dephasing for the also comes from differences in the respective masses of the secondary (which is shown in the fourth column of Table~\ref{tab:dephase_VACICII}). 
The behavior of $\delta m_\fy^{\mathrm{2_{II}}}$ is similar to the two cases of $\delta m_\fy^{\mathrm{(A)}}$ given in Table~\ref{tab:dephase_VAPLCI}, but the largest value for $\delta m_\fy^{\mathrm{2_{II}}}$ occurs at a mass of $m_1 = 10^4\,\Msolar$.
The shift in the peak value and the smaller values of $\delta m_\fy^{\mathrm{2_{II}}}$ at less extreme mass ratios arises because of the larger changes to the densities that occur during the first inspiral.
This leaves less dark matter that can be accreted onto the secondary during the second-generation inspiral.
In the cases with the more-extreme mass ratios, such as when $m_1 = 10^5\, \Msolar$, the accreted mass $\delta m_\fy^{\mathrm{2_{II}}}$ is more similar to $\delta m_\fy^{\mathrm{2_{I}}}$, because of the smaller changes to the density during the first-generation inspiral.
In addition, because $m_2$ is everywhere smaller during the second inspiral (given that we always choose $m_{2,0}=10\,\Msolar)$, radiation reaction is slightly smaller, which contributes an additional source of dephasing between the first- and second-generation inspirals.
These three effects together combine to create the trends shown in Table~\ref{tab:dephase_VACICII}.
In addition, a comparison of the third column of Table~\ref{tab:dephase_VACICII} and the fourth of Table~\ref{tab:dephase_VAPLCI} shows that for smaller primary masses, the changes in the initial densities have a larger effect on the GW dephasing than including a nonzero angular-momentum cutoff does.

\begin{figure*}[t!]
    \centering
    \includegraphics[width=0.49\textwidth]{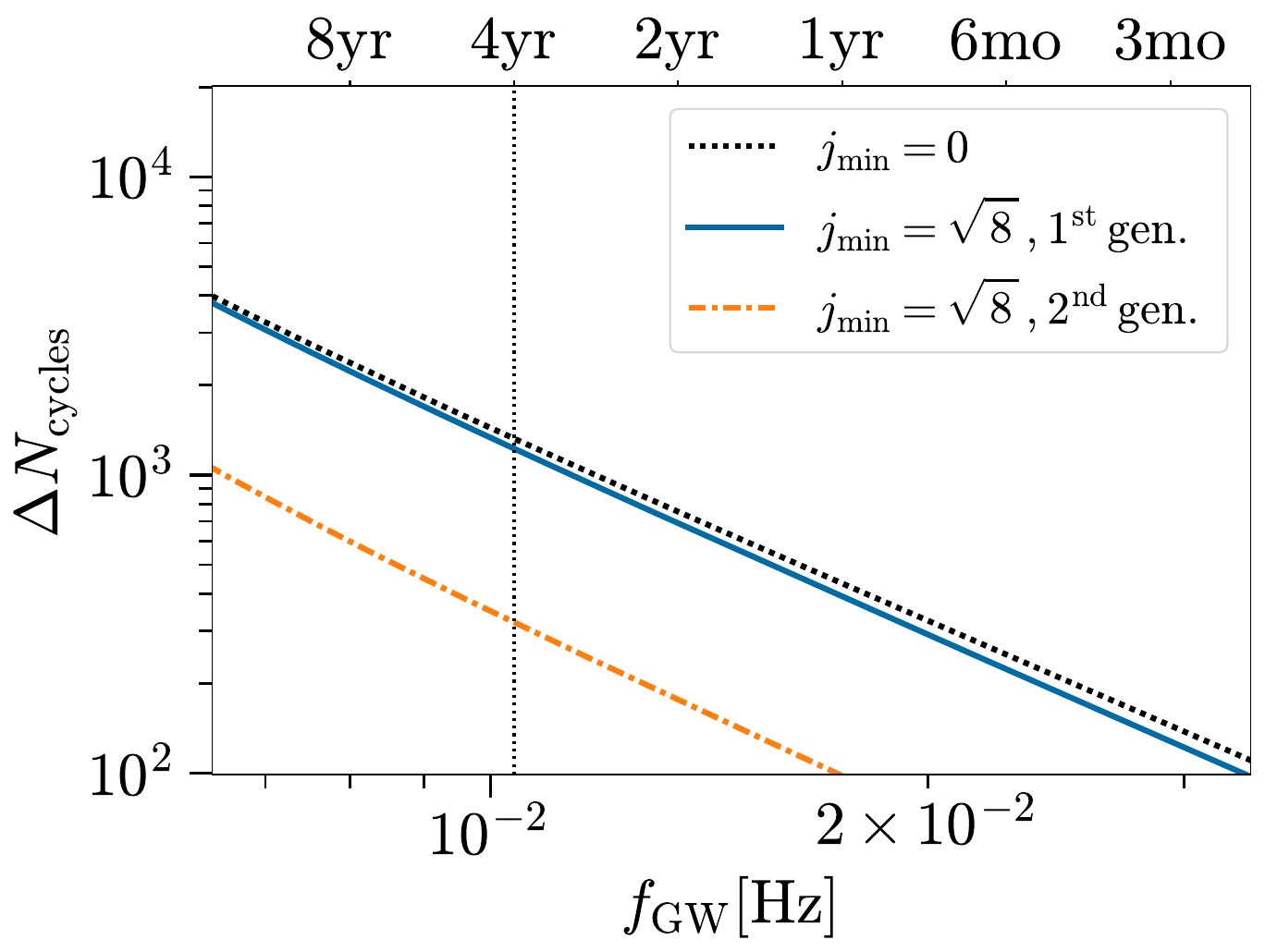} \
    \includegraphics[width=0.49\textwidth]{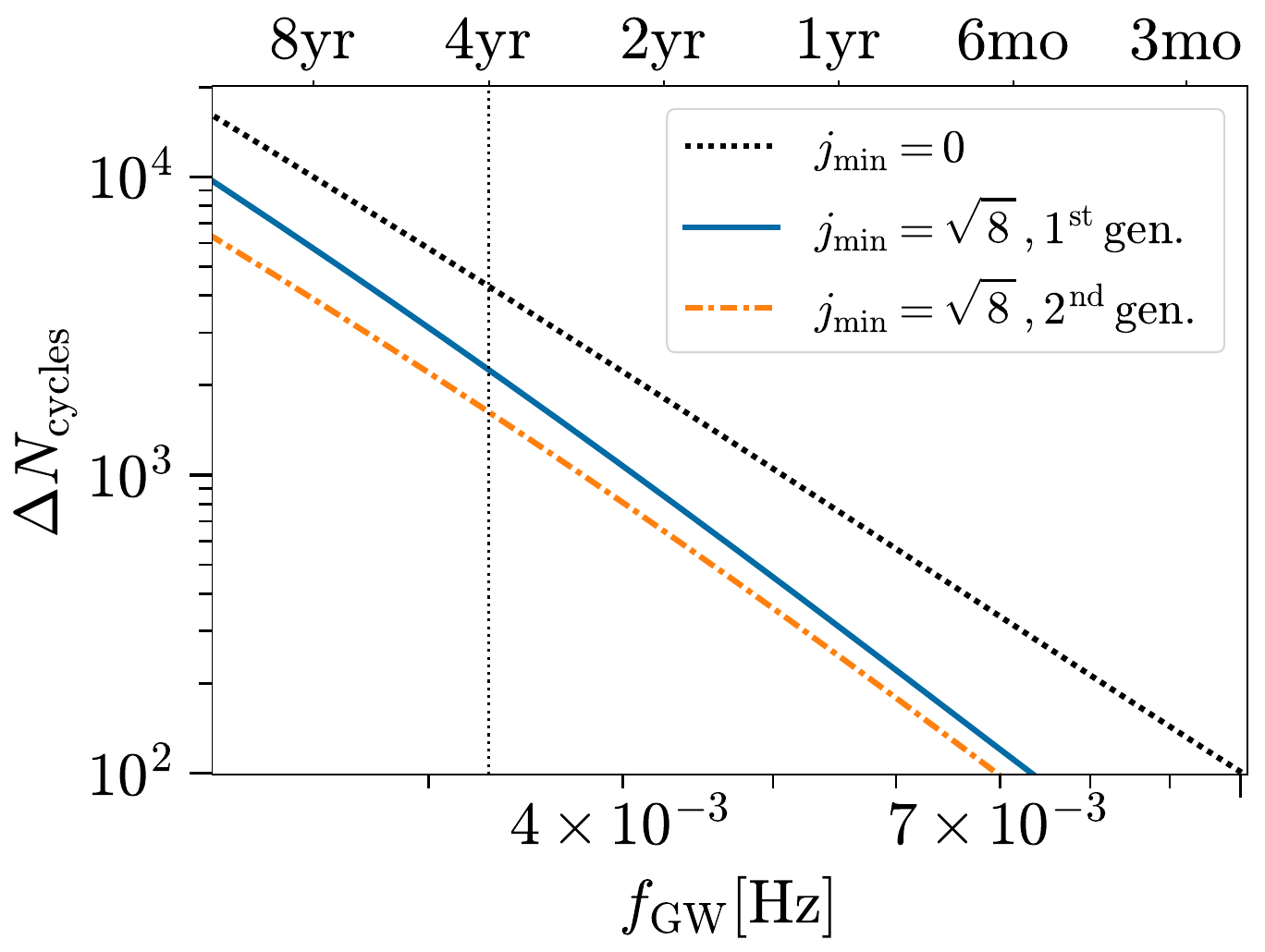}
    \caption{\textbf{Cycles of dephasing versus gravitational-wave frequency for different initial dark-matter distributions and angular-momentum cutoffs}. 
    The dark-matter parameters ($\rho_\SP = 200\,\Msolar/\mathrm{pc}^{3}$ and $\gamma_\SP = 7/3$) are the same as in Fig.~\ref{fig:dephase_PL1_100K} for both panels and the primary mass is chosen to be $m_1 = 10^3\,\Msolar$ (\emph{left}) and $m_1 = 10^5\,\Msolar$ (\emph{right}). 
    The dotted, black lines are the GW dephasing between vacuum binaries and those in a density profile without enforcing a cut to angular momentum ($j_\MIN = 0$), while the solid blue and dash-dotted orange correspond to the GW dephasing from vacuum for systems with a nonzero angular-momentum cutoff ($j_\MIN = \sqrt{8}$) for first- and second-generation inspirals, respectively. 
    The vertical, black, dotted lines are plotted at the frequency for which the IMRI will merger in four years. 
    More detailed discussion of the results is given in the text of Sec.~\ref{sec:successive_dephasing}.}
    \label{fig:dephase_PL12_1K_100K}
\end{figure*}

Figure~\ref{fig:dephase_PL12_1K_100K} shares some similarities with Fig.~\ref{fig:dephase_PL1_100K} (the black and blue curves in the right panel are the same as in Fig.~\ref{fig:dephase_PL1_100K}). 
Specifically, it shows the dephasing of different IMRIs relative to vacuum systems as a function of the GW frequency. 
However, it considers two different primary black-hole masses ($m_1 = 10^3\,\Msolar$ on the left and $m_1 = 10^5\,\Msolar$ on the right), and it also shows the dephasing of second-generation inspirals (not just first-generation systems with zero and nonzero cutoffs in angular momentum). 
There is a qualitative difference between the two mass-ratios, which we discuss in more detail.

In the left panel ($m_1 = 10^3\,\Msolar$), there is a relatively small difference between the curves corresponding to the dephasing against vacuum for the case of $j_\MIN=0$ (dotted black) and against vacuum for the case of $j_\MIN=\sqrt{8}$ (solid blue).
This implies that both the changes in the density and the changes to the feedback formalism arising from the nonzero cutoff in angular momentum have small effects. 
Because the secondary spends most of the inspiral at radii at which the difference in the densities with and without an angular-momentum cutoff is small, this is reasonable.
The second-generation inspiral, denoted by the dash-dotted orange line, has a much smaller magnitude of the dephasing at all frequencies (though a similar qualitative shape of the dephasing curve).
As Fig.~\ref{fig:density_PL1_1K} shows that the initial density for the second-generation merger is slightly under an order of magnitude lower than that of the first-generation merger, this greatly decreases the dark-matter effects on the inspiral and produces less dephasing from vacuum IMRIs.

We now turn to the right panel of Fig.~\ref{fig:dephase_PL12_1K_100K} (the results for a primary mass of $m_1 = 10^5\,\Msolar$).
For this more massive primary, the secondary spends a much larger portion of the last four years of inspiral at radii closer to the ISCO, where the effects of the nonzero angular-momentum cutoff on the density are more significant (see Fig.~\ref{fig:Jcut_density}, and note that $r_\fy^\mathrm{(A)} \approx 17\radsch$).
At all frequencies shown, the dephasing of the second-generation inspiral from vacuum is a larger fraction of that of the first-generation inspiral for the $m_1=10^5\,\Msolar$ than it is for the $m_1 = 10^3\,\Msolar$ case. 
This is consistent with the fact that the DM distribution following the first-generation inspiral is much more similar to the initial distribution for this merger. 
The dephasing curves for the first- and second-generation inspirals converge towards each other at larger frequencies, which is indicative of a smaller difference in the densities at smaller radii.
 
As in Sec.~\ref{sec:cut_result_dephase}, we comment about the choice of the initial binary separation and its effects on the dephasing during the last four years of inspiral.
Given that there are some artifacts in the density at radii greater than $r_\fy^\mathrm{(A)}$ (as noted in Sec.~\ref{sec:cut_result_density}) it is worth revisiting whether they might affect the results for the dephasing.
We perform a similar analysis to that in Sec.~\ref{sec:cut_result_dephase}, where we performed simulations starting at $r_2 = 2 r_\fy$ and $r_2 = 4 r_\fy$, though now for second-generation inspirals.
We use the densities from first-generation simulations starting at that same radius, and we compare the dephasing to that listed in Table~\ref{tab:dephase_VACICII}. 
We observed the same behavior of dephasing with the initial separation as described Sec.~\ref{sec:cut_result_dephase} (that is, a difference in dephasing of less than our numerical uncertainty when starting at $r_2 = 4 r_\fy$, but a significant difference when starting at $r_2 = 2 r_\fy$). 
We conclude that the GW dephasing, including that for second-generation inspirals, is robust to the choice of the initial separation, when that separation is sufficiently large.

Note that while the dephasing curves for first- and second-generation mergers shown in Fig.~\ref{fig:dephase_PL12_1K_100K} are distinguishable from each other, we have not attempted to determine if there exists a first-generation merger with a different initial density that could mimic the dephasing curve of the second-generation merger.
It would be interesting to explore this possibility in future work.

\subsection{Dark-matter density following the second inspiral} \label{subsec:2gdensity}

\begin{figure*}[t!]
    \centering
    \includegraphics[width=0.49\textwidth]{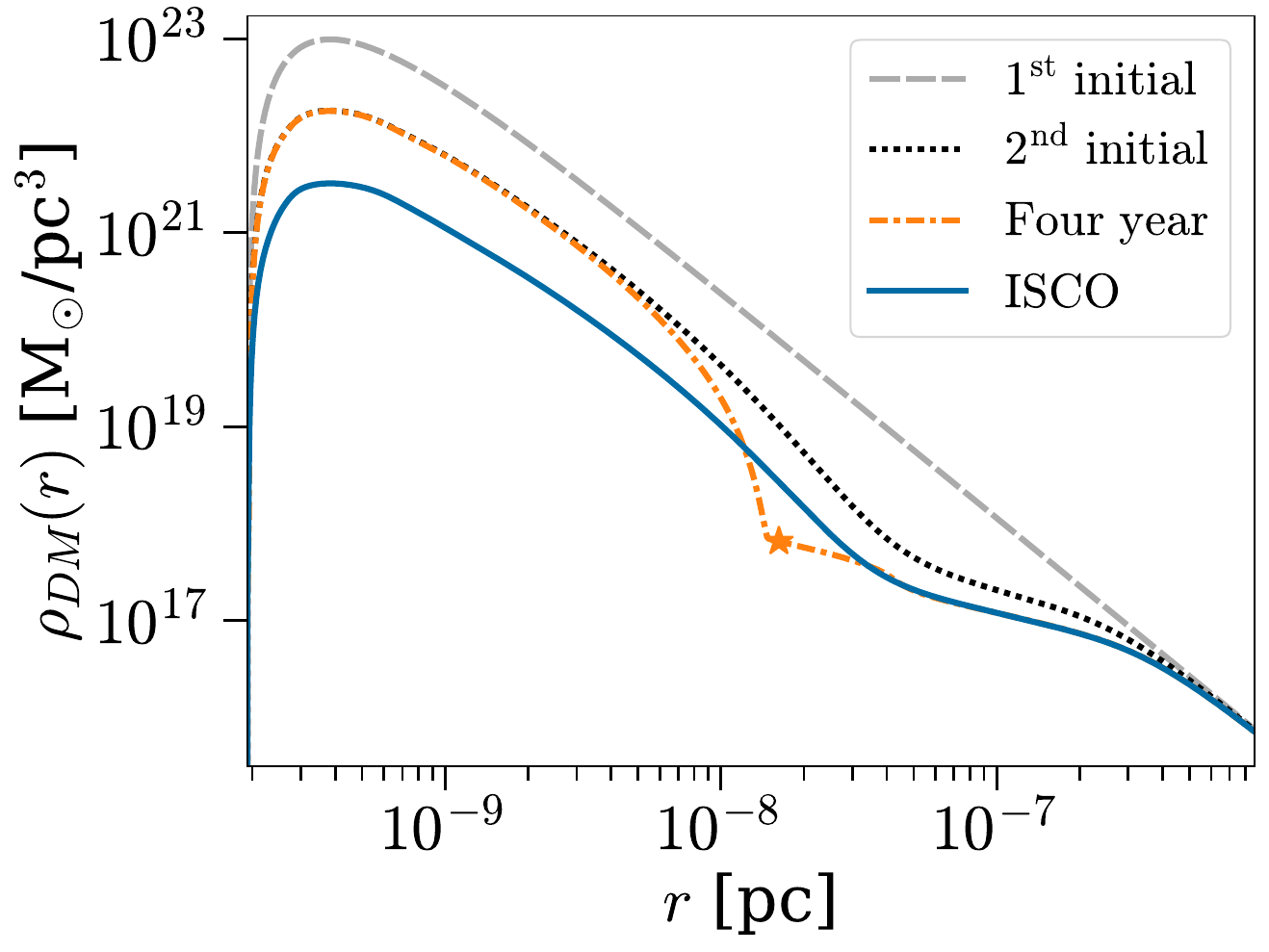} \
    \includegraphics[width=0.49\textwidth]{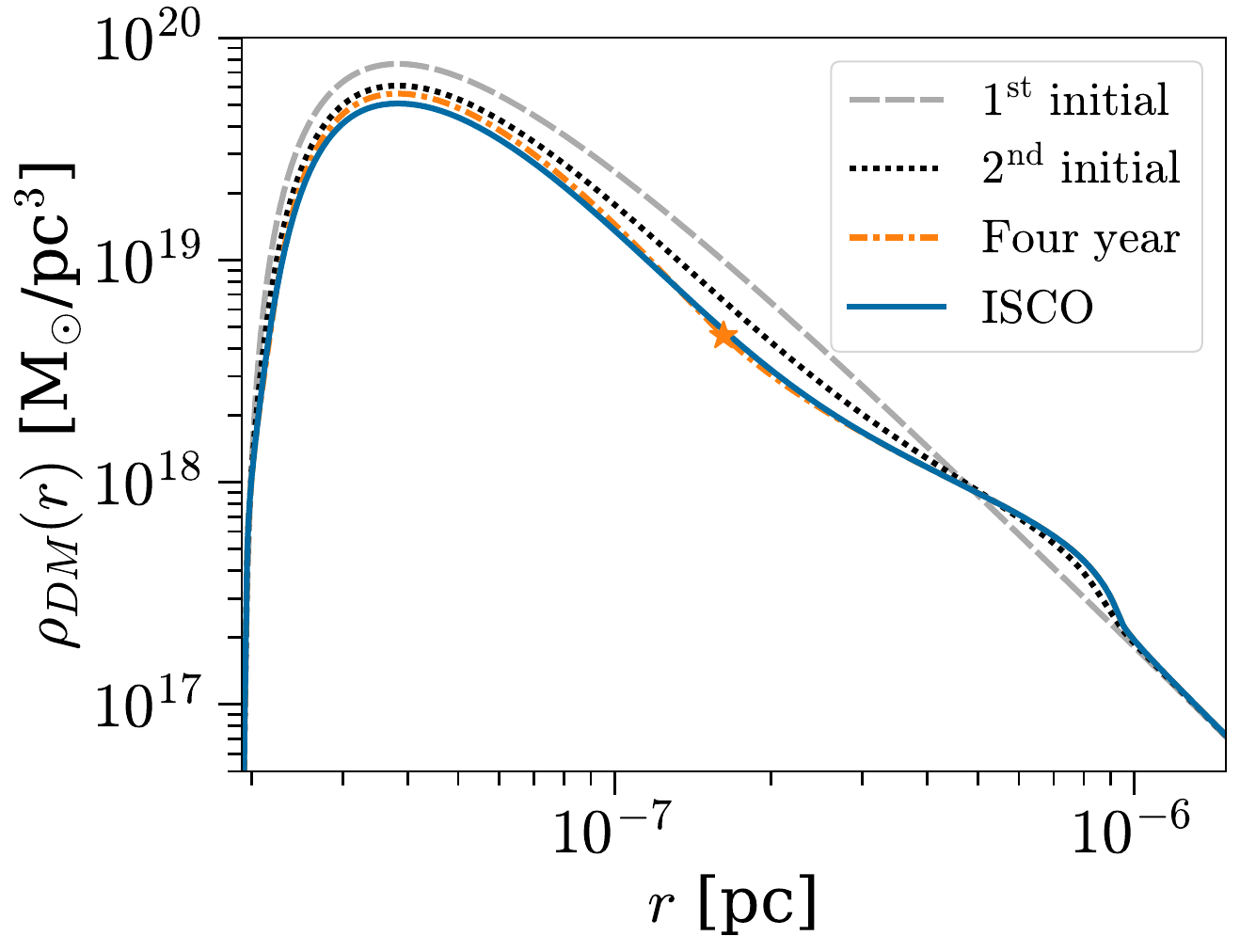}
    \caption{\textbf{Dark-matter density at three stages of the inspiral for a second-generation inspiral}. 
    The dark matter and binary parameters are the same as in the Fig.~\ref{fig:density_PL1_1K}.
    We show the density of dark matter versus radius for several orbital separations during the inspiral of a second-generation merger.
    The dotted black line shows the initial density profile when $r_2 = 3 r_\fy$, the dash-dotted orange the density for $r_2 = r_\fy$, and the solid blue is the density for $r_2=r_\ISCO$. 
    The light-gray dashed line is the initial density from the first-generation inspiral, which is given for comparison. 
    More detailed discussion about the figure appears in the text of Sec.~\ref{subsec:2gdensity}.}
    \label{fig:density_PL12_1K}
\end{figure*}

In Fig.~\ref{fig:density_PL12_1K}, we plot the density at three stages of the inspiral for a second-generation inspiral. 
The initial parameters for the binaries are the same as those used in Fig.~\ref{fig:density_PL1_1K}, which have $m_{2,0} =$ $10\,\Msolar$, $\gamma_\SP = 7/3$, and $\rho_\SP$ = $200\,\Msolar/ \mathrm{pc}^{3}$.
The panel on the left shows the $m_1 =$ $10^3\,\Msolar$ case, while the panel on the right is for $m_1 =$ $10^5\,\Msolar$.
The densities of the first-generation inspirals when the secondary reaches the ISCO (the solid blue curves in Fig.~\ref{fig:density_PL1_1K}) serve as the initial densities for the second inspirals (they are depicted as the dotted, black curves in the two panels of Fig.~\ref{fig:density_PL12_1K}). 
The initial densities from Fig.~\ref{fig:density_PL1_1K} are reproduced as the dashed gray curves in Fig.~\ref{fig:density_PL12_1K} for comparison. 
The curves not contained in Fig.~\ref{fig:density_PL1_1K} are the dashed-dotted orange curves, which represent the densities during the second-generation inspiral when $r_2=r_\fy$, and the solid blue curves, which are the corresponding densities when $r_2=r_\ISCO$.

Comparing the blue and black curves shows that the densities decrease with each subsequent inspiral, but the amount of decrease depends strongly on the primary mass.
In particular, for the $m_1 = 10^3\,\Msolar$ case, the peak of the density near $r \approx 4 \times 10^{-9}\,\mathrm{pc}$ is decreased by slightly less than an order of magnitude after each inspiral.
This leads to an overall flatter radial density profile at small radii.
A stronger flattening is predicted to occur in other scenarios as well, such as in the ``annihilation plateau'' for self-annihilating dark-matter models~\cite{Bertone:2016nfn}, the shallower power law in self-interacting DM models (see, e.g.,~\cite{Banik:2025fnc}), or some DM profiles formed from collapse scenarios (e.g.,~\cite{Bertone:2024wbn}).
The change in density for the $m_1 =$ $10^5\,\Msolar$ case is less significant, but there is still a noticeable redistribution of DM particles. 
For example, the density decreases for $r \lesssim 5 \times 10^{-7}\,\mathrm{pc}$ and increasing it at radii larger than $\sim 5\times 10^{-7}$\,pc.

As described in Sec.~\ref{sec:cut_result}, it is important to start the secondary at $r_2 = 3 r_\fy$, so that the density when $r_2 = r_\fy$ is more consistent with an adiabatic quasicircular inspiral from much larger radii. 
The local density when $r_2 = r_\fy$ (the orange star) is lower than it would have been had we the inspiral started with $r_2 = r_\fy$ with the density from the past inspiral (the black, dotted curve). 
As we did in Sec.~\ref{sec:cut_result_density}, we again comment that the density in Fig.~\ref{fig:density_PL12_1K} at radii less than $r_\fy$ is reasonably robust to the choice of the initial separation, but further out it is not. 
Therefore, the detailed profile at radii larger than $r_\fy$ is not necessarily consistent with a quasicircular inspiral beginning from a much larger initial separation.

\section{Conclusions} \label{sec:conclusions}

In this paper, we investigated two aspects of environmental effects of dense dark-matter spikes on intermediate mass-ratio inspirals. 
We first discussed how dark-matter particles with a sufficiently low angular momentum would be accreted by the primary black hole, and that enforcing a sharp cutoff in the density at a given radius would include such low angular-momentum particles in the distribution function.
Instead, it was more natural to impose a cutoff on the angular momentum of dark-matter particles, which produces a smooth truncation of the dark-matter density in position space.
The dark-matter densities with angular-momentum cutoffs were lower at smaller radii than those with a position-space cutoff.
We investigated the effects of IMRIs evolving in these two dark-matter environments.
We showed that using the position-space cutoff can lead to an overestimation of the gravitational-wave dephasing of such an IMRI system against a similar IMRI in vacuum.
By running numerical simulations for several different mass ratios, we found that the largest overestimation of the dephasing occurred for more extreme mass ratios, and its cause was related primarily to the higher densities for the position-space cutoff. 
The changes in the feedback rates that arise from introducing a lower bound in angular momentum, produced a dephasing that was lower than the accuracy of the results produced by our simulations (one-hundred gravitational-wave cycles).

These results indicated that the initial density is important for determining the amount and the frequency dependence of the gravitational-wave dephasing from vacuum systems. 
There have been studies of IMBH formation scenarios that predict enough stellar-mass black holes form near or migrate towards the IMBH to produce an IMRI merger rate of $10^{-8}$--$10^{-7}$\,yr$^{-1}$ (which is shorter than the relaxation time of the dark-matter spike).
Thus, a plausible initial condition for an IMRI system is that it may not be the first IMRI to merge with the IMBH, and that the density is consistent with the density immediately following an earlier IMRI. 
We described such systems as second-generation mergers, and we studied the amount of dephasing and the dark-matter density in such mergers. 
For all the mass ratios we studied, there was a significant decrease in dephasing against vacuum.
For less extreme mass ratios that the dark-matter density was the most depleted during the prior first-generation inspiral, which correspondingly reduced the effects of the dark-matter environment.
Additionally, since the secondary accreted less dark matter during the second inspiral (as a result of the environment being depleted) and had a smaller mass, radiation reaction was less efficient, leading to additional dephasing between the two generations.

There are several directions in which this work could be extended or generalized.
First, we worked to leading order in mass ratio and in post-Newtonian order for all aspects of the evolution equations for the inspiral, but higher-order calculations would be necessary for modeling these systems at the accuracy required for LISA data analysis. 
Second, we assumed that the secondary always follows a quasicircular orbit about the primary; depending on its formation scenario, an IMRI could have nonzero eccentricity. 
The effects of eccentricity have been considered in~\cite{Karydas:2024fcn,Kavanagh:2024lgq}, though not for the initial densities considered in this work.
While Fig.~14 of~\cite{Karydas:2024fcn} showed that the dephasing for eccentricities below roughly 0.7 were qualitatively similar to the circular case, it would be useful to verify that this continues to hold for initial data with a cutoff in angular momentum. 
Third, we assumed that the primary remained in the center of the dark-matter spike at all times.
At leading order in the mass ratio, the center of mass of the system coincides with the primary's position, but at subleading order, it does not.
How the primary moves with respect to the center of the dark-matter distribution would also be interesting to incorporate.
Following a first-generation merger, the resulting black hole (which would serve as the primary for the second-generation merger) would experience a gravitational-wave recoil (a ``kick'') with a speed that is proportional to the mass ratio of the system. 
This could also displace the primary from the center of the dark-matter density, and depending on the relaxation time, the second-generation merger could be affected by this. 
Any such effects would be largest for the least-extreme mass-ratio IMRIs.

Fourth, there has been recent work~\cite{Mukherjee:2023lzn,Kavanagh:2024lgq} which indicated that there are other dynamical processes that affect the evolution of the IMRI and the dark matter.
Specifically, Ref.~\cite{Kavanagh:2024lgq} showed how the time-changing perturbing potential from the secondary as it orbits induces some ``stirring'' of the dark matter, which aids in redistributing energy within the spike. 
Also, Ref.~\cite{Mukherjee:2023lzn} found that the effects of precession and bulk rotation of the spike can accelerate the inspiral, which could come from the transfer of angular momentum between the IMRI and dark matter.
Although it was argued in~\cite{Kavanagh:2020cfn} that the angular momentum transferred to the dark-matter distribution would not significantly modify the evolution of these systems, it is worth revisiting this in light of the results in~\cite{Mukherjee:2023lzn,Kavanagh:2024lgq}.
Transfer of angular momentum could provide a new channel for additional accretion onto the primary during the inspiral, if these processes decrease the angular momentum of dark matter particles below the minimum value.
Investigating these effects in more detail could serve as the bases for future works.

\acknowledgments

D.A.N.\ and B.A.W.\ acknowledge support from the NSF grant No.\ PHY-2309021. D.A.N.\ also acknowledges support from the NSF CAREER Award PHY-2439893.
The authors acknowledge Research Computing at The University of Virginia for providing computational resources and technical support that have contributed to the results reported within this publication.

\appendix

\section{Computing the dynamical-friction feedback scattering rate} \label{sec:DF_feedback_app}

To evaluate the integral in Eq.~\eqref{eq:DF_rate_dr3}, we will put further constraints on the integration bounds for each DM particle with energy $\mathcal E - \Delta\mathcal E$ that scatters to an energy $\mathcal E$.
The bounds of integration in Eq.~\eqref{eq:DF_rate_dr3} define a spherical shell of inner radius $r_\MIN$ and outer radius $r_+$.
However, the delta function restricts this region to the intersection of this shell with the surface of a torus of major radius $r_2$ and minor radius $b_\star(\Delta \mathcal E)$, for each $\Delta \mathcal E$.
Furthermore, by assumption the interval of possible impact parameters that can scatter is $b_{90} = q r_2 \leq b_\star(\Delta \mathcal E) \leq \sqrt{q} r_2$.
This assumption and Eq.~\eqref{eq:Delta-cal-E} give additional constraints on the possible values of energy changes $\Delta\mathcal E$ as well as the energies $\mathcal E$ of dark-matter particles that can scatter with the secondary.

We evaluate the integral~\eqref{eq:DF_rate_dr3} in toroidal coordinates $(b,\alpha,\varphi)$, where $\alpha$ is the poloidal angle and $\varphi$ is the toroidal angle.
Because the integrand is independent of $\varphi$, the three-dimensional integral reduces to $2\pi$ times a two-dimensional integral over $b$ and $\alpha$.
Specifically, after this integration over $\varphi$ the volume element $\ud^3r$ reduces to $2\pi(r_2+b\cos\alpha) b \, \ud b \, \ud\alpha$.
The integrand is symmetric about the equatorial plane, so $\alpha$ can be restricted to the interval $(0,\pi)$ and the integral is doubled.
We denote the integration bounds for $\alpha$ in this range by $\alpha_1$, and $\alpha_2$.
The values of these parameters will be determined by the fact that the surface of the torus must also lie in the spherical shell bounded between $r_\MIN$ and $r_+$.
The integral over $b$ can now be evaluated.
It gives a nonzero value for $\Delta\mathcal E \in [-v_2^2, -v_2^2 q/(1+q)]$; it will also restrict the range of energies $\mathcal E$ for which $\mathcal R_{\mathcal E}(\Delta\mathcal E)$ is nonzero (this will be discussed in more detail shortly):
\begin{align} \label{eq:calR-alpha-int}
    \mathcal R_{\mathcal E}(\Delta \mathcal E) = & {} 
    \dfrac {4\pi^2 b_{90}^2}{T_2 g(\mathcal E) v_2^2} \left( 1 + \dfrac{b_\star^2}{b_{90}^2} \right)^2 \! \! \displaystyle\int_{\alpha_1}^{\alpha_2}\! \! \! \ud \alpha  (r_2 + b_\star\cos\alpha) \nonumber \\
    & \times \sqrt{2 \left(\dfrac{G m_1}{r(b_\star,\alpha)} - \mathcal E \right) - \dfrac{\mathcal J_{\MIN}^2}{r(b_\star,\alpha)^2}} .
\end{align}
Here we introduced the notation
\begin{equation}
    r\boldsymbol(b_\star(\Delta\mathcal E),\alpha\boldsymbol)=\sqrt{(r_2)^2 + b_\star^2 + 2r_2 b_\star\cos\alpha} ,
\end{equation} 
so that the integrand should be considered to be a function of $\alpha$ (which is itself a function of $r_2$, $\mathcal E$ and $\Delta \mathcal E$) and $\Delta \mathcal E$, as we describe in more detail below.

The upper and lower limits of the $\alpha$ integral need to be consistent with the fact that the integral takes place over regions on the surface of the torus with $r \leq r_+$ and with $r \geq r_\MIN$.
These give lower and upper limits in $\alpha$, respectively, which are given by
\begin{subequations} \label{eq:alpha-plus-cut}
\begin{align}
    \alpha_1 = {} & \cos^{-1} [\MAX\boldsymbol( \MIN\boldsymbol( (r_+^2 - r_2^2 - b_\star^2)/(2 r_2 b_\star) , 1 \boldsymbol), -1 \boldsymbol) ] , \\
    \alpha_2 = {} & \cos^{-1} [ \MIN\boldsymbol( \MAX\boldsymbol( (r_\MIN^2 - r_2^2 - b_\star^2)/(2 r_2 b_\star) , -1 \boldsymbol), 1 \boldsymbol) ] .
\end{align}
\end{subequations}
Note that $\alpha_1$ and $\alpha_2$ are equal when $r_+ = r_\MIN$, which is consistent with the fact that there is no region in position space over which scattering can occur.
These integration bounds are functions of both $\mathcal E$ and $\Delta\mathcal E$ given the $\mathcal E$ dependence of $r_\MIN$ and $r_+$, as well as the $\Delta\mathcal E$ dependence of the function $b_\star$.
The dependence on $r_2$ is more explicit.

The requirement in Eq.~\eqref{eq:alpha-plus-cut} that $r_+ \geq r_2-b_\star$ is equivalent to the statement that the lower bound of the integral satisfies $\alpha_1 \leq \pi$ (i.e., the integration domain is nontrivial).
The condition $r_\MIN \leq r_2 + b_\star$ is an analogous requirement that the upper bound of the integral is $\alpha_2 \geq 0$ (again, the integration domain is nontrivial).
Given the energy dependence of $r_+$ and $r_\MIN$, these statements can be translated into bounds on the energies $\mathcal E$ that can scatter (for a given $b_\star$ and $r_2$).

Because $b_\star$ is restricted to be between $q r_2$ and $\sqrt q r_2$, $\alpha_1$ differs from zero only for radii for which $r_+$ is within $\sqrt q r_2$ of $r_2$.
Similarly, $\alpha_2$ will differ from $\pi$ for radii that differ from $r_\MIN$ by an order $\sqrt q r_2$ amount.
A further restriction on the angle $\alpha$ can come from the fact that the quantity inside the square root in the integral~\eqref{eq:calR-alpha-int} needs to be positive.
We discuss these conditions in more detail below.

To analyze these cases (and to evaluate the integral), it will be useful to use the fact that the impact parameter satisfies $b_\star/r_2 \leq \sqrt{q}$, so that $b_\star/r_2$ is a small parameter.
This permits us to Taylor expand the function $r(b_\star,\alpha)$ in this small parameter.
To linear order in $b\cos\alpha$, the square root in Eq.~\eqref{eq:calR-alpha-int} can be written in the form $\sqrt{2[\mathcal E_\MAX(r_2) - \mathcal E + \mathcal E'_\MAX(r_2) b_\star\cos\alpha]}$, where $\mathcal E'_\MAX(r_2)$ is the derivative of $\mathcal E_\MAX(r)$ with respect to $r$ (and evaluated at $r_2$).
The square root could also be Taylor expanded for energies for which $\mathcal E_\MAX(r_2) -\mathcal E$ is large compared to $b_\star \mathcal E'_\MAX(r_2)$. 
When $\mathcal E_\MAX(r_2) -\mathcal E$ is of order $b_\star \mathcal E'_\MAX(r_2)$ this approximation becomes inaccurate, so we do not expand the square root.
For the term in the integral of the form $r_2$ times the square root, we will keep terms linear in $b_\star$ in the root.

In~\cite{Kavanagh:2020cfn}, the term $b_\star \cos\alpha$ multiplying the square root in Eq.~\eqref{eq:calR-alpha-int} was dropped.
However, it is also reasonable, when expanding in $b_\star/r_2$ to linear order, to keep this term and evaluate the square root at $r=r_2$ (i.e., $b_\star = 0$).
When $\alpha_1=0$ and $\alpha_2=\pi$, this term will vanish because the integral will be proportional to $\sin\alpha_2 - \sin\alpha_1$; however, there will be a nonzero contribution when either $\alpha_1$ or $\alpha_2$ differ from $0$ or $\pi$, respectively.
Thus, we expect this term to contribute near the upper and lower allowed energies.

It is now possible to evaluate the integral in Eq.~\eqref{eq:calR-alpha-int} (using the approximations described above) in terms of incomplete elliptic integrals of the second kind, $E(\phi,m)$.
The form of their arguments depends on whether $\mathcal E'_\MAX(r_2)$ is positive or negative [which corresponds to when $r_2$ is less than or greater than $r_{\mathcal J}=\Jmin^2/(Gm_1)$, respectively].
When $\mathcal E'_\MAX(r_2)$ is positive, one can write it in this form using the cosine double-angle formula.
When it is negative (the case in~\cite{Kavanagh:2020cfn}), it is useful to perform a coordinate transformation of the form $\alpha \rightarrow \pi -\alpha$ before applying the double-angle formula.

In both cases, the modulus parameter $m$ is given by
\begin{equation} \label{eq:modulus}
    m = \frac{2b_\star|\mathcal E'_\MAX(r_2)|}{\mathcal E_\MAX(r_2) - \mathcal E + b_\star |\mathcal E'_\MAX(r_2)|} .
\end{equation}
where one can write $\mathcal E_\MAX(r_2) - \mathcal E$ as
\begin{equation}
    \mathcal E_\MAX(r_2) - \mathcal E = \mathcal E \left( \frac{r_+}{r_2} - 1 \right) \left( 1 - \frac{r_-}{r_2} \right) 
\end{equation}
and $\mathcal E'_\MAX(r_2)$ as
\begin{equation}
    \mathcal E'_\MAX(r_2) = -\frac{Gm_1}{r_2^2} + \frac{\Jmin^2}{r_2^3} . 
\end{equation}
The relevant integral is given by
\begin{align} \label{eq:I-integral}
    & I_{\mathcal E}(\Delta\mathcal E) \equiv \nonumber \\ 
    & \int_{\alpha_1}^{\alpha_2} \! \! d\alpha  \sqrt{2[\mathcal E_\MAX(r_2) - \mathcal E + \mathcal E'_\MAX(r_2) b_\star\cos\alpha]} = \nonumber \\
    & \begin{cases}
    4 \sqrt{\dfrac{b_\star|\mathcal E'_\MAX(r_2)|}{m} } \left[
    E\left(\phi_1 , m \right) - E\left(\phi_2 , m \right) \right] & \mbox{for} \ \ r_2 > r_{\mathcal J} , \\
    \sqrt{2[\mathcal E_\MAX(r_2) - \mathcal E]}(\alpha_2 - \alpha_1) & \mbox{for} \ \ r_2 = r_{\mathcal J} , \\
    4 \sqrt{\dfrac{b_\star|\mathcal E'_\MAX(r_2)|}{m} } \left[
    E\left(\phi_2 , m \right) - E\left(\phi_1 , m \right) \right] & \mbox{for} \ \ r_2 < r_{\mathcal J} .
    \end{cases}
\end{align}
The angles $\phi_1$ and $\phi_2$ will depend on whether the modulus parameter $m$ is greater or less than one.
From Eq.~\eqref{eq:modulus}, to have $m \leq 1$, the energy needs to satisfy 
\begin{equation} \label{eq:mEcondition}
    \mathcal E \leq \mathcal E_\MAX(r_2) - b_\star|\mathcal E'_\MAX(r_2)|  .
\end{equation}
For these energies, the angles are given by $\phi_1 = \psi_1$ and $\phi_2 = \psi_2$ where we defined 
\begin{equation} \label{eq:psi_i}
    \psi_i = 
    \begin{cases}
        \dfrac{\pi-\alpha_i}{2} & \mbox{for} \ \ r_2 \geq r_{\mathcal J} , \\
        \dfrac{\alpha_i}{2} & \mbox{for} \ \ r_2 < r_{\mathcal J} , 
    \end{cases}
\end{equation}
for $i=1,2$.

When $m > 1$, it is useful to perform a reciprocal modulus transformation (see~\cite[\S{}19.7]{NIST:DLMF}), as was done in~\cite{Kavanagh:2020cfn}.
As we discuss in Appendix~\ref{app:efficiency}, in~\cite{Kavanagh:2020cfn}, given the definition of the analogues of $\alpha_1$ and $\alpha_2$ used when $\Jmin=0$, it can be shown that the conditions required of the reciprocal modulus transformation (discussed below) always hold.
This transformation, when written in terms of the modulus $m$ (not $k$, as in~\cite[\S{}19.7]{NIST:DLMF}), allows an elliptic integral of the second kind with $m > 1$ to be recast in terms of elliptic integrals with $1/m < 1$ of the first and second kinds as follows:
\begin{equation} \label{eq:ellipe_inc}
    E(\phi, m) = \sqrt m E(\beta, 1/m) + \frac{(1-m)}{\sqrt m} F(\beta, 1/m) .
\end{equation}
Here $F(\beta, 1/m)$ is the incomplete elliptic integral of the first kind.
The angle $\beta$ is defined by $\sin\beta = \sqrt m \sin\phi$.
To have a real $\beta$ (i.e., to have $\sin\beta \leq 1$), then this requires that that $\sin\phi \leq 1/\sqrt{m}$.
While this condition does not introduce any new constraints on the integral when $\Jmin = 0$ (see Appendix~\ref{app:efficiency}), we were not able to show the same holds true for $\Jmin > 0$.
Thus, we further restrict the bounds of integration in $\alpha$ to ensure that the integrand in Eq.~\eqref{eq:I-integral} is always real-valued.
This is equivalent to defining the angles $\phi_1$ and $\phi_2$ by
\begin{equation} \label{eq:phi_i}
    \phi_i = 
    \begin{cases}
        \psi_i & \mbox{for} \ \ m \leq 1 , \\
        \MIN\boldsymbol(\psi_i, \sin^{-1}(1/\sqrt m) \boldsymbol) & \mbox{for} \ \ m > 1 .
    \end{cases}
\end{equation}
For a given $r_2$ and $\Delta\mathcal E$, this is necessary when $\mathcal E > \mathcal E_\MAX(r_2) - b_\star |\mathcal E'_\MAX(r_2)|$.

In summary, we can write the expression for $\mathcal R_{\mathcal E}(\Delta \mathcal E)$ as
\begin{align} \label{eq:calR-final}
    \mathcal R_{\mathcal E}(\Delta \mathcal E) = {} &
    \dfrac {4\pi^2 b_{90}^2}{T_2 g(\mathcal E) v_2^2} \left( 1 + \dfrac{b_\star^2}{b_{90}^2} \right)^2 \bigg[r_2 I_{\mathcal E}(\Delta\mathcal E) \nonumber \\
    & + b_\star (\sin\alpha_2 - \sin\alpha_1)\sqrt{2[\mathcal E_\MAX(r_2) - \mathcal E]} \bigg] .
\end{align}
The rate coefficient is nonzero for $\Delta \mathcal E \in [-v_2^2, -2 v_2^2 q/(1+q)]$.
The values of $\mathcal E$ for which $\mathcal R_{\mathcal E}(\Delta \mathcal E)$ is nonzero depend on $\mathcal E$, $\Delta\mathcal E$ and $r_2$.
We will discuss them in more detail in Appendix~\ref{app:efficiency}.

\section{Improving the \textsc{HaloFeedback} code efficiency} \label{app:efficiency}
 
For less extreme mass ratios (for example, the  $m_1=10^3\,\Msolar$ and $m_2 = 10\,\Msolar$ inspiral performed in~\cite{Nichols:2023ufs}), an inspiral run using the \textsc{HaloFeedback} code (with $\Jmin=0$) took about $\sim 70$ hours to run on a single-core machine running at about 3\,GHz.
By profiling the code, we determined that much of the computation time was spent evaluating elliptic integrals analogous to those in Eq.~\eqref{eq:I-integral}, and the slowest case occurred when the modulus parameter was greater than 1.

The \textsc{HaloFeedback} code uses the \textsc{SciPy} library to compute the incomplete elliptic integrals of the second kind $E(\phi, m)$, but this package requires that the modulus parameter satisfy $m \leq 1$.
Thus, the reciprocal modulus transformation in Eq.~\eqref{eq:ellipe_inc} was used to evaluate the $m>1$ with the \textsc{SciPy} functions.

We identified the cause of this longer evaluation time and how it can be resolved, which we describe in the first part of this appendix.
For $\Jmin > 0$, we found that there was not the same degree of slowdown when evaluating the integral $I_{\mathcal E}(\Delta\mathcal E)$ in Eq.~\eqref{eq:I-integral}.
Nevertheless, we describe a procedure in the second part of this appendix that speeds up the evaluation of the integral $I_{\mathcal E}(\Delta\mathcal E)$ to a lesser extent.

\subsection{No angular-momentum cutoff}

When $\Jmin=0$, the results in Appendix~\ref{sec:DF_feedback_app} significantly simplify.
The modulus parameter is given by 
\begin{equation} \label{eq:m_nocut}
    m = \dfrac{2}{(r_2 - r_2^2/r_\mathcal{E})/b_\star + 1} \equiv \dfrac{2}{\chi + 1} ,
\end{equation}
where we defined $\chi = r_2(1 - r_2/r_\mathcal{E})/b_\star$ and $r_{\mathcal E} = Gm_1/\mathcal E$. 
The constraints on $\alpha_1$ and $\alpha_2$ come from $r \leq r_{\mathcal E}$ and $r \geq r_\cut$.
These constraints were solved to linear order in $b_\star$ using the approximate expression for $r$ that 
\begin{equation}
    r \approx r_2 + b_\star \cos \alpha + \mathcal O(b_\star^2) \approx \dfrac{r_2}{1-(b_\star/r_2) \cos \alpha} .
\end{equation}
These give rise to the conditions
\begin{equation} \label{eq:DF_scatterangle}
    \begin{split}
        \alpha_1 &= \cos^{-1} [ \min(\chi, 1) ] ,\\
        \alpha_2 &= \cos^{-1} [ \max\boldsymbol(r_2(1 - r_2/r_\mathrm{cut})/b_\star, -1\boldsymbol) ] .
    \end{split}
\end{equation}
When $\Jmin=0$, then $\mathcal E_\MAX'(r_2)$ is negative, which implies that the arguments entering the elliptic integral are $\phi_{i} = (\pi - \alpha_{i})/2$ (for $i=1$,2) and any $m > 0$.

One can see from Eq.~\eqref{eq:m_nocut} that the condition $m \leq 1$ requires $\chi \geq 1$ and therefore $\alpha_1 = 0$. 
Thus, the incomplete elliptic integral $E\left( \phi_1, m \right)$ reduces to a complete elliptic integral of the second kind, $E(m)$. 
Conversely, when the modulus parameter satisfies $m>1$, then $\chi$ satisfies $\chi<1$, which implies $\cos(\alpha_1) = \chi$. 
Using the double-angle formula, one can show that $\sin(\phi_1) = 1/\sqrt{m}$.
This implies that the angle $\beta$ which enters into the reciprocal modulus formula, Eq.~\eqref{eq:ellipe_inc}, is $\beta = \pi/2$. 
In this case, too, the incomplete integrals in Eq.~\eqref{eq:ellipe_inc} reduce to complete elliptic integrals. 
This last result was not used in \textsc{HaloFeedback}; instead, the code restricted $\beta$ to be \textit{nearly} $\pi/2$, and computed the incomplete elliptic integrals using that value of $\beta$. 

The \textsc{SciPy} elliptic integral functions are wrappers around the \textsc{Cephes} library functions \textit{ellie} and \textit{ellik}.
When $\beta$ is very near $\pi/2$, these functions warn that are slow to converge, and thus to compute. 
When we implemented the complete elliptic integrals in our version of the \textsc{HaloFeedback} code for the example of the $m_1=10^3\,\Msolar$ and $m_2 = 10\,\Msolar$ inspiral, the computation was sped up by a factor of about ten.

\subsection{Nonzero angular-momentum cutoff}

Given the previous discussion that the \textsc{SciPy} elliptic integrals take longer to evaluate for angles near $\pi/2$, it is useful to identify, when $\Jmin$ is nonzero, the cases for which the elliptic integrals in Eq.~\eqref{eq:I-integral} reduce to complete integrals or for which they vanish.
These cases will depend on the energy $\mathcal E$, the scattering energy transfer $\Delta\mathcal E$, and the binary separation $r_2$.

We have already discussed in Appendix~\ref{sec:DF_feedback_app} how $r_{\mathcal J}$ plays an important role in distinguishing different cases for the integral $I_{\mathcal E}(\Delta\mathcal E)$ in Eq.~\eqref{eq:I-integral}.
It will be useful to introduce notation that will play a role in understanding the different cases of the integral for different energies $\mathcal E$ and scattering energy transfers $\Delta\mathcal E$.
For the energy, we introduce the notation
\begin{equation}
    \mathcal E_\MAX^{(\pm b_\star)}(r_2) \equiv \mathcal E_\MAX(r_2) \pm b_\star |\mathcal E'_\MAX(r_2)|
\end{equation}
and 
\begin{equation} \label{eq:E-cut-J}
    \mathcal E_\cut^{(\pm b_\star)}(r_2)  \equiv \frac{Gm_1}{2r_2}\left(1 \pm \frac{2b_\star}{r_2} \right) , 
\end{equation}
For the scattering energy transfer, an impact parameter of $b_{90}$ gives rise to a change in energy of $-v_2^2$, which in some cases will be the lower bound. 
The upper bound (in some cases) takes place for an impact parameter of $\sqrt q r_2$, which corresponds to an energy change of
\begin{equation}
    \Delta \mathcal E_\mathrm{up} \equiv -\frac{2v_2^2q}{1+q} \approx - 2 q v_2^2.
\end{equation}
When $r_2-r_{\mathcal J}$ is of order $b_\star$, it is helpful to determine whether $\Delta \mathcal E$ is associated with an impact parameter that is smaller or larger than $r_2-r_{\mathcal J}$.
This leads us to define one additional change in energy given by
\begin{equation} \label{eq:Delta-E-Jq}
     \Delta \mathcal E_{\mathcal J}^{(q)} \equiv - \frac{2 (q v_2 r_2)^2}{(q r_2)^2 + (r_2 - r_{\mathcal J})^2 } ,
\end{equation}
which will also arise in the expressions below.

First, it is useful to determine the energies (for a given $r_2$ and $\Delta\mathcal E$) for which the reciprocal modulus transform does, or does not need to be applied (i.e., when $m > 1$ or $m \leq 1$, respectively).
A short calculation using Eq.~\eqref{eq:modulus} shows that $m = 1$ when $\mathcal E = \mathcal E_\MAX^{(-b_\star)}$.
Given that the maximum energy that scatters when the secondary is at $r_2$ is the smaller of $\mathcal E_\MAX^{(+b_\star)}(r_2)$ and $\mathcal E_\MAX(r_{\mathcal J})$, the reciprocal modulus transformation needs to be applied when $\mathcal E$ satisfies $\mathcal E > \mathcal E_\MAX^{(-b_\star)}$, but is less than the maximum allowed energy at that $r_2$.
The smallest energy that can scatter is $\mathcal E_\cut^{(-b_\star)}$, so the modulus parameter satisfies $m \leq 1$ for $\mathcal E \in [\mathcal E_\cut^{(-b_\star)},\mathcal E_\MAX^{(-b_\star)}]$.
We will not explicitly apply the reciprocal modulus transform to write our result, but when the energy satisfies $\mathcal E > \mathcal E_\MAX^{(-b_\star)}$, it is implied that the reciprocal transform should be applied.

Next, we analyze the angles $\alpha_1$ and $\alpha_2$ in Eq.~\eqref{eq:alpha-plus-cut}.
Here we work to linear order in $b_\star$, so that the equations reduce to
\begin{subequations} \label{eq:alpha-plus-cut-linear}
\begin{align}
    \alpha_1 = {} & \cos^{-1} [\MAX\boldsymbol( \MIN\boldsymbol( r_2[(r_+/r_2)^2 - 1]/(2 b_\star) , 1 \boldsymbol), -1 \boldsymbol) ] , \\
    \alpha_2 = {} & \cos^{-1} [ \MIN\boldsymbol( \MAX\boldsymbol( r_2[(r_\MIN/r_2)^2 - 1]/(2 b_\star) , -1 \boldsymbol), 1 \boldsymbol) ] .
\end{align}
\end{subequations}

The condition to have $\alpha_1=0$ is $(r_+/r_2)^2 - 1 \geq 2 b_\star/r_2$.
Analyzing this condition shows that it implies $\alpha_1 = 0$ for $\mathcal E \leq \mathcal E^{(-b_\star)}_\MAX(r_2)$ for $r_2 \geq r_{\mathcal J}$ and $\mathcal E \leq \mathcal E^{(+b_\star)}_\MAX(r_2)$ for $r_2 < r_{\mathcal J}$.
From Eqs.~\eqref{eq:psi_i} and~\eqref{eq:phi_i}, this leads to different values of $\phi_1$ depending on the value of $r_2$ and the range of energy, $\mathcal E$.
This implies that
\begin{equation} \label{eq:phi1-0}
    \phi_1|_{\alpha_1=0} = 
    \begin{cases}
        \pi/2 & \mbox{for } \ r_2 \geq r_{\mathcal J}, \ \mathcal E \leq \mathcal E_\MAX^{(-b_\star)}(r_2) ,\\
        0 & \mbox{for } \ r_2 <r_{\mathcal J},\ \mathcal E \leq \mathcal E_\MAX^{(+b_\star)}(r_2) .
    \end{cases}
\end{equation}
Only for $r_2 \geq r_{\mathcal J}$ and $\mathcal E > \mathcal E_\MAX^{(-b_\star)}(r_2)$ is $\phi_1$ neither zero or $\pi$.
Next, we analyze when $\alpha_1 = \pi$ [namely, the condition $(r_+/r_2)^2 - 1 \leq -2b_\star/r_2$].
It gives that $\mathcal E \geq \mathcal E^{(+b_\star)}_\MAX(r_2)$ for $r_2 \geq r_{\mathcal J} + b_\star$ and it cannot occur for $r_2 < r_{\mathcal J} + b_\star$.
Thus, there are no permitted energies or radii $r_2$ for which it can occur.

We can perform a similar analysis to determine when the angles $\alpha_2=0$ and $\alpha_2 = \pi$. 
To determine when $\alpha_2 = 0$, one must solve the condition $r_\MIN^2 - r_2^2 \geq 2r_2 b_\star$ for the energy.
This calculation shows that the energy must be less than $\mathcal E_\cut^{(-b_\star)}$ for $r_2 > r_{\mathcal J}$ and greater than $\mathcal E_{\MAX}^{(+b_\star)}$ for $r_2 \leq r_{\mathcal J}$.
This allows us to determine that the angle $\phi_2$ when $\alpha_2=0$ takes on the following values:
\begin{equation} \label{eq:phi2-0}
    \phi_2|_{\alpha_2=0} = 
    \begin{cases}
        \pi/2 & \mbox{for } \ r_2 \geq r_{\mathcal J}, \ \mathcal E < \mathcal E_\cut^{(-b_\star)}(r_2) \\
        0 & \mbox{for } \ r_2 \leq r_{\mathcal J},\ \mathcal E \geq \mathcal E_{\MAX}^{(+b_\star)}(r_2)
    \end{cases}
\end{equation}
The condition to have $\alpha_2 = \pi$, comes from solving the inequality $r_\MIN^2 - r_2^2 \leq -2r_2 b_\star$ for the energy.
A similar calculation to that for $\alpha_2=0$ allows us to determine that when $\alpha_2=\pi$ the angle $\phi_2$ takes on the following values:
\begin{equation} \label{eq:phi2-pi}
    \phi_2|_{\alpha_2=\pi} = 
    \begin{cases}
        0 & \mbox{for } \ r_2 > r_{\mathcal J}, \ \mathcal E \geq \mathcal E_\cut^{(+b_\star)}(r_2) \\
        \pi/2 & \mbox{for } \ r_2 \leq r_{\mathcal J},\ \mathcal E \leq \mathcal E_{\MAX}^{(-b_\star)}(r_2)
    \end{cases}
\end{equation}
The angle $\phi_2$ takes on a value between 0 and $\pi/2$ for energies between $\mathcal E^{(\pm b_\star)}_\cut$ and $\mathcal E^{(\pm b_\star)}_{\MAX}$ for $r_2$ greater than or less than $r_{\mathcal J}$, respectively.

The results in Eqs.~\eqref{eq:phi1-0}--\eqref{eq:phi2-pi} show the radii $r_2$ and energies $\mathcal E$ for which each of the two elliptic integrals in Eq.~\eqref{eq:I-integral} become complete, vanish, or must be evaluated as incomplete.
We can now analyze these two cases in tandem, so as to determine where they overlap or not. 

For example, when $\phi_1=\pi/2$ and $\phi_2=0$ for $r_2 > r_{\mathcal J}$, then the integral in Eq.~\eqref{eq:I-integral} becomes proportional to a single complete elliptic integral $E(m)$.
This occurs for energies for which $\mathcal E_\cut^{(+b_\star)} \leq \mathcal E \leq \mathcal E_\MAX^{(-b_\star)}$.
This range of energies is nonzero when $\mathcal E_\cut^{(+b_\star)} \leq \mathcal E_\MAX^{(-b_\star)}$, which occurs when $r_2 \geq r_{\mathcal J} + 2 b_\star$.
Because $b_\star$ is a function of $\Delta\mathcal E$ and $r_2$ (the latter via the relationship that $b_{90} = q r_2$), it is helpful to instead solve the inequality in terms of $\Delta \mathcal E$ using the definition of $b_\star$ in Eq.~\eqref{eq:b-star-def}.
Doing so (to linear order in $b_\star/r_2$) gives the constraint that $\Delta \mathcal E \leq \Delta \mathcal E^{(2q)}_{\mathcal J}$, where the upper bound is defined in Eq.~\eqref{eq:Delta-E-Jq}.
The smallest radius for which this case arises is $r_2 = (1+2q)r_{\mathcal J}$.
Combining all these results gives rise to the first case in Eq.~\eqref{eq:I-int-cases}.

\begin{widetext}
    \begin{equation} \label{eq:I-int-cases}
        I_{\mathcal E}(\Delta\mathcal E) = 
        \begin{cases}
          4 \sqrt{\dfrac{b_\star|\mathcal E'_\MAX(r_2)|}{m} } E\left( m \right)   & \mbox{for} \ \ r_2 \geq r_{\mathcal J}(1 + 2 q), \ \ -v_2^2 \leq \Delta \mathcal E \leq \MIN(\Delta \mathcal E_\mathrm{up},\Delta \mathcal E_{\mathcal J}^{(2q)}) , \\ 
          & \phantom{for } \mathcal E_\cut^{(+b_\star)}(r_2) \leq \mathcal E \leq \mathcal E_\MAX^{(-b_\star)}(r_2) , \\
          4 \sqrt{\dfrac{b_\star|\mathcal E'_\MAX(r_2)|}{m} } E\left(\phi_1, m \right)   & \mbox{for} \ \ r_2 \geq r_{\mathcal J}(1 + 2 q), \ \ -v_2^2 \leq \Delta \mathcal E \leq \MIN(\Delta \mathcal E_\mathrm{up},\Delta \mathcal E_{\mathcal J}^{(2q)}) , \\
           & \phantom{for } \mathcal E_\MAX^{(-b_\star)}(r_2) < \mathcal E \leq \MIN[\mathcal E_\MAX^{(+b_\star)}(r_2), \mathcal E_\MAX(r_\mathcal J)], \\ 
          4 \sqrt{\dfrac{b_\star|\mathcal E'_\MAX(r_2)|}{m} } [E\left( m \right) - E(\phi_2,m) ]   & \mbox{for} \ \ r_2 \geq r_{\mathcal J}(1 + 2 q), \ \ -v_2^2 \leq \Delta \mathcal E \leq \MIN(\Delta \mathcal E_\mathrm{up},\Delta \mathcal E_{\mathcal J}^{(2q)}) , \\ 
          & \phantom{for } \mathcal E_\cut^{(-b_\star)}(r_2) \leq \mathcal E < \mathcal E_\cut^{(+b_\star)}(r_2) , \ \mbox{ and } \\ 
          & \phantom{for} \ r_{\mathcal J} < r_2 \leq r_{\mathcal J}(1 + 2\sqrt{q}), \ \ \MAX(-v_2^2, \Delta \mathcal E_{\mathcal J}^{(2q)}) \leq \Delta \mathcal E \leq \Delta \mathcal E_\mathrm{up} , \\
          & \phantom{for } \mathcal E_\cut^{(-b_\star)}(r_2) \leq \mathcal E \leq \mathcal E_\MAX^{(-b_\star)}(r_2) , \\
           4 \sqrt{\dfrac{b_\star|\mathcal E'_\MAX(r_2)|}{m} } [E\left(\phi_1, m \right) - E(\phi_2, m) ]   & \mbox{for} \ \ r_{\mathcal J} < r_2 \leq r_{\mathcal J}(1 + 2\sqrt{q}), \ \ \MAX(-v_2^2, \Delta \mathcal E_{\mathcal J}^{(2q)}) \leq \Delta \mathcal E \leq \Delta \mathcal E_\mathrm{up} , \\
           & \phantom{for } \mathcal E_\MAX^{(-b_\star)}(r_2) \leq \mathcal E \leq \MIN[\mathcal E_\MAX^{(+b_\star)}(r_2), \mathcal E_\MAX(r_\mathcal J)] , \\ 
          \sqrt{2[\mathcal E_\MAX(r_2) - \mathcal E]} \, (\alpha_2-\alpha_1) & \mbox{for} \ \ r_2 = r_{\mathcal J}, \ \ -v_2^2 \leq \Delta \mathcal E \leq \Delta \mathcal E_\mathrm{up} , \\
          & \phantom{for } \mathcal E_\cut^{(-b_\star)}(r_2) \leq \mathcal E \leq \mathcal E_\MAX(r_2) , \\
          4 \sqrt{\dfrac{b_\star|\mathcal E'_\MAX(r_2)|}{m} } E\left(\phi_2, m \right) & \mbox{for} \ \ r_{\mathcal J}(1 - 8\sqrt q/3) \leq r_2 < r_{\mathcal J}, \ \ \MAX(-v_2^2, \Delta \mathcal E_{\mathcal J}^{(8q/3)}) \leq \Delta \mathcal E \leq \Delta \mathcal E_\mathrm{up} \\
          & \phantom{for } \mathcal E_\MAX^{(-b_\star)}(r_2) \leq \mathcal E \leq \MIN[\mathcal E_\MAX^{(+b_\star)}(r_2), \mathcal E_\MAX(r_\mathcal J)], \\
          0 & \mbox{otherwise} .
        \end{cases}
    \end{equation}
\end{widetext}

The remaining cases in Eq.~\eqref{eq:I-int-cases} can be obtained through similar arguments.
The second line of Eq.~\eqref{eq:I-int-cases} is the case where $\phi_1$ is between zero and $\pi/2$, and $\phi_2 = 0$.
This requires that $\mathcal E > \mathcal E_\MAX^{(-b_\star)}$, but all other conditions are the same as in the previous case.
In the region $r_2 \geq r_{\mathcal J} + 2b_\star$ where $\mathcal E_\cut^{(+b_\star)} \leq \mathcal E_\MAX^{(-b_\star)}$, then the energy range has a lower bound of $\mathcal E_\MAX^{(-b_\star)}$, but the allowed region of $\Delta\mathcal E$ remains the same.

The first set of conditions in the third case are when $\phi_1 = \pi/2$ and $\phi_2$ is between zero and $\pi/2$.
This requires that the energy falls between $\mathcal E_\cut^{(\pm b_\star)}$, so that the incomplete elliptic integral with $\phi_2$ should be evaluated.
The second set of conditions correspond to the permitted energies and radii (for $r_2 > r_{\mathcal J}$) when $\mathcal E_\cut^{(+b_\star)} \geq \mathcal E_\MAX^{(-b_\star)}$, where $r_2$ satisfies $r_2 \leq r_{\mathcal J} + 2b_\star$.
We must also ensure that $\mathcal E_\cut^{(-b_\star)} \leq \mathcal E_\MAX^{(-b_\star)}$.
The latter constraint requires $r_2 \geq r_{\mathcal J} - 2b_\star$, which is weaker than the assumption of $r_2 > r_{\mathcal J}$. 
This restricts the energies to be between $\mathcal E_\cut^{(-b_\star)}$ and $\mathcal E_\MAX^{(-b_\star)}$.
The condition $r_2 \leq r_{\mathcal J} + 2b_\star$ now gives a lower bound on the scattering energy $\Delta \mathcal E$.

The fourth case is similar is when both $\phi_1$ and $\phi_2$ are between zero and $\pi/2$.
The energy satisfies $\mathcal E \geq \mathcal E^{(-b_\star)}_\MAX$ in this case, and the analysis of the range of scattering energies and orbital separations is similar to that in the second and third cases.
The fifth case is the special case of $r_2 = r_{\mathcal J}$.
At this point, $\mathcal E'_\MAX$ vanishes and the integral reduces to a square root times the difference of the angles $\alpha_2$ and $\alpha_1$.
Because $\mathcal E'_\MAX(r_{\mathcal J}) = 0$, the upper limit of the energy range reduces to $\mathcal E_\MAX(r_2)$ and the lower limit reduces to $\mathcal E_\cut^{(-b_\star)} = (1-2b_\star/r_2) \mathcal E_\MAX(r_2)$.
No further restrictions on the scattering energies arise at this particular value.

The sixth case treats when $r_2$ is smaller than $r_{\mathcal J}$ and $\phi_1 = 0$, whereas $\phi_2$ is between zero and $\pi/2$.
As noted in Eq.~\eqref{eq:r_min} the energy is bounded below by $\mathcal E_{2,\mathcal J}$.
In this region of $r_2$, one can show that $\mathcal E_{2,\mathcal J}$ is larger than $\mathcal E_\MAX^{(-b_\star)}$, so that $\phi_2$ will not be equal to $\pi/2$ and the incomplete elliptic integral involving $\phi_2$ must be evaluated.
One can also determine that when $r_2 < r_{\mathcal J}$, then $\mathcal E_{2,\mathcal J}$ is less than $\mathcal E_\MAX^{(+b_\star)}$ when $r_2$ is also greater than $r_{\mathcal J} -(8/3) b_\star$.
This condition on $r_2$ can be translated into a lower limit on the scattering energy of the larger of $-v_2^2$ and $\Delta \mathcal E^{(8q/3)}_{\mathcal J}$.
This gives rise to the final nontrivial line in Eq.~\eqref{eq:I-int-cases}.

Note that $\Delta \mathcal E_{\mathcal J}^{(2q)}$ is between $-v_2^2$ and $\Delta \mathcal E_\mathrm{up}$ for $r_2$ between $r_{\mathcal J}(1+2 q)$ and $r_{\mathcal J}(1+2 \sqrt q)$. 
Although there is overlap in the regions of $r_2$ in Eq.~\eqref{eq:I-int-cases}, where there is overlap in $r_2$ there is no overlap in the scattering energy transferred $\Delta\mathcal E$, or the energies $\mathcal E$, so that all the cases there correspond to a unique triple of $(r_2, \Delta\mathcal E, \mathcal E)$.

A similar analysis can be performed for the other part of the solution for $\mathcal R_{\mathcal E}(\Delta \mathcal E)$, which is proportional to $\sin\alpha_2 - \sin\alpha_1$. 
However, we do not give the result here.

We conclude this appendix with a discussion of how $\mathcal R_{\mathcal E}(\Delta \mathcal E)$ behaves in the limit that $\mathcal E$ approaches $\mathcal E_\MAX(r_{\mathcal J}) = v_{\mathcal J}^2/2$.
Note that in all but the first two cases (and one part of the third case) in Eq.~\eqref{eq:I-int-cases}, the orbital separation $r_2$ is restricted to be in a region of order $r_{\mathcal J}$ plus or minus order $b_\star$ corrections. 
The corresponding range of energies where the integral $I_{\mathcal E}(\Delta\mathcal E)$ is nonzero spans energies of that are at most $\mathcal E_\MAX(r_{\mathcal J})$, or $(b_\star/r_{\mathcal J}) \mathcal E_\MAX(r_{\mathcal J})$ smaller than this maximum energy.
For these energies and radii, the modulus parameter $m$ is of order $b_\star/r_{\mathcal J}$ (namely, it is small), so to good approximation one can expand the elliptic integrals in the small $m$ limit where $E(\phi_i,m) \approx \phi_1 + O(m)$. 
Thus, it is a good approximation to use the expression for $I_{\mathcal E}(\Delta\mathcal E)$ when $r_2 = r_{\mathcal J}$---namely,
$\sqrt{2[\mathcal E_\MAX(r_2) - \mathcal E]} (\alpha_2 - \alpha_1)$---for any $r_2 = r_{\mathcal J} + O(b_\star)$ and energy $\mathcal E = \mathcal E_\MAX(r_2)[1-O(b_\star/r_{\mathcal J})]$.

In this approximation, it is possible to show that as $\mathcal E$ approaches $\mathcal E_\MAX(r_{\mathcal J})$, the angle $\alpha_1 = \cos^{-1}[(r_+^2-r_2^2)/(2r_2 b_\star)]$ is given by $\pi/2 - (r_+^2-r_2^2)/(2r_2 b_\star)$.
For $r_2 < r_{\mathcal J}$, the difference $\alpha_2 - \alpha_1$ is proportional to $(r_+ - r_-)/b_\star$, which is proportional to $\sqrt{1-2\mathcal E/v_{\mathcal J}^2}$.
When this is multiplied by the factor of $\sqrt{2[\mathcal E_\MAX(r_2) - \mathcal E]}$, it cancels with the corresponding term $1 - \mathcal E/\mathcal E_\MAX(r_{\mathcal J})$ that appears in $g(\mathcal E)$ in the denominator of the expression for $\mathcal R_{\mathcal E}(\Delta \mathcal E)$.
Note however, that $r_+ - r_-$ is smaller than $b_\star$ just for energies greater than $\mathcal E_\MAX(r_{\mathcal J})[1-b_\star^2/(2r_{\mathcal J})^2]$: namely, those that are within an $O(b_\star/r_{\mathcal J})^2$ difference from the maximum possible energy of dark-matter particles.
While the differential rate coefficient $\mathcal R_{\mathcal E}(\Delta \mathcal E)$ will remain finite, for energies that scale as $\mathcal E_\MAX(r_{\mathcal J})[1-O(b_\star/r_{\mathcal J})]$ the rate will be enhanced by a factor of $\sqrt{r_{\mathcal J}/b_\star}$ from energies not near $\mathcal E_\MAX(r_{\mathcal J})$.
Thus, scattering becomes relatively more efficient for these energies.

The expression for $I_{\mathcal E}(\Delta \mathcal E)$ in Eq.~\eqref{eq:I-int-cases} was obtained by solving to linear order in $b_\star$ for the regions of $r_2$, $\mathcal E$ and $\Delta\mathcal E$ consistent with when the angles $\phi_1$ and $\phi_2$ are either zero, $\pi/2$, or between the two values.
Terms of order $b_\star^2$ were neglected in this process.
The discussion in the previous two paragraphs, however, showed that it is important to consider order $b_\star^2$ corrections to ensure that the integral remains finite as $\mathcal E$ approaches $\mathcal E_\MAX(r_{\mathcal J})$ for $r_2$ near $r_{\mathcal J}$.
Thus, the expressions in Eq.~\eqref{eq:I-int-cases} that are accurate to linear order in $b_\star$ will not be sufficient to capture the behavior of the differential rate coefficient as $r_2$ approaches $r_{\mathcal J}$ and $\mathcal E$ approaches $\mathcal E_\MAX(r_{\mathcal J})$.

To have the benefit of the speed up when $r_2$ is greater than $r_{\mathcal J} + O(b_\star)$ but still have the accuracy and correct limiting behavior when $r_2$ is near $r_{\mathcal J}$ and $\mathcal E$ approaches $\mathcal E_\MAX(r_{\mathcal J})$, we use the expressions in the first three lines of Eq.~\eqref{eq:I-int-cases} for $r_2 \geq r_{\mathcal J}(1+4\sqrt{q})$ and the appropriate ranges of $\mathcal E$ and $\Delta \mathcal E$, and the full expression in Eq.~\eqref{eq:I-integral} for $r_2$ smaller than $r_{\mathcal J}(1+4\sqrt{q})$.
This gives
\begin{widetext}
    \begin{equation} \label{eq:I-int-cases2}
        I_{\mathcal E}(\Delta\mathcal E) = 
        \begin{cases}
          4 \sqrt{\dfrac{b_\star|\mathcal E'_\MAX(r_2)|}{m} } E\left( m \right)   & \mbox{for} \ \ r_2 \geq r_{\mathcal J}(1 + 4 \sqrt q), \ \ -v_2^2 \leq \Delta \mathcal E \leq \MIN(\Delta \mathcal E_\mathrm{up},\Delta \mathcal E_{\mathcal J}^{(4q)}) , \\ 
          & \phantom{for } \mathcal E_\cut^{(+b_\star)}(r_2) \leq \mathcal E \leq \mathcal E_\MAX^{(-b_\star)}(r_2) , \\
          4 \sqrt{\dfrac{b_\star|\mathcal E'_\MAX(r_2)|}{m} } E\left(\phi_1, m \right)   & \mbox{for} \ \ r_2 \geq r_{\mathcal J}(1 + 4 \sqrt q), \ \ -v_2^2 \leq \Delta \mathcal E \leq \MIN(\Delta \mathcal E_\mathrm{up},\Delta \mathcal E_{\mathcal J}^{(4q)}) , \\
           & \phantom{for } \mathcal E_\MAX^{(-b_\star)}(r_2) < \mathcal E \leq \MIN[\mathcal E_\MAX^{(+b_\star)}(r_2), \mathcal E_\MAX(r_\mathcal J)], \\ 
          4 \sqrt{\dfrac{b_\star|\mathcal E'_\MAX(r_2)|}{m} } [E\left( m \right) - E(\phi_2,m) ]   & \mbox{for} \ \ r_2 \geq r_{\mathcal J}(1 + 4 \sqrt q), \ \ -v_2^2 \leq \Delta \mathcal E \leq \MIN(\Delta \mathcal E_\mathrm{up},\Delta \mathcal E_{\mathcal J}^{(4q)}) , \\ 
          & \phantom{for } \mathcal E_\cut^{(-b_\star)}(r_2) \leq \mathcal E < \mathcal E_\cut^{(+b_\star)}(r_2) , \\
           4 \sqrt{\dfrac{b_\star|\mathcal E'_\MAX(r_2)|}{m} } [E\left(\phi_1, m \right) - E(\phi_2, m) ]   & \mbox{for} \ \ r_{\mathcal J} < r_2 \leq r_{\mathcal J}(1 + 4\sqrt{q}), \ \ \MAX(-v_2^2, \Delta \mathcal E_{\mathcal J}^{(4q)}) \leq \Delta \mathcal E \leq \Delta \mathcal E_\mathrm{up} , \\
           & \phantom{for } \mathcal E_\cut^{(-b_\star)}(r_2) \leq \mathcal E \leq \MIN[\mathcal E_\MAX^{(+b_\star)}(r_2), \mathcal E_\MAX(r_\mathcal J)] , \\ 
          \sqrt{2[\mathcal E_\MAX(r_2) - \mathcal E]} \, (\alpha_2-\alpha_1) & \mbox{for} \ \ r_2 = r_{\mathcal J}, \ \ -v_2^2 \leq \Delta \mathcal E \leq \Delta \mathcal E_\mathrm{up} , \\
          & \phantom{for } \mathcal E_\cut^{(-b_\star)}(r_2) \leq \mathcal E \leq \mathcal E_\MAX(r_2) , \\
          4 \sqrt{\dfrac{b_\star|\mathcal E'_\MAX(r_2)|}{m} } [E\left(\phi_2, m \right)-E\left(\phi_1, m \right)] & \mbox{for} \ \ r_{\mathcal J}(1 - 4\sqrt q) \leq r_2 < r_{\mathcal J}, \ \ \MAX(-v_2^2, \Delta \mathcal E_{\mathcal J}^{(4q)}) \leq \Delta \mathcal E \leq \Delta \mathcal E_\mathrm{up} , \\
          & \phantom{for } \mathcal E_\MAX^{(-b_\star)}(r_2) \leq \mathcal E \leq \MIN[\mathcal E_\MAX^{(+b_\star)}(r_2), \mathcal E_\MAX(r_\mathcal J)], \\
          0 & \mbox{otherwise} .
        \end{cases}
    \end{equation}
\end{widetext}

Similar considerations can show that the part of the differential rate coefficient proportional to $\sin\alpha_2 -\sin\alpha_1$ remains finite as $r_2$ approaches $r_{\mathcal J}$ and $\mathcal E$ approaches $\mathcal E_\MAX(r_{\mathcal J})$.
We use the expression in the second line of Eq.~\eqref{eq:calR-final} for all permitted values of $r_2$, $\mathcal E$ and $\Delta \mathcal E$, rather than consider different cases as in Eq.~\eqref{eq:I-int-cases2}.

\bibliography{Refs}

\end{document}